\documentclass{iopart}

\usepackage{graphicx}
\usepackage{epsfig}
\usepackage{color}
\usepackage{bm}
\usepackage{amssymb}
\usepackage{mathrsfs}
\usepackage{dcolumn}
\usepackage{iopams}
\usepackage{setstack}
\usepackage{pifont}
\usepackage[normalem]{ulem}
\usepackage{type1cm}
\usepackage{tabularx}
\usepackage{subfigure}
\usepackage{hyperref}

\newcommand{\beq}{\begin{equation}}
\newcommand{\eeq}{\end{equation}}
\newcommand{\bea}{\begin{eqnarray}}
\newcommand{\eea}{\end{eqnarray}}

\newcommand{\nn}{\nonumber} 

\begin{document}

%%%%%%%%%%%%%%%%%%%%%%%%%%%%%%%%%%%%%%%%%%%%%%%%%%%%%%%%%
%

{\vbox{
\hbox{JLAB-THY-14-1901}}
{\vbox{
\hbox{INT-PUB-14-015}
}}

\title{Nuclear Reactions from Lattice QCD}

\author{Ra\'ul A. Brice\~no$^{1}$, Zohreh Davoudi$^{2,3}$, Thomas C. Luu$^{4}$}

\address{$^1$Thomas Jefferson National Accelerator Facility, 12000 Jefferson Avenue, Newport News, VA 23606, USA}
\address{$^2$ Department of Physics, University of Washington,  Box 351560, Seattle, WA 98195, USA}
\address{$^3$ Institute for Nuclear Theory, Box 351550, Seattle, WA 98195-1550, USA}
\address{$^4$ Institute for Advanced Simulation, Institut f\"ur Kernphysik and J\"ulich Center for Hadron Physics, Forschungszentrum J\"{u}lich, D--52425 J\"{u}lich, Germany}
\eads{\mailto{rbriceno@jlab.org}, \mailto{davoudi@uw.edu}, \mailto{t.luu@fz-juelich.de}}
\date{\today}

\begin{abstract}\\
One of the overarching goals of nuclear physics is to rigorously compute properties of hadronic systems directly from the fundamental theory of strong interactions, Quantum Chromodynamics (QCD). In particular, the hope is to perform reliable calculations of nuclear reactions which will impact our understanding of environments that occur during big bang nucleosynthesis, the evolution of stars and supernovae, and within nuclear reactors and high energy/density facilities. Such calculations, being truly \emph{ab initio}, would include all two-nucleon and three-nucleon (and higher) interactions in a consistent manner.  Currently, lattice QCD provides the only reliable option for performing calculations of some of the low-energy hadronic observables. With the aim of bridging the gap between lattice QCD and nuclear many-body physics, the Institute for Nuclear Theory held a workshop on \emph{Nuclear Reactions from Lattice QCD} on March 2013. In this review article, we report on the topics discussed in this workshop and the path planned to move forward in the upcoming years. \end{abstract}
\tableofcontents
\title[Nuclear Reactions from Lattice QCD]
\maketitle

%%%%%%%%%%%%%%%%%%%%%%%%%%%%%%%%%%%%%%%%%%%%%%%%%%%%%%%%%
\section{Introduction \label{Sec:Introduction}}

A truly \emph{ab initio} method, based on the Standard Model of particles and interactions, gives further insight to hadronic systems that are not experimentally accessible or whose experimental programs are plagued by systematic errors. 
Making reliable predictions with such an \emph{ab initio} method, with controlled uncertainties, are particularly crucial for an accurate description of the evolution of stars, Big Bang/supernovae nucleosynthesis, the composition of neutron stars and the fusion processes in terrestrial high-density facilities.  Studying nuclear reactions from the underlying theory of strong interactions, Quantum Chromo-Dynamics (QCD), requires understanding how forces among nucleons emerge from the fundamental interactions among quarks and gluons. It is known that these interactions are described by a local, non-Abelian, SU(3) gauge theory, within which all hadronic phenomena can, in principle, be predicted once a few input parameters are set to their physical values. These parameters include the masses of the quarks and the strength of the QCD bare coupling constant, or in turn the QCD scale, $\Lambda_{\rm QCD}$.\footnote{Electromagnetism also plays a crucial role in nuclear physics. Its associated quantum field theory, quantum electrodynamics (QED), has a coupling that remains weak at all relevant energy regimes. Its effect can be included in calculations perturbatively for most quantities of interest. For recent developments in directly including QED interactions in LQCD calculations, see Refs. \cite{Duncan:1996xy,Hayakawa:2005eq,Blum:2007cy,Basak:2008na,Blum:2010ym,Portelli:2010yn,Portelli:2012pn,Aoki:2012st,deDivitiis:2013xla,Borsanyi:2013lga,Drury:2013sfa, Borsanyi:2014jba}, and a brief discussion of QED effects in the conclusion.}

%~\cite{Yang:1954ek}

Given the asymptotic freedom of QCD, only high-energy processes can be studied via perturbation theory. At low energies, quarks and gluons form clusters of hadrons -- mesons and baryons. This remarkable feature, along with the running of the QCD coupling towards larger values, prohibits the use of standard perturbative methods and requires techniques that can solve the theory exactly. To date, the only fully predictive non-perturbative method for studying QCD at low energies is lattice QCD (LQCD) which is based on a numerical evaluation of the QCD path integral using Monte Carlo techniques \cite{Wilson:1974sk}. LQCD has already begun to pave the road that connects our theoretical understanding of the fundamental forces with experimental nuclear programs around the world. The progress in LQCD studies of the low-lying spectrum of QCD, hadronic structure and interactions has been significant in recent years. In this review, we only focus on recent developments in LQCD calculations of \emph{multi-hadron} systems as they are directly related to the main objective of this workshop. As will be discussed, a combination of LQCD, effective field theories (EFTs) and nuclear few/many-body calculations will enable first-principle calculations of nuclear reaction cross sections.  

%%%%%%%%%
\begin{figure}[b]
\begin{center}  
\includegraphics[scale=0.695]{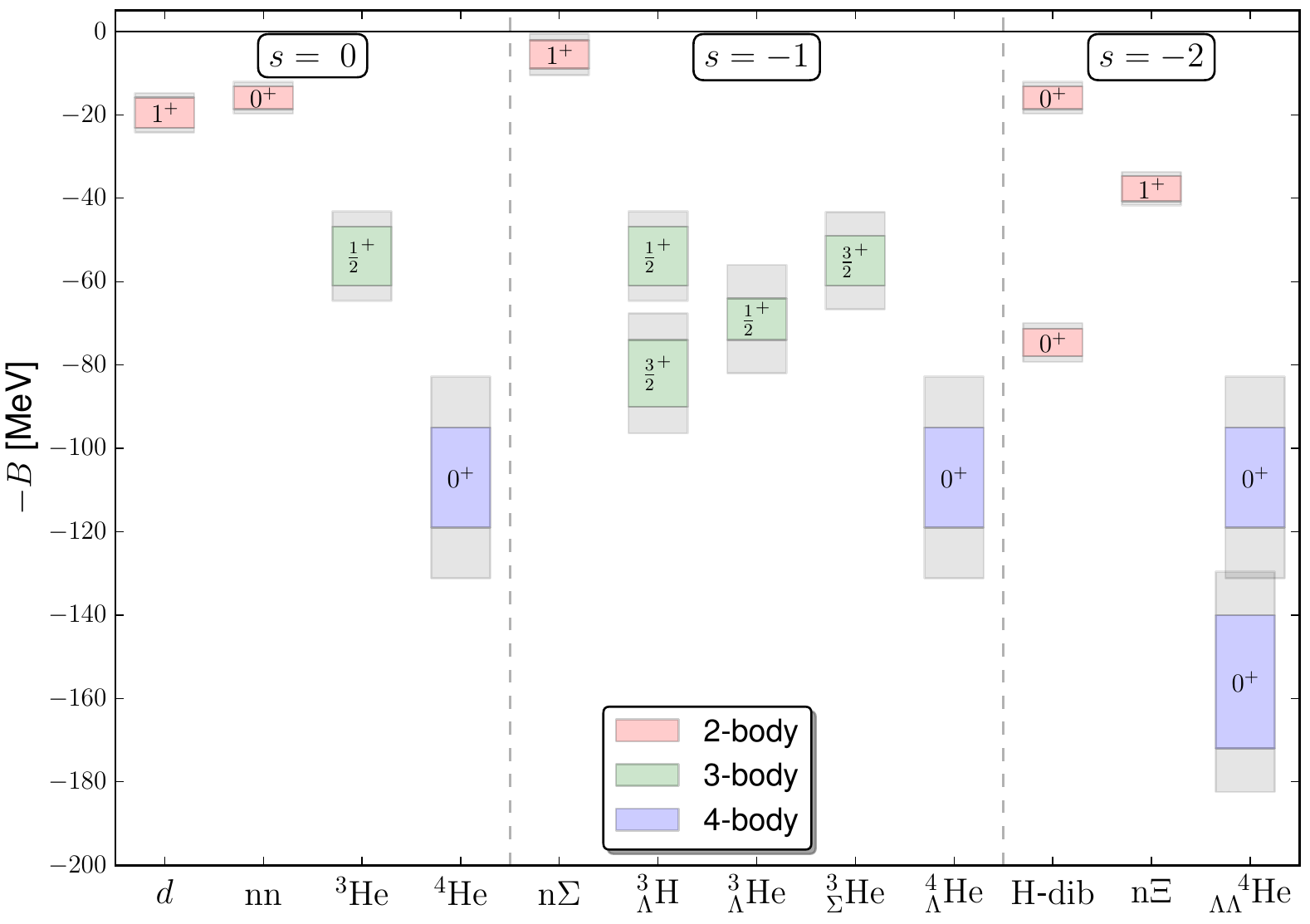}
\caption{The binding energy of light nuclei and hyper nuclei with $A<5$ at the $SU(3)$ symmetric point ($m_\pi\approx 800$ MeV), extracted from LQCD calculations with $n_f = 3$ dynamical light quarks using an isotropic clover discretization of the quark action in three lattice volumes of spatial extent $L \sim 3.4~{\rm fm}$, $4.5~{\rm fm}$ and $6.7~{\rm fm}$, and with a single lattice spacing $b \sim 0.145 ~{\rm fm}$ \cite{Beane:2012vq}. $s$ denotes the strangeness. The $J_{P}$ quantum number of each state is included in the representative boxes of the states. Figure is reproduced with the permission of the NPLQCD collaboration.}
\label{Nuclei}
\end{center}
\end{figure}
%%%%%%%%%

The relevant energy scales in nuclear few-body systems range from a few GeVs for the masses of the light nuclei down to sub-MeVs for their excitation energies, requiring precisions for LQCD calculations that could only be achieved with significant computational resources. LQCD calculations of multi-nucleon systems come with further difficulties that are not present, or are less prominent, in the calculations of single-nucleon and multi-meson systems. The rapid growth in the number of the required \emph{Wick contractions} in the evaluation of nucleonic n-point correlation functions, the exponential degradation of the signal at large times, as well as the need for moderately large volumes due to significant finite-volume (FV) effects in nuclear observable are among the challenges to be overcome. Nonetheless, there has been a great deal of algorithmic and computational progress in the past few years which  has led to major accomplishments in determining several few-nucleon quantities directly from LQCD.

One of the building blocks for performing reliable LQCD calculations is to construct lattice operators that maximize the overlap onto the multi-hadron states with particular quantum numbers (for recent development on this topic see Refs.~\cite{Beane:2005rj, Beane:2006mx, Beane:2006gf, Beane:2007es, Detmold:2008fn, Beane:2009py, Thomas:2011rh, Beane:2011sc, Basak:2005ir, Peardon:2009gh, Dudek:2010wm, Edwards:2011jj, Dudek:2012ag, Dudek:2012gj, Yamazaki:2009ua, Yamazaki:2012hi}). Having constructed the appropriate operators, Wick contractions are performed to build correlation functions. The large number of contractions associated with few-body correlation functions has historically limited the number of hadrons for which LQCD calculations can be performed. This shortcoming was first circumvented for multi-meson systems \cite{Shi:2011mr, Detmold:2012wc, Detmold:2010au} and more recently for nuclear systems~\cite{Doi:2012xd, Detmold:2012eu} and multi-meson systems with a single baryon~\cite{Detmold:2013gua}. In particular, Detmold and Orginos were able to considerably reduce the time required to perform contractions for %the s-shell 
nuclei as large as $^{28}{\rm Si}$~\cite{Detmold:2012eu}-- an algorithm that was subsequently used in a LQCD calculation of multi-nucleon and hyper-nucleon systems by the NPLQCD collaboration \cite{Beane:2012vq}. The bound-state spectra of several light nuclei and hyper nuclei (up to $A<5$ where $A$ is the atomic number) were computed by Beane, \emph{et al.} \cite{Beane:2012vq} at the SU(3) flavor symmetric point with $m_{\pi}=m_{K}\approx 800~{\rm MeV}$ (see Fig. \ref{Nuclei}), and shortly followed by another determination of the binding energies of the light nuclei at a slightly lighter pion mass, $m_{\pi}\approx 500~{\rm MeV}$ by Yamazaki, \emph{et al.} \cite{Yamazaki:2012hi}. These calculations are major accomplishments in truly first-principle multi-nucleon studies and serve as a proof of principle that once the required computational resources become available in the upcoming years, calculations at the physical pion mass and with sufficiently high statistics can be immediately performed.  

\color{black}
It is important to emphasize that for a LQCD result to be directly compared to experiment, appropriate limits must be taken: quantities must be extrapolated to the continuum and infinite-volume limit, and the quark masses must be tuned or extrapolated to their physical values. Of course, this is a common practice for the vast majority of lattice calculations in the single-hadron sector where the productions of multiple hadrons are kinematically forbidden. For these classes of observables, LQCD has proven to be a remarkably powerful tool in reaching precision and accuracy that sometimes has surpassed those of experiments (see e.g., Ref. \cite{Aoki:2013ldr} for a review of LQCD result concerning low-energy particle physics). For observables involving two hadrons or more, the path towards the final answer is still under construction, with valuable progress in the mesonic sector having already been made, see e.g., Refs. \cite{Beane:2007xs, Feng:2009ij, Yamazaki:2004qb, Dudek:2010ew, Beane:2011sc}. This is due to the many challenges associated with studying these systems as listed briefly in the conclusion of this review. Before designing a procedure that leads to extrapolating few-body physical observables to the physical point, one needs to first address the following questions: ``\emph{What do we need to calculate on the lattice?}'' and  ``\emph{How is this LQCD quantity connected to a physical observable?}''. Here we aim to outline some of the answers to these questions. %that have emerged or are beginning to emerge.
% while further progress is needed to complete the gap between the multi-hadron observables and their lattice counterparts.

\color{black}

%%%%%%%%%
\begin{figure}[t]
\begin{center}  
\includegraphics[scale=0.875]{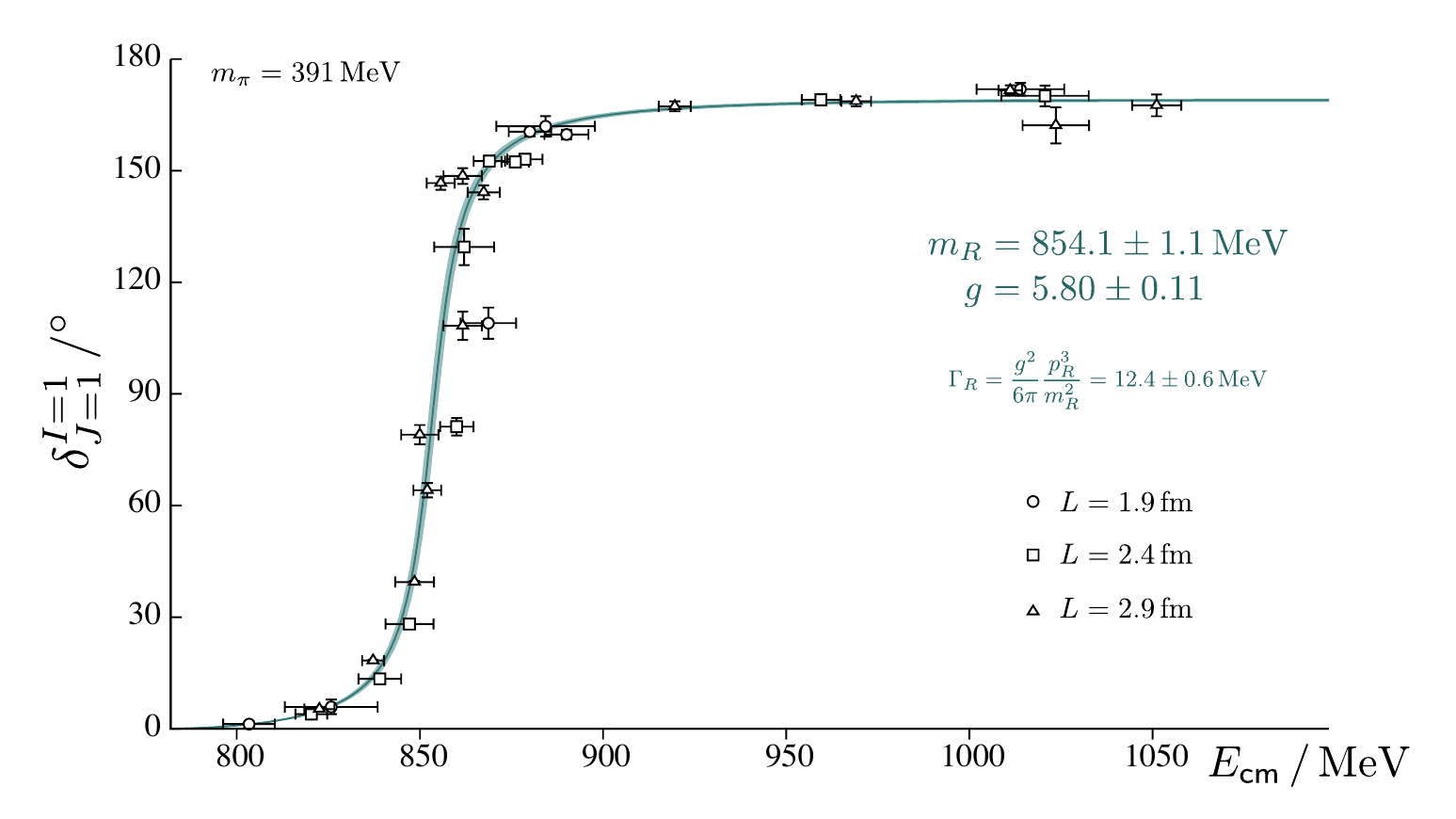}
\caption{$I=1$, $P$-wave $\pi \pi$ elastic scattering phase shift for $m_{\pi}=391~{\rm MeV}$ below the $K\bar{K}$ threshold, featuring the presence of the $\rho$ resonance, obtained by the Hadron Spectrum Collaboration \cite{Dudek:2012xn}. Figure is reproduced with the permission of Jozef Dudek.}
\label{Rho}
\end{center}
\end{figure}
%

% and Ref.~\cite{Guo:2013vsa} for an exploratory numerical calculation of a coupled-channel system in 1 + 1 dimensional lattice model.
The most easily-determined quantities via LQCD are the low-energy spectra. One however needs to take further steps to understand these FV spectra and be able to connect them to physical amplitudes and hadronic interactions. One method that has proven to be successful is the \emph{L\"uscher} method,  which allows one to obtain scattering amplitudes indirectly through the FV spectrum. This method utilizes the calculated energy eigenvalues of the interacting two-particle states in a finite volume to extract the scattering phase shifts, as long as the multi-particle inelastic channels are not kinematically accessible (see Refs. \cite{Luscher:1986pf, Luscher:1990ux} for the original development of this method by M. L\"uscher, and Refs. \cite{Rummukainen:1995vs,  Beane:2003yx, Beane:2003da, Li:2003jn, Detmold:2004qn, Feng:2004ua, Liu:2005kr, Bedaque:2004kc, Christ:2005gi, He:2005ey, Kim:2005gf, Bernard:2008ax,Lage:2009zv, Bour:2011ef, Davoudi:2011md, Fu:2011xz, Leskovec:2012gb, Gockeler:2012yj, Hansen:2012tf, Briceno:2012yi,  Li:2012bi, Guo:2012hv, Ishizuka:2009bx, Bernard:2010fp, Briceno:2013rwa, Briceno:2013lba, Briceno:2013bda, Briceno:2013hya, Li:2014wga} for various generalizations). A recent example of the application of this method by Dudek, \emph{et al.} \cite{Dudek:2012xn} in the two-body sector, as shown in Fig. \ref{Rho}, is the successful extraction of the $\rho$ resonance (albeit yet at unphysical quark masses). Another example in this sector is the extraction of the $S$-wave NN scattering length and effective range by Beane, \emph{et al.} \cite{Beane:2013br} which has enabled them to study fundamental questions regarding the naturalness of the NN interactions and fine tunings with respect to the light-quark masses, as shown in Fig. \ref{a-to-r}. As LQCD calculations are constantly providing the spectra of multi-hadron states with ever-increasing precision~\cite{Dudek:2009qf, Dudek:2010wm, Dudek:2011tt, Edwards:2011jj, Dudek:2011bn, Dudek:2013yja}, it is crucial to be able to extract physical amplitudes for these systems in the same way as is done extensively for two-hadron systems \cite{Li:2007ey, Aoki:2007rd, Beane:2010hg, Beane:2011xf, Beane:2011sc, Beane:2011iw, Beane:2012ey,Yamazaki:2012hi, Beane:2013br,  Pelissier:2011ib,  Lang:2011mn, Pelissier:2012pi, Ozaki:2012ce, Buchoff:2012ja, Dudek:2012xn, Dudek:2012gj, Mohler:2013rwa, Lang:2014tia}. 
\begin{figure}[h]
\begin{center}
\label{a-to-r}
\includegraphics[scale=0.35]{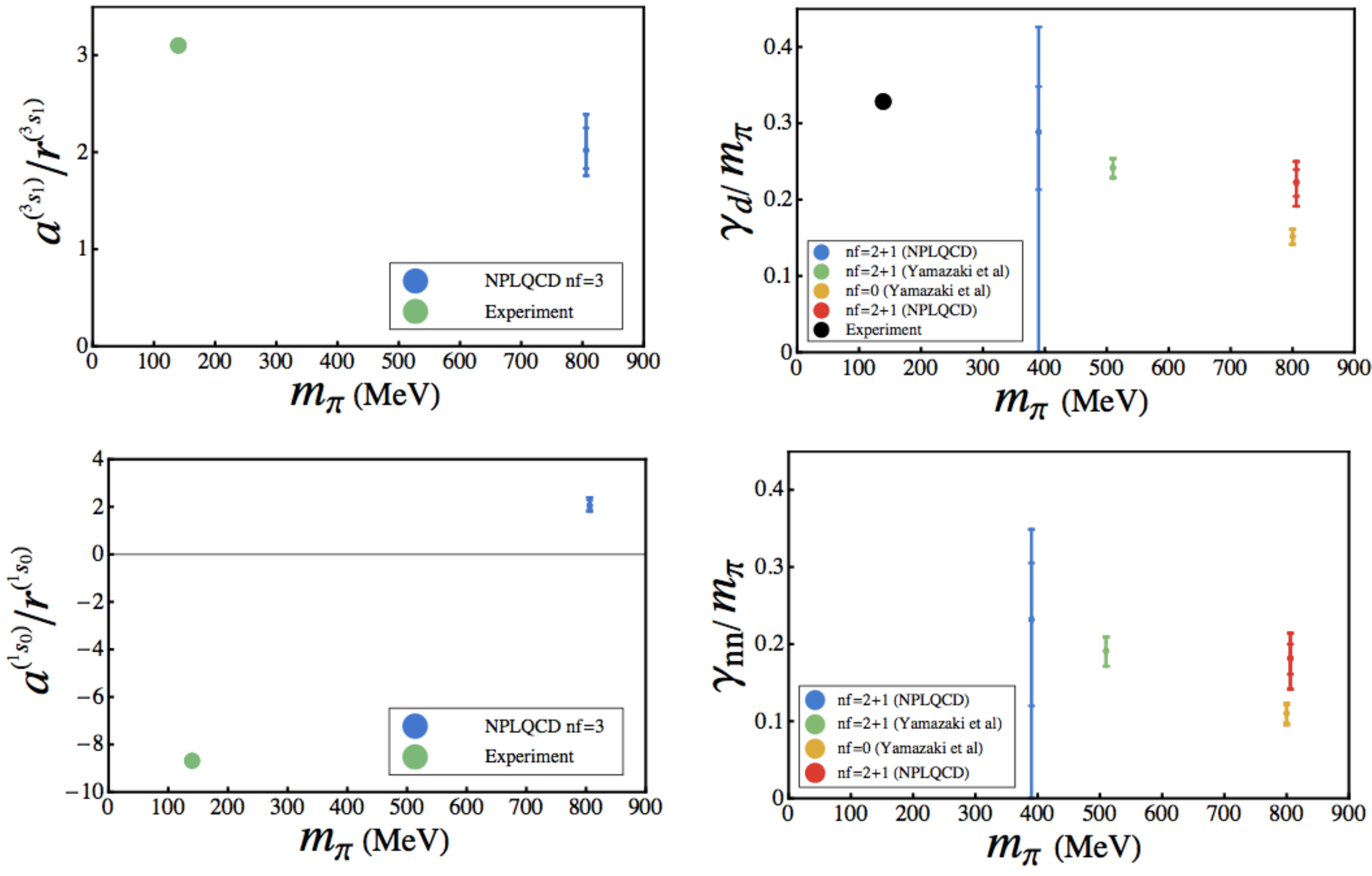}
\caption{The left panel represents the ratio of the two-nucleon scattering length, $a$, to the effective range, $r$, in the ${^3}S_1$ (top) and ${^1}S_0$ (bottom) channels at the physical point as well at the $SU(3)$ symmetric point with $m_{\pi}\approx 800~{\rm MeV}$ \cite{Beane:2013br}. As can be inferred from the plots, the NN interactions remain unnatural over a wide range of pion masses. The right panel represents the plots of the binding energy as a function of pion mass. These indicate that the size of the deuteron and the $nn$ bound state remain large compared with the range of interactions at heavier pion masses. The figure is reproduced with the permission of the NPLQCD collaboration.}
\label{a-to-r}
\end{center}
\end{figure}
In this workshop, the generalization of this formalism to two-nucleon systems~\cite{Briceno:2013lba, Briceno:2013hya, Briceno:2013pda, Briceno:2013rwa} as well as three-particle systems~\cite{Roca:2012rx,  Polejaeva:2012ut,Briceno:2012rv, Hansen:2013dla, Hansen:2014eka} were presented and will be reviewed in Sec. \ref{subsec: FV-formalism}. A complimentary program to the well-stablished L\"uscher formalism is the \textit{potential} method developed and implemented by the HAL QCD Collaboration~\cite{Aoki:2009ji, HALQCD:2012aa, Aoki:2012tk, Murano:2013xxa, Aoki:2013cra}. Reported in the workshop was an extension of this formalism to multi-hadron systems~\cite{Aoki:2013cra}. In this review we briefly comment on this method and report on critical viewpoints concerning the range of validity of potential methods in accessing scattering parameters of multi-(hyper)nucleon systems.

Beyond a direct determination of physical observables, LQCD calculations will provide the input to modern-day few- and many-body nuclear physics calculations, such as no-core shell model (see Refs.~\cite{Zheng:1993qx, Jaqua:1993zz, Zheng:1994zza, Navratil:1996vm, Navratil:1998uf, Navratil:2000gs,Zhan:2004ct,Barrett:2012dr,Barrett:2013nh}) with continuum (NCSMC) \cite{Baroni:2013fe,Baroni:2012su}, coupled-cluster (CC) methods \cite{Hagen:2007hi,Hagen:2008iw}), quantum Monte Carlo (QMC) with local  \cite{Gezerlis:2013ipa, Gezerlis:2014zia, Lynn:2014zia} and non-local \cite{Roggero:2014lga} chiral EFT interactions, shell-model calculations based on chiral NN and 3N interactions \cite{Otsuka:2009cs, Holt:2010yb, Gallant:2012as, Wienholtz:2013nya, Holt:2014aya, Holt:2012fr}, Green's function \cite{Pudliner:1997ck,Wiringa:2000gb,Pieper:2004qw} methods for medium-mass nuclei \cite{Cipollone:2013zma, Soma:2013xha}, in-medium similarity renormalization group method for medium-mass nuclei \cite{Tsukiyama:2010rj, Tsukiyama:2012sm, Hergert:2012nb, Hergert:2013uja, Bogner:2014baa}, and nuclear lattice effective field theories (LEFTs)~\cite{Lee:2004si, Borasoy:2005yc, Borasoy:2006qn}. LQCD inputs aim to reduce those systematic uncertainties of these calculations that are due to the poor knowledge of nuclear few-body forces, or in certain cases to provide the only way to constrain nuclear parameters. Such systematic matching to EFTs would extend the impact of LQCD to heavier nuclei, where a direct LQCD calculation is currently believed to require an unrealistic amount of computational resources. In Sec.~\ref{few_many_body}, we review partly the \emph{ab initio} (continuum and non-continuum) many-body nuclear physics program with regard to nuclear reactions and discuss where LQCD is expected to have an impact. Finally in Sec.~\ref{Sec:ConclusionFutureOutlook} the two-year and five-year outlook of nuclear physics from LQCD will be summarized.

%%%%%%%%%%%%%%%%%%%%%%%%%%%%%%%%%
\section{LQCD and Developments in the Finite-volume Formalism \label{Sec:LQCD}}
%%%%%%%%%%%%%%%%%%%%%%%%%%%%%%%%%%%%%%%%%%%%%%%%%%%%%%%%%%
\noindent 
The progress towards studying nuclear physics quantities directly from LQCD has been a major motivation for this workshop. The hope is that by applying the recently developed numerical and formal tools, and by continuously developing new techniques and overcoming formal challenges, the road towards \emph{direct} determinations of key nuclear reaction cross sections from QCD will be paved in short term. Obviously, the success of this program has been, and will be, crucially dependent upon the availability of ever-increasing large-scale computational resources (both capability and capacity) to the community. However, as discussed in the introduction, without further formal developments that enable interpreting the output of LQCD calculations of multi-nucleon systems, the future of this program remains somewhat obscure. To overcome this roadblock, several groups and individuals have put significant efforts in these types of development, the most recent of which were presented in this workshop. In the following section, after introducing the origin of the problem, we summarize the past and recent progress in tackling this problem.

In LQCD matter fields are evaluated on a discrete set of spacetime points (sites). Due to finite computational resources, the volume of spacetime is truncated to finite extents, represented by $L_x\times L_y\times L_z \times T$ for lattices with spatial extent $L_j$ along the $j^{th}$ Cartesian axis and temporal extent $T$ ($L\equiv L_x= L_y= L_z$ for cubic volumes). The most commonly used boundary conditions (BCs) are periodic BCs imposed on the fields in spatial directions and (anti-)periodic BCs imposed on the (matter) gauge fields in the temporal direction.  \emph{Twisted} BCs have also been proven to be advantageous in LQCD studies of hadronic form factors in the low-momentum transfer region \cite{Tiburzi:2005hg,Jiang:2006gna,Boyle:2007wg,Simula:2007fa,Boyle:2008yd,Aoki:2008gv,Boyle:2012nb, Brandt:2013mb}, and also in FV studies of two-nucleon systems \cite{Bedaque:2004kc, Briceno:2013hya}. To be able to evaluate expectation values in the background of the QCD vacuum (the path integral approach) using a Monte Carlo sampling method, it is necessary to Wick rotate from  Minkowski spacetime to Euclidean spacetime. Consequently lattice correlation functions do not immediately correspond to the physical correlation functions. The connection between these two quantities is nontrivial for some physically interesting cases such as scattering processes. Constructing this connection is the subject of the so-called FV formalism for LQCD.

%%%%%%%%%%%%%%%%%%%%%%%%%%%%%%%%%%%%%%%%%%%%%%%%%%%%%%%%%%
\subsection{Finite-volume formalism for two-hadron systems
\label{subsec: FV-formalism}}
\noindent
Euclidean correlation functions with the reflection-positivity property can be Wick rotated back to the Minkowski spacetime~\cite{Osterwalder:1973dx}. Therefore, if one was able to fully reconstruct the continuum correlation functions from their Euclidean lattice counterparts, such analytic continuation would be formally possible. However, lattice correlation functions are evaluated at a discrete set of spacetime points and are not exact. Although it has been previously pointed out that the Euclidean nature of the calculations imposes challenges on the determination of few-body scattering quantities in the infinite-volume limit (unless at the kinematic threshold) \cite{Maiani:1990ca}, LQCD correlation functions are evaluated in a finite volume, and the scattering amplitudes of the infinite volume can be reconstructed from the FV spectrum. Motivated by the quantum-mechanical problem of a two-body system in a finite volume interacting via a hard spherical potential (see Huang and Yang \cite{Huang:1957im}), L\"uscher derived a non-perturbative relation between the two-body scattering amplitudes and the FV energy eigenvalues for scalar bosons with zero total momentum using a field theoretic approach~\cite{Luscher:1986pf, Luscher:1990ux}. 

The original applications of the L\"uscher formula were mostly focused on an approximated form, where the energy eigenvalue equation, or quantization condition (QC), is expanded in powers of $a/L$. $a$ denotes the two-body scattering length defined  as $-1/a=\lim_{p\rightarrow0} p \cot \delta$, where $\delta$ is the scattering phase shift of the two-particle system and $p$ is the relative momentum of the two particles in the center of mass (c.m.) frame. This is an unnecessary approximation, as recognized by L\"uscher and later revisited in Ref. \cite{Beane:2003da}. By rederiving the S-wave limit of the L\"uscher QC using an EFT approach, Beane, \emph{et al.} \cite{Beane:2003da} pointed out that studies of two-nucleon systems with large scattering lengths do not require large volumes -- volumes whose spatial extents are large compared with the scattering length. In fact the finite range of the nuclear force remains the only relevant scale that governs the range of applicability of this formalism, giving rise to exponential corrections to the L\"uscher formula that scale as $e^{-m_{\pi}L}$ at leading order (LO), with $m_{\pi}$ being the mass of the pion \cite{Luscher:1985dn, Luscher:1986pf, Luscher:1990ux, Bedaque:2006yi, Sato:2007ms}.  Although the exact QC was already presented in L\"uscher's papers, its applicability to the S-wave NN interactions was explicitly proposed in Ref. \cite{Beane:2003da} and shortly followed by the first fully dynamical LQCD calculations of nucleon-nucleon and hyperon-nucleon systems \cite{Beane:2006mx, Beane:2006gf}. 

An important generalization by Rummukainen and Gottlieb, \cite{Rummukainen:1995vs} and later by Kim, \emph{et al.} \cite{Kim:2005gf} and Christ, \emph{et al.} \cite{Christ:2005gi}, extends the L\"uscher formula to scalar systems with a non-zero c.m. momentum. The  significance of such QCs is in providing further kinematic inputs that can better constrain the scattering parameters. The effect of boosting the two-particle system in a finite volume is not only to shift the total energy of the system, but also to make a nontrivial shift in the system's c.m. (interaction) energy as given by the boosted QC. This can be easily understood by noting that the cubic symmetry of the volume is reduced when viewed in the c.m. frame of the moving system. Consequently the spatial extents of the cubic volume will be effectively different for different boost vectors. This also suggests further partial-wave mixings in the QC compared with systems at rest. Such partial-wave mixing is due to the broken rotational symmetry in a finite cubic volume (see discussions following Eq. (\ref{eq:deltaGPBCs})), and presents a source of systematics in the analyses of present LQCD calculations of two-hadron scattering. Recently, by analyzing the LQCD calculations of two-pion systems with several c.m. boosts using the moving frame QC, the $l=0,2$ phase shifts of $I=2$ $\pi\pi$ scattering have been extracted by Dudek, \emph{et al.} \cite{Dudek:2012gj}.

An alternative and equivalent method to boosting is to perform calculations in an asymmetric lattice to obtain extra energy levels~\cite{Li:2003jn, Feng:2004ua}. This however requires generating gauge-field configurations at multiple volumes with several asymmetry parameters, which is computationally costly compared to boosting. The other closely related possibility is changing the boundary conditions on the quark fields such that the hadronic fields effectively gain a non-zero c.m. momentum. This can be done by introducing twisted BCs~\cite{PhysRevLett.7.46, Bedaque:2004kc}. Periodic BCs are a subset of twisted BCs, which in general require that quark fields are proportional to their images up to an overall phase,  
${\psi}(\mathbf{x}+\mathbf{n}{{L}})=e^{i{\bm{\theta}} \cdot \mathbf{n}}{\psi}(\mathbf{x}),$ 
where $0\leq\theta_i<2\pi$ is the twist angle in the $ith$ Cartesian direction. As a result the free FV momenta satisfy $\mathbf{p}=\frac{2\pi}{L}\mathbf{n}+\frac{\bm{\theta}}{L}$,
where $\mathbf{n}$ is an integer triplet. Periodic BCs are recovered when the twist angle is set to zero. As it is evident by dialling the twist angle in the one-body sector, one can in principle access a continuous set of momenta. This is advantageous when performing calculations in a finite volume where spectra are necessarily discretized, as has been explored extensively in the one-body sector~\cite{deDivitiis:2004rf, Sachrajda:2004mi, Tiburzi:2005hg,Jiang:2006gna,Boyle:2007wg,Simula:2007fa,Boyle:2008yd,Aoki:2008gv, Jiang:2008ja, Brandt:2013mb} as well as the two-body sector~\cite{Bedaque:2004kc, Bedaque:2004ax, Bernard:2010fp, Ozaki:2012ce, Briceno:2013hya, Agadjanov:2013wqa, Briceno:2014oea}. In a recent publication, the use of twisted BCs in improving the volume-dependence of the masses of nucleons and the binding energy of the deuteron was investigated \cite{Briceno:2013hya}. Partial twisting, where the sea quarks retain their periodic BCs while the valence quarks are twisted, will induce different FV corrections to the single- and two-nucleon observables compared with full twisting, but these corrections are exponentially suppressed \cite{Bedaque:2004ax, Agadjanov:2013wqa, Briceno:2013hya}. Consequently these BCs can be practically implemented in future LQCD calculations of these systems. For three- and multi-nucleon calculations the possibility of similar volume-effects improvement in the binding energies remains to be investigated.

%%%%%%%%%
\begin{figure}[b]
\begin{center}  
\includegraphics[scale=0.380]{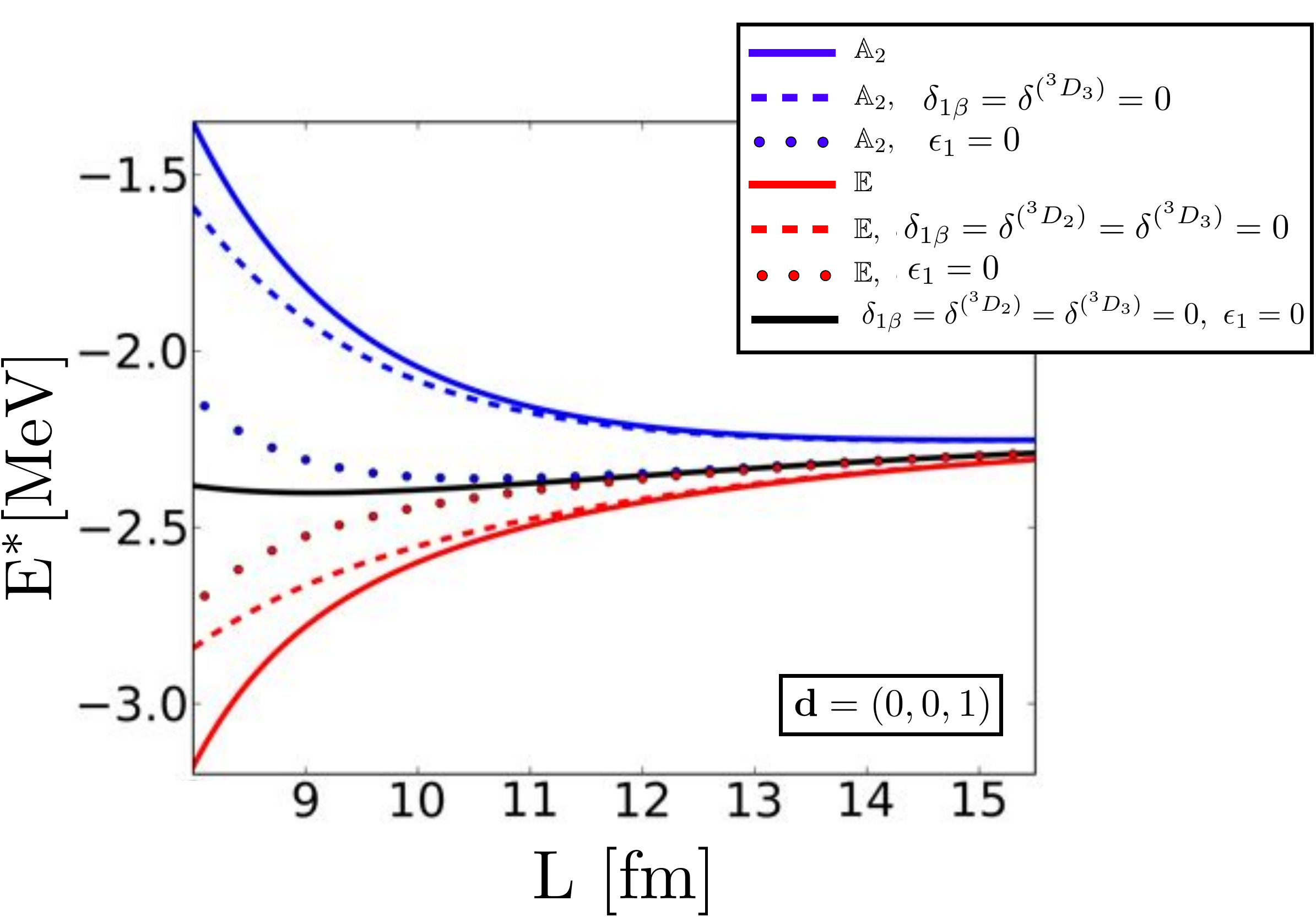}
\caption{The energy of two nucleons in the
 positive-parity
isoscalar channel with $\mathbf{d}=\frac{\bm{P}L}{2\pi}=(0,0,1)$ as a function of the spatial extent of the cubic volume $L$, 
extracted from the two QCs corresponding to the $\mathbb{A}_2$ (red) and $\mathbb{E}$ (blue) irreducible representation (irrep) of the tetragonal symmetry group. 
The scattering parameters, including the phase shifts $\delta$ and the mixing parameter $\epsilon_1$ are obtained by fitting 
different phenomenological analyses of the experimental data, extrapolated to negative values of energies, and are used as inputs to QCs. As is seen, the energy gap between two energy eigenstates is highly sensitive to the nonzero value of the mixing parameter. The FV-induced mixing with the D-wave states is negligible for $L \gtrsim 10~{\rm fm}$ \cite{Briceno:2013bda}.
}
\label{deut_tet}
\end{center}
\end{figure}

The L\"uscher formula for scalar particles must be extended to particles with spin to become applicable to general nucleonic systems. For pion-nucleon scattering, the necessary FV formalism for systems composed of a spin zero and a spin $1/2$ particle was developed by Gockeler \emph{et al.} \cite{Gockeler:2012yj}. Presented in this workshop was the extension of the L\"uscher QC to the two-nucleon systems with arbitrary total spin, isospin and c.m. motion~\cite{Briceno:2013lba}. In particular due to the physical partial-wave mixing in the two-nucleon systems in the spin-triplet channels, restricting to the S-wave QC in analyzing the LQCD calculations is, in general, a poor approximation. In a subsequent paper~\cite{Briceno:2013bda}, the expected energy spectra of the deuteron -- which is a bound state in the coupled $^3{S_1}-{^3{D_1}}$ channels -- are obtained using these newly developed QCs for several c.m. momenta. Different spin orientations of the deuteron with respect to its boost vector in a finite volume can enhance its deformed shape, giving rise to spectral quantities that are highly sensitive to the S-D mixing parameter in the deuteron channel, see Fig.~\ref{deut_tet}. This gives confidence that not only the deuteron binding energy but also the mixing angle can be extracted with high precision in upcoming calculations with physical quark masses.

The L\"uscher formula is not only applicable to coupled partial-wave channels, but can be further generalized to coupled channels in flavor space. This is the first step in extending the L\"uscher methodology above inelastic thresholds. Since the spectrum of QCD contains a wealth of resonances that sit above multi-particle thresholds, analyzing the FV spectra, in particular those of the excited spectra of QCD as produced by various LQCD collaborations (e.g., Refs.~\cite{Dudek:2009qf, Dudek:2010wm, Dudek:2011tt, Edwards:2011jj, Dudek:2011bn, Dudek:2013yja}), requires applying a multi-coupled channel formalism. This necessary step can provide the theoretical guidance for the forthcoming JLab GlueX experiment~\cite{Shepherd:2009zz, Zihlmann:2010zz, Somov:2011zz, Smith:2012ch} as well as other spectroscopy experiments worldwide. The coupled-channel extension of the L\"uscher formula was first derived by He, \emph{et al.} using a quantum mechanical approach \cite{He:2005ey}, by Lage, \emph{et al.} using a NR EFT approach \cite{Lage:2009zv}, by Bernard, \emph{et al.} using a relativistic EFT approach in the c.m. frame \cite{Bernard:2010fp}, and by Hansen, Sharpe \cite{Hansen:2012tf, Hansen:2012bj} and Briceno, Davoudi \cite{Briceno:2012yi} using a relativistic field theory approach for systems with arbitrary momenta.

A recent paper by one of the authors \cite{Briceno:2014oea} reviews all recent formal FV developments in multi-channel two-particle systems and provides the most general energy QC for systems with arbitrary spin, which can be presented in the following form
\begin{eqnarray}
\!\!\!\!\!\!\!\!\!\!
\det~[\mathcal{M}^{-1}+\delta \mathcal{G}^V]~\equiv~{\det}_{\rm{oc}}\left[{\det}_{{lsJM_J}}~[\mathcal{M}^{-1}+\delta \mathcal{G}^V]\right]=0,
\label{eq:QC}
\end{eqnarray}
where the determinant ${\det}_{\rm{oc}}$ is over N open coupled channels and the determinant $\det_{{lsJM_J}}$ is realized in the $|ls,JM_J\rangle$ basis. $l,s$ denote the orbital angular momentum and the total spin of the two-particle systems, respectively, and $J,M_J$ denote the total angular momentum and its projection into the $z$ axis. $\mathcal{M}$ is the c.m. scattering amplitude which is diagonal in the $|J,M_J\rangle$ basis but not necessarily in the $|l,s\rangle$ or the flavor channel basis. The scattering amplitude is more readily defined in terms of the $S$-matrix, which requires introducing a matrix that is diagonal over the N open channels $\mathbb{P}={\rm{diag}}(\sqrt{n_1p_1},\sqrt{n_2p_2},\ldots,\sqrt{n_Np_N})/\sqrt{4\pi E^*}$, where $n_j$ is the symmetry factor for the $j^{th}$ channel and is equal to $1/2$ if the two-particles are identical and 1 otherwise. $E^*$ denotes the c.m. energy of the two-particle system, and $p_j$ is the on-shell relative momentum in the c.m. frame in the $j^{th}$ channel. For a channel composed of two particles with masses $m_{j,1}$ and $m_{j,2}$,
\begin{eqnarray}
\label{momentumcc}
p_j=\left(\frac{E^{*2}}{4}-\frac{(m_{j,1}^2+m_{j,2}^2)}{2}+\frac{
(m_{j,1}^2-m_{j,2}^2)^2}{4E^{*2}}\right)^{1/2}.~~
\end{eqnarray}
The $S$-matrix is diagonal in the total angular momentum basis. Then for a system with total angular momentum $J$, the scattering amplitude $\mathcal{M}$ reads,~\cite{Hansen:2012tf}
\begin{eqnarray}
\qquad \qquad ~~ i\mathcal{M}=\mathbb{P}^{-1}~{(S-\mathbb{I})}~\mathbb{P}^{-1}.
\label{Smatrix}
\end{eqnarray}

Adopting the notation introduced in Refs.~\cite{Li:2003jn, Feng:2004ua, Detmold:2004qn}, let $L$ be the spatial extent of the volume along the z-axis and $\eta_i$ be the asymmetric factor along the $ith$ axis, i.e., $L_x=\eta_xL$ and  $L_y=\eta_yL$. Then the matrix elements of the $\delta \mathcal{G}^V$ matrix, which is a diagonal matrix in the flavor channel basis, for the $j^{th}$ channel can be written as
\begin{eqnarray}
\!\!\!\!\!\!\!\!\!\!\!\!\!\!\!\!\!\!\!\!\!\!\!\!\!\!\!\!\!\!\!\!\!\!\!\!\!\!\!\!\!\!
\left[\delta\mathcal{G}^V_j\right]_{Jm_J,ls;J'M_{J'},l's'}=
\nn\\
\frac{ip_jn_j}{8\pi E^*}\delta_{ss'}\left[\delta_{JJ'}\delta_{M_JM_{J'}}\delta_{ll'} +i\sum_{l'',m''}\frac{(4\pi)^{3/2}}{p^{{l''}+1}_j}c^{\mathbf{d},\bm{\phi}_{j,1},\bm{\phi}_{j,2}}_{l''m''}(p_j^{2};{L};\eta_x,\eta_y) \right.
\nonumber\\
\!\!\!\!\!\!\!\!\!\!\!\!\!\!\!\!\!\!\!\!\!\!\!\!\!\!\!\!\!\!\!\!\!\!
\left .  \times \sum_{m_l,m_{l'},m_{s}}\langle ls,Jm_J|lm_l,sm_{s}\rangle \langle l'm_{l'},sm_{s}|l's,J'M_{J'}\rangle \int d\Omega~Y^*_{ l,m_l}Y^*_{l'',m''}Y_{l',m_{l'}}\right].
\label{eq:deltaGPBCs}
\end{eqnarray}
The FV dependence of the QC, including its dependence on the BCs as well as the shape of the volume is present in the $c^{\mathbf{d},\bm{\phi}_{j,1},\bm{\phi}_{j,2}}_{lm}(p^{2};{L};\eta_x,\eta_y)$ function,
\begin{eqnarray}
\!\!\!\!\!\!\!\!\!\!\!\!\!\!\!\!\!\!\!\!\!\!\!\!\!\!\!\!\!\!\!\!\!\!\!\!\!
c^{\mathbf{d},\bm{\phi}_{j,1},\bm{\phi}_{j,2}}_{lm}(p^{2};{L};\eta_x,\eta_y)
= \frac{\sqrt{4\pi}}{\eta_x\eta_y\gamma {L}^3}\left(\frac{2\pi}{{L}}\right)^{l-2}
\mathcal{Z}^{\mathbf{d},\bm{\phi}_{j,1},\bm{\phi}_{j,2}}_{lm}[1;(p{L}/2\pi)^2;\eta_x,\eta_y],~~
\nn\\
\!\!\!\!
\mathcal{Z}^{\mathbf{d},\bm{\phi}_{j,1},\bm{\phi}_{j,2}}_{lm}[s;x^2;\eta_x,\eta_y]
= \sum_{\mathbf r \in \mathcal{P}_{\mathbf{d};\eta_x,\eta_y}^{\bm{\phi}_1,\bm{\phi}_2;}}
\frac{ |{\bf r}|^l \ Y_{l,m}(\mathbf{r})}{(\mathbf{r}^2-x^2)^s},~~~
\label{Clm}
\end{eqnarray}
where $\gamma=E/E^*$ with $E$ being the total energy of the two particles in the lab (lattice) frame. The sum is performed over $\mathcal{P}_{\mathbf{d};\eta_x,\eta_y}^{\bm{\phi}_1,\bm{\phi}_2}$ where
\begin{eqnarray}
\mathcal{P}_{\mathbf{d};\eta_x,\eta_y}^{\bm{\phi}_1,\bm{\phi}_2}
=\left\{\mathbf{r}\in \mathbb{R}^3\hspace{.1cm} | \hspace{.1cm}\mathbf{r}={\hat{\gamma}}^{-1}\left(\tilde{\mathbf m}-\alpha_j \tilde{\mathbf d} +\frac{\tilde{\bm{\Delta}}^{(j)}}{2\pi}\right)\right\}.
\end{eqnarray}
Here $\hat{\gamma}^{-1}\mathbf{x}\equiv{\gamma}^{-1}\mathbf{x}_{||}+\mathbf{x}_{\perp}$, with $\mathbf{x}_{||}~(\mathbf{x}_{\perp})$ denoting the component of $\mathbf{x}$ that is parallel (perpendicular) to the total momentum $\mathbf{P}$. Tilde over vectors is defined as $\tilde{\bm{\chi}}=(\chi_x/\eta_x,\chi_y/\eta_y,\chi_z)$.  The boost vector $\mathbf d$ is defined to be equal to ${\mathbf P}L/2\pi$, and $\mathbf{m}$ is an integer triplet. When twisted BCs are imposed on the quark fields, the net twist of each particle is given by $\bm{\phi}_{i,j}$ for $i=1,2$, and the dependence on these phases is manifest in the vector $\bm{\Delta}^{(j)}$ defined as $\bm{\Delta}^{(j)}=-({\alpha}_{j}-\frac{1}{2})(\bm{\phi}_{j,1}+\bm{\phi}_{j,2})+\frac{1}{2}(\bm{\phi}_{j,1}-\bm{\phi}_{j,2})$. $\alpha_j$ is sensitive the mass difference of the particles in channel $j$, and is given by $\alpha_j=\frac{1}{2}\left[1+\frac{m_{j,1}^2-m_{j,2}^2}{E^{*2}}\right]$~\cite{Davoudi:2011md, Fu:2011xz, Leskovec:2012gb}.

As is seen, the $\delta \mathcal{G}^V_j$ matrix is not diagonal in the $l$ and $J,M_J$ basis and mixes different partial-wave channels in the QC. The form of the mixing can be inferred by studying the FV point symmetry group of the calculation, and after a truncation of the QC made in the space of angular momentum, it can be decomposed into different QCs corresponding to the irreps of the particular symmetry group of the problem. Such decompositions have been performed in great detail for different scenarios as can be found in Refs.~\cite{Luscher:1986pf, Luscher:1990ux, Feng:2004ua, Dresselhaus, Thomas:2011rh, Moore:2005dw, Moore:2006ng, Dudek:2010wm, Luu:2011ep, Gockeler:2012yj, Dudek:2012gj}. For example, for systems with zero twist in cubic volumes, and with $\tilde{\mathbf{d}}\rightarrow \mathbf{d}=\{(0,0,0)$, $(0,0,n)$, $(n,n,0)$, $(n,n,n)$, $(n,m,0)$, $(n,n,m)$, $(n,m,p)\}$ -- as well as any cubic rotation of these -- the symmetry point groups are the double cover of the octahedral ($\mathrm{O}^\mathrm{D}_h$)  and the dicyclic groups $\mathrm{Dic}_4$, $\mathrm{Dic}_2$, $\mathrm{Dic}_3$, $\mathrm{C}_4$, $\mathrm{C}_4$ and $\mathrm{C}_2$, respectively~\cite{Moore:2005dw, Moore:2006ng}. Table~\ref{table:irreps} lists the decomposition of the irreps of these groups onto continuum states that have overlap with both half-integer and integer spin systems up to $J=4$, and can be used in performing the reduction of the master QC to different irreps of the calculation.  

\begin{table}[h]
\begin{center}
\subtable[]{ \label{table:irrepsa}
\begin{tabular}{lc}
\hline\hline
$J^P$&$\mathrm{O^D_h}$ \\\hline
$0^\pm$&${A}_1^\pm$\\
 $\frac{1}{2}^\pm$&$G_1^\pm$\\
 $1^\pm$&${T}_1^\pm$\\
$\frac{3}{2}^\pm$&$H^\pm$\\
$2^\pm$&$E^\pm\oplus T_2^\pm$ \\
$\frac{5}{2}^\pm$&$G_2^\pm \oplus H^\pm  $\\
$3^\pm$&$A_2^\pm\oplus T_1^\pm \oplus T_2^\pm$\\
$\frac{7}{2}^\pm$&$ G_1^\pm \oplus G_2^\pm \oplus H^\pm $\\
$4^\pm$&$A_1^\pm\oplus E ^\pm\oplus T_1^\pm \oplus T_2^\pm$\\
\hline\hline
\end{tabular}
} \subtable[]{ \label{table:irrepsb}
\begin{tabular}{lllllllllll}
\hline\hline
$|\lambda|^{\tilde{\eta}}$ & $\mathrm{Dic}_4$ & $\mathrm{Dic}_2$& $\mathrm{Dic}_3$  & $\mathrm{C_{4}}$ & $\mathrm{C_{2}}$\\
  \hline
  $0^+$         & $A_1$            & $A_1$            & $A_1$            & $A$ & $A$ \\
  $0^-$         & $A_2$            & $A_2$            & $A_2$            & $B$ & $A$ \\
  $\frac{1}{2}$ & $E_1$            & $E$            & $E_1$              & $E$ & $2B$ \\
  $1$           & $E_2$            & $B_1	\oplus B_2$			& $E_2$              & $A \oplus B$ & $2A$  \\
  $\frac{3}{2}$ & $E_3$            & $E$             & $B_1 \oplus B_2$  & $E$ & $2B$ \\
  $2$           & $B_1 \oplus B_2$ & $A_1 \oplus A_2$& $E_2$             & $A \oplus B$ & $2A$ \\
  $\frac{5}{2}$ & $E_3$            & $E$            & $E_1$              & $E$ & $2B$ \\
  $3$           & $E_2$            & $B_1 \oplus B_2$ & $A_1 \oplus A_2$ & $A \oplus B$& $2A$  \\
  $\frac{7}{2}$ & $E_1$            & $E$            & $E_1$              & $E$ & $2B$ \\
  $4$           & $A_1 \oplus A_2$ & $A_1 \oplus A_2$  & $E_2$            & $A \oplus B$& $2A$\\
\hline\hline
\end{tabular}
}
\caption{ \label{table:irreps} (a)~The decomposition of the irreps of the SO(3) group up to $J=4$ in terms of the irreps of the $\mathrm{O^D_h}$~\cite{Luscher:1986pf, Luscher:1990ux, Mandula:1983wb, Johnson:1982yq, Basak:2005aq, Dresselhaus}.
(b) The decomposition of the helicity states to the irreps of five of the little groups of $\mathrm{O^D_h}$: $\mathrm{Dic}_4$, $\mathrm{Dic}_2$, $\mathrm{Dic}_3$, $\mathrm{C}_4$ and $\mathrm{C}_2$~\cite{Moore:2005dw, Moore:2006ng, Dudek:2010wm, Thomas:2011rh, Dudek:2012gj}. $\lambda$ labels the helicity of the state and $\tilde{\eta}=\mathcal{P}(-1)^J$, where $\mathcal{P}$ is the parity of the state. 
}
\end{center}
\end{table}
% 

%%%%%%%%%%%%%%%%%%%%%%%%%%%
\begin{figure}[h!] 
\begin{center} 
\subfigure[]{
\includegraphics[scale=0.26]{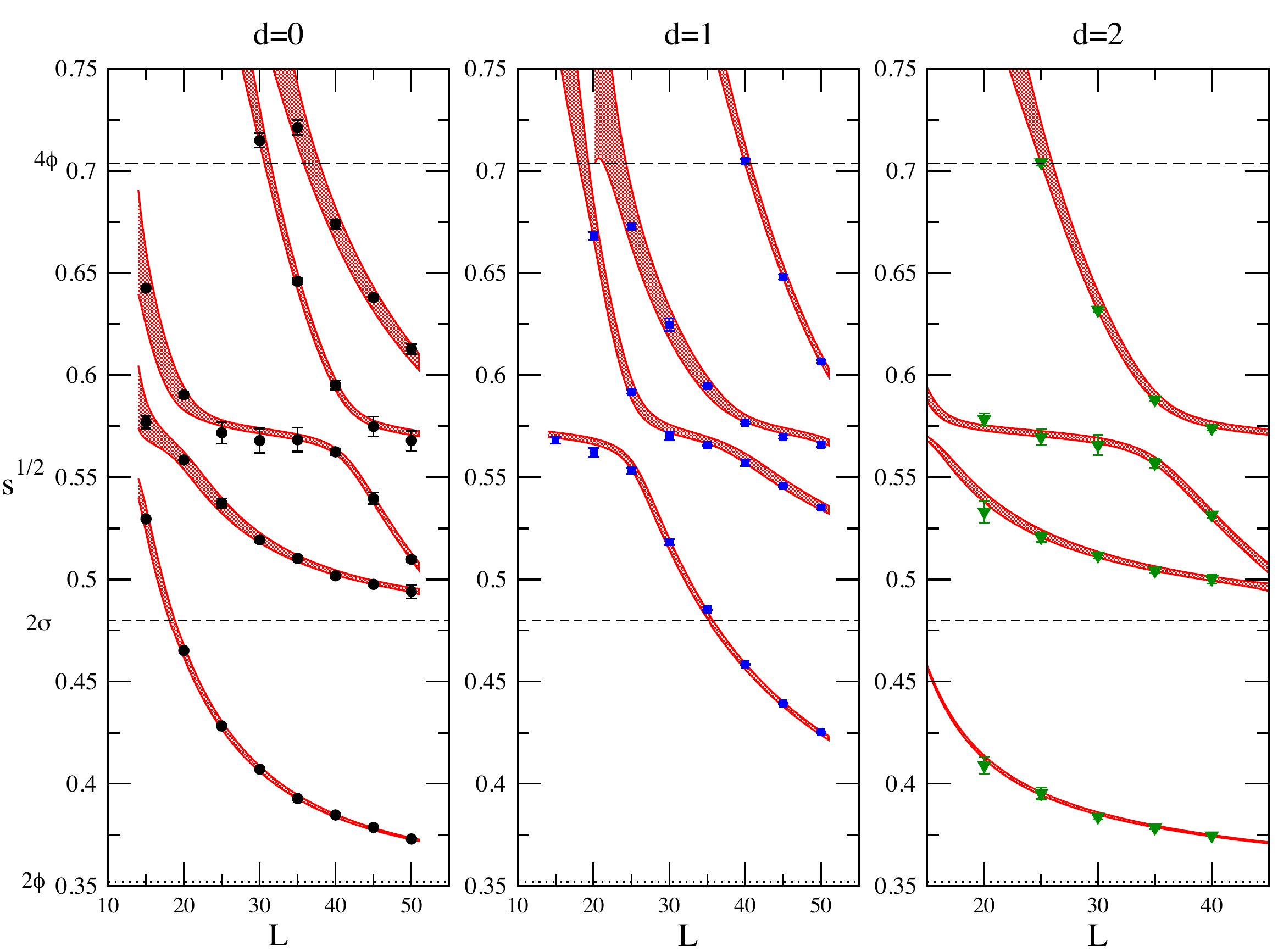}
\label{fig:spec_fits_ds} }
\subfigure[]{
\includegraphics[scale=0.26]{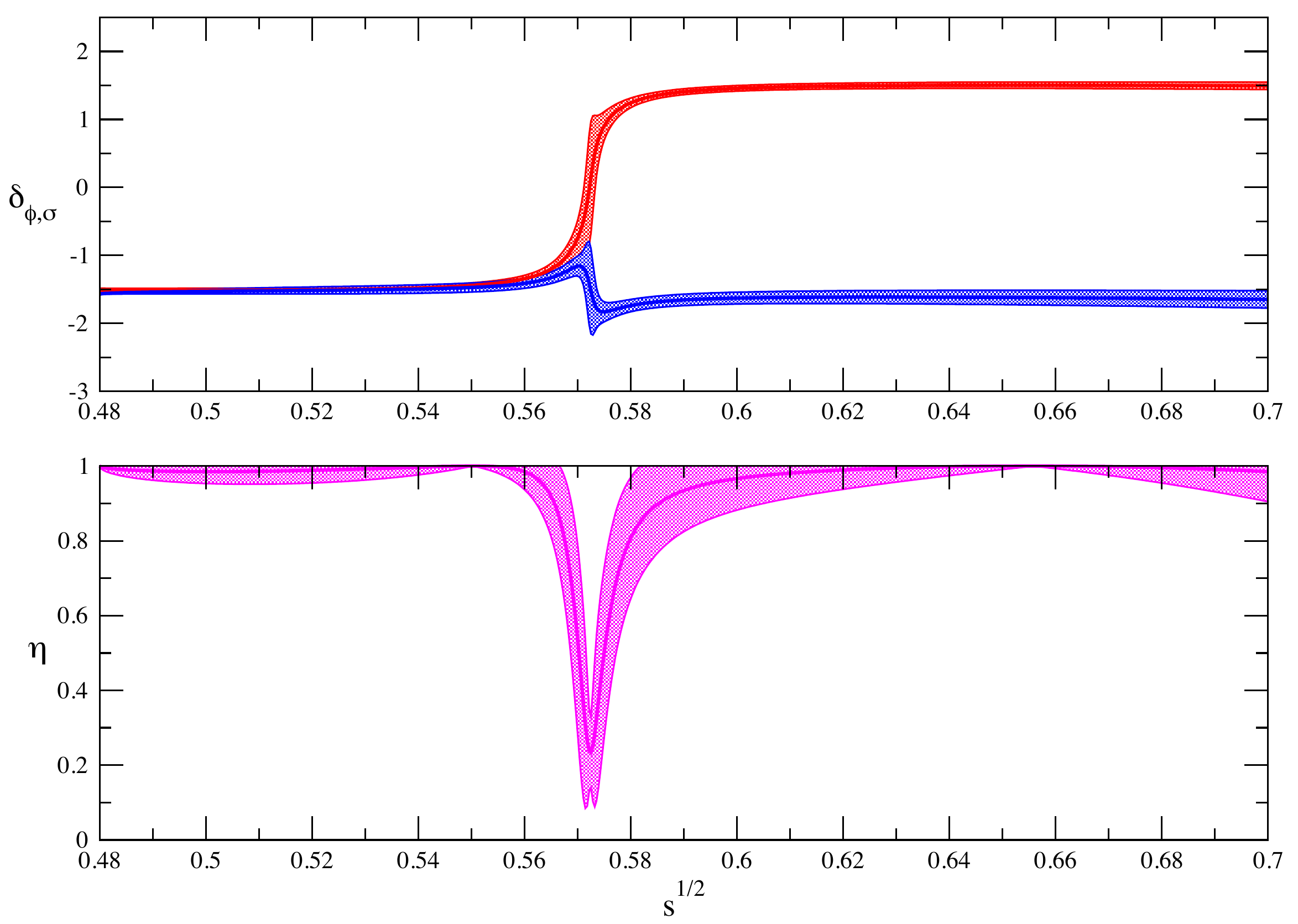}
\label{fig:phaseinelast} }
\caption{(a) The FV spectra of a 1+1 model for a coupled-channel system, $\phi\phi-\sigma\sigma$, as a function of volume. The spectra are calculated for d=0,1,2, where d=LP/$2\pi$ is a scalar boost for a 1+1 system. The red band denotes the fits of the spectrum using a parametrization of the K-matrix below the $4\phi$ threshold. The fit was performed using the 1+1 analogue of Eq.~\ref{eq:QC} for coupled-channel systems, which was first presented by Berkowitz \emph{et al.}~\cite{Berkowitz:2012xq} for unbooted systems and then generalized by Guo \cite{Guo:2013vsa} for systems with arbitrary total momentum. (b) The extracted phase shifts $\delta_\phi$ (red), $\delta_\sigma$ (blue) and mixing parameter $\eta$ (purple) as a function of the total c.m. energy, $\sqrt{s}$ \cite{Guo:2013vsa}. The figures are reproduced with the permission of Peng Guo.
\label{fig:Guo:2013vsa}
} \end{center} \end{figure}
%%%%%%%%%%%%%%%%%%%%%%%%%%%

The FV two-body formalism enjoys a fair level of maturity at this point and further developments in this direction will be focused on implementing it in practice. In particular, given the limited high-performance computational resources, it is the subject of an ongoing investigation to determine how to optimally extract as many scattering parameters as plausible from a few number of low-lying  energy levels, and in many instances in a single lattice volume. This is particularly crucial in the studies of the coupled-channel systems, see e.g., Refs. \cite{Doring:2011vk, Doring:2011ip, Doring:2012eu, Guo:2012hv, Briceno:2013bda, Wu:2014vma}. The first implementation of this formalism was an exploratory numerical calculation of a coupled-channel system in a 1 + 1 dimensional lattice model with two coupled-channels by Guo~\cite{Guo:2013vsa} which includes nontrivial FV and discretization effects. As is depicted in Fig.~\ref{fig:Guo:2013vsa}, this benchmark calculation demonstrates that the scattering phase shifts and mixing angle of a $\phi\phi-\sigma\sigma$ coupled system can be unambiguously determined directly from the FV spectrum, where $\phi$ and $\sigma$ are two scalar particles. The first LQCD study of coupled channels, performed for the $\pi K-K\eta$ $I=1/2$ system at $m_\pi\approx 390$~MeV by Hadron Spectrum Collaboration, was recently published ~\cite{Dudek:2014qha}.

Finally, LQCD multi-nucleon calculations will eventually include QED effects, and as a result the relevant FV formalism must be developed. For efforts in identifying FV QED corrections to the masses of hadrons and the mass splittings between the neutral and charged pseudo-scalar meson and baryon octets, see Refs. \cite{Duncan:1996xy, Blum:2007cy, Hayakawa:2008an, Davoudi:2014qua, Borsanyi:2014jba}.

 \subsection{Finite-volume matrix elements for $1\rightarrow 2$ and $2\rightarrow 2$ processes
\label{subsec:FVMatrixelemts}}

In their seminal paper, Maiani and Testa showed that for energies where two or more particles can go \emph{on-shell} all Euclidean correlation functions not only depend on the on-shell scattering amplitude but also on the amplitudes evaluated on the \emph{off-shell} kinematics~\cite{Maiani:1990ca}. This did not, however, prevent the evaluation of scattering phase shifts from the two-point correlation functions because the FV formalism presented by L\"uscher~\cite{Luscher:1986pf} (and discussed in the previous section) had already been in place a few years earlier. Note that although the correlation function may depend on both the on-shell and off-shell  amplitudes, the spectrum only depends on the on-shell amplitude, see Eq.~(\ref{eq:QC}). Where this observation did have a major impact was in the studies of electroweak current matrix elements for processes involving two or more particles in either the initial and/or final states. Maiani and Testa observed that the off-shell kinematic dependence of the correlation functions exactly vanishes at the kinematic threshold. This is to be expected since at the kinematic threshold the scattering amplitude is momentum independent. 

It was not until 10 years after the original observation by Maiani and Testa that Lellouch and L\"uscher showed how to study $1 \rightarrow 2$ processes in a finite volume above the kinematic threshold~\cite{Lellouch:2000pv}.  Their original motivation was in the context of $K \rightarrow \pi \pi$ weak decays where they found  a relation between FV and infinite-volume matrix elements of the external weak current. This relation was obtained by restricting the final state to be composed of two particles in an S-wave with zero total momentum, and shows that the FV and infinite-volume matrix elements are proportional to each other and the proportionality factor depends on the energy, volume, phase shift and derivative of the phase shift of the final state with respect to energy. In arriving at their result, they used degenerate perturbation theory to force the initial kaon state and the final two-pion state to be exactly degenerate.\footnote{This formalism has proven to be remarkably successful in the studies of $K\rightarrow\pi\pi$ decay, see Refs.~\cite{Blum:2011pu, Blum:2011ng, Blum:2012uk, Boyle:2012ys} for recent calculations.}  This formalism worked particularly well for the $K \rightarrow \pi \pi$ system, but was not easily generalizable to other systems and a more universal approach was needed for more complex systems. 

The original result by Lellouch and L\"uscher was extended to systems with nonzero total momentum in Refs. \cite{Kim:2005gf, Christ:2005gi}, as well as processes where the final state is composed of two coupled channels such as is the case of $D\rightarrow\{\pi\pi,K\bar{K}\}$ by Hansen and Sharpe~\cite{Hansen:2012tf}, as well as for the decay process $\pi^0\rightarrow \gamma\gamma$ by Meyer~\cite{Meyer:2013dxa}. Some limited progress has been made towards generalizing this framework for $2 \rightarrow 2$ processes~\cite{Detmold:2004qn, Meyer:2012wk, Bernard:2012bi, Briceno:2012yi}, but much more work is needed along this direction. In particular, for processes of the form $2\rightarrow 2$ where the external current has both one-body and two-body contributions, no universal expression has been developed~\cite{Detmold:2004qn, Bernard:2012bi, Briceno:2012yi}. Instead what is found is a relationship between the FV matrix elements and the low-energy constants (LECs) of the Lagrangian written for the given process of interest. For processes when there are no one-body contributions, there does seem to be a universal form~\cite{Meyer:2012wk, Briceno:2012yi}.

As discussed in the previous section and manifested by the master equation describing the FV spectrum, Eq.~(\ref{eq:QC}), angular momentum is not a good quantum number in a finite volume. Therefore, it is expected that for sufficiently high energies or seemingly fine-tuned systems such as the deuteron~\cite{Briceno:2013bda}, this admixture would lead to large corrections to the formalism discussed above. Additionally, in general the energy and momentum injected by the external current is not vanishingly small. Furthermore, there are many physical systems of interest where the final two-particle state is not in an S-wave but rather in a non-zero  angular momentum configuration. One example that demonstrates the need to extend this formalism is a system involving heavy quark decays, e.g., $B^0 \rightarrow K^*\ell^+\ell^-\rightarrow \pi K\ell^+\ell^-$. This is a particularly interesting process as it has been speculated to be a venue for observing new physics~\cite{Descotes-Genon:2013wba, Altmannshofer:2013foa, Bobeth:2012vn, vanDyk:2013uaa, Hambrock:2013zya, Beaujean:2013soa, Lyon:2014hpa, Lyon:2013gba}. This has led LQCD efforts in studying form factors for $B^0 \rightarrow h_1 h_2$, where $h_1$ and $h_2$ label generic hadrons, using the quenched approximation~\cite{Bowler:1993rz,Bernard:1993yt,Burford:1995fc,Abada:2002ie,Bowler:2004zb,Becirevic:2006nm,Abada:1995fa} and most recently using dynamical gauge configurations~\cite{Horgan:2013hoa, Horgan:2013pva}.  The $K^*(892)$ is a P-wave resonance which may be stable for heavy pion masses but is unstable for moderately light pion masses around 300-400~MeV~\cite{Horgan:2013hoa, Horgan:2013pva}. For such light pion masses, no calculation has been able to properly account for this significant subtlety. %, in part due to the fact that the formalism was not previously in place. 
Furthermore, a considerable source of systematics for heavy meson decay processes arises from long distance effects. Of these, probably the most important is the production of intermediate charmonium resonances, which has been proven to be a necessary piece in order to be able to describe experimental transition rates~\cite{Lyon:2014hpa}. Except for some special cases, such as $K_L-K_S$ mass difference at heavy pion masses~\cite{Christ:2012se, Bai:2014cva}, the determination of long-distance effects to weak decay processes from LQCD still remains a challenge.

Most recently, there have been two relevant works for studies of $1\rightarrow 2$ transition amplitudes. The first work is by Agadjanov, \emph{et al.}~\cite{Agadjanov:2014kha} where the process $\gamma N\rightarrow N\pi$ in the $\Delta$ channel is studied. To allow for the insertion of energy and momentum by the current, but to minimize the FV-induced mixing at the same time, the $\pi N$ system is assumed to be at rest, as in that scenario the odd and even partial waves do not mix. Furthermore, the contributions from the $P_{31}$ wave is neglected in this formalism. With these simplifications, one can in fact determine the infinite-volume transition form factor of this process from the corresponding FV matrix element. Although the authors focus their attention to $\gamma N\rightarrow N\pi$ in the $\Delta$ channel, this formalism will also impact LQCD calculations of parity violation in the nuclear sector. In particular, this formalism can be applied to obtain the $N\rightarrow N\pi$ transition amplitude with a $P$-wave final state as calculated via LQCD by Wasem in Ref.~\cite{Wasem:2011zz}.

The other recent work by one of the authors, in collaboration with Hansen and Walker-Loud, demonstrates how to circumvent limitations associated with partial-wave mixing and coupled-channel systems for such transition form factors in the scalar sector~\cite{Briceno:2014uqa}. 
Following the work by Thomas, \emph{et al.}~\cite{Thomas:2011rh}, the external currents are allowed to inject arbitrary energy and momentum while being correctly projected onto the irrep of their corresponding little groups. 
The QC found in Ref. \cite{Briceno:2014uqa} for the FV matrix elements of currents involving $1\rightarrow 2$ processes will be applicable to a wide range of systems, such as $B\rightarrow\pi K$, where S- and P-wave can mix in flight, as well as for $\pi\gamma\rightarrow\pi\pi$. Furthermore, this formalism allows one to study processes where the final state is not only an admixture of angular momentum states but also when more than one channel is present.  This is necessary for future studies of $B\rightarrow K^*$ form factors, for example, where the final state will be an admixture of $\pi K$ and $ K\eta$. 
The extension of this formalism to nuclear systems and properly understanding the subtleties associated with the partial-wave mixing will be crucial in studies of nucleonic matrix elements as described in Sec. \ref{Matrix elements}.

%%%%%%%%%%%%%%%%%%%%%%%%%%%%%%%%%%%%%%%%%%
\subsection{Finite-volume formalism for multi-hadron systems
\label{subsec: FV-formalism}}
\noindent
A major development in the FV formalism for lattice nuclear calculations will be the extension of the two-body formalism to three-(multi-) particle systems. The main drive  to put such a formalism in place from a nuclear physics perspective is to be able to extract, from first-principle lattice calculations, the three-(multi-) nucleon forces. As will be discussed in detail in the following section, many-body nuclear calculations are awaiting tighter constraints on the multi-nucleon force parameters to be provided by LQCD calculations. However, little is yet known about multi-nucleon interactions from first-principle calculations. To be able to interpret multi-nucleon spectra (see Fig. \ref{Nuclei}) and make any conclusion about the multi-nucleon interactions, a multi-particle counterpart of the L\"uscher formula has to be developed. This, in principle, can be also directly applied to multi-particle reaction processes once the connection between the scattering amplitudes and the FV spectra is known. The impact of such a formalism on the determination of the excited spectrum of QCD will be immediate too, giving further insight into, e.g., the nature of the Roper resonance, as well as many more resonances in the proximity of multi-particle thresholds. Given the importance of this type of formalism, several talks and extensive discussions were devoted to the subject in this workshop. 
To date all of the formalism developed so far within this problem have been constrained to the multi-mesonic sector. However, the multi-meson problem includes the main formal challenges of multi-baryon systems, but without the complications regarding the spin structure of baryons. 

One approach in obtaining a FV three-particle QC (below the four-particle thresholds) is to take advantage of the non-relativistic \emph{dimer} formalism \cite{Kaplan:1996nv, Beane:2000fi, Bedaque:1997qi, Bedaque:1998mb, Bedaque:1998kg, Bedaque:1998km, Gabbiani:1999yv, Bedaque:1999vb, Bedaque:1999ve, Bedaque:2000ft} to effectively reduce the three-particle problem to a 2+1 problem as is described in Ref. \cite{Briceno:2012rv}. The dimer field sums up the two-body interactions to all orders. For energies below the four-particle threshold, the two-body and three-body interactions can be described by contact interactions.\footnote{When interested in the nuclear analogue of this problem, parametrizing the interactions by contact terms only, will restrict the applicability of the formalism below the \emph{t-channel cut}, which for the two-nucleon systems is $E\sim 5~$MeV. This is an unnecessary approximation that can be easily circumvented by simply parametrizing interactions by the Bethe-Salpeter kernel that encodes all the kinematical and symmetric properties of the system of interest.}  The three-point correlation function, $C_{L}(E,\bm{P})$ -- with $(E,\bm{P})$ being the total energy and momentum of the system -- is then formed by a geometric expansion of the three-body kernels where the three momenta flowing in the intermediate loops are only allowed to be quantized due the periodic BCs,
\begin{eqnarray}
\int \frac{d^4k}{(2 \pi)^4} \rightarrow \frac{1}{L^3}\sum_{\bm{k} \in \frac{2\pi \bm{n}}{L}}\int \frac{dk^0}{(2\pi)} ~~ {\rm with} ~~ \bm{n} \in \mathbb{Z}^3.
\end{eqnarray}
\begin{figure}[h!]
\begin{center}
\subfigure[]{
\includegraphics[scale=0.45]{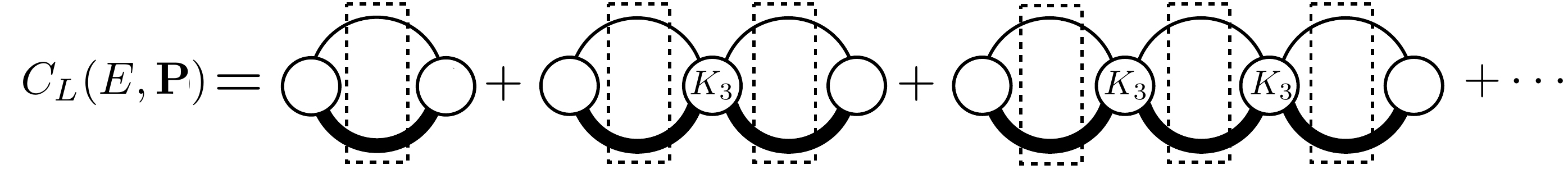}}
\subfigure[]{
\includegraphics[scale=0.285]{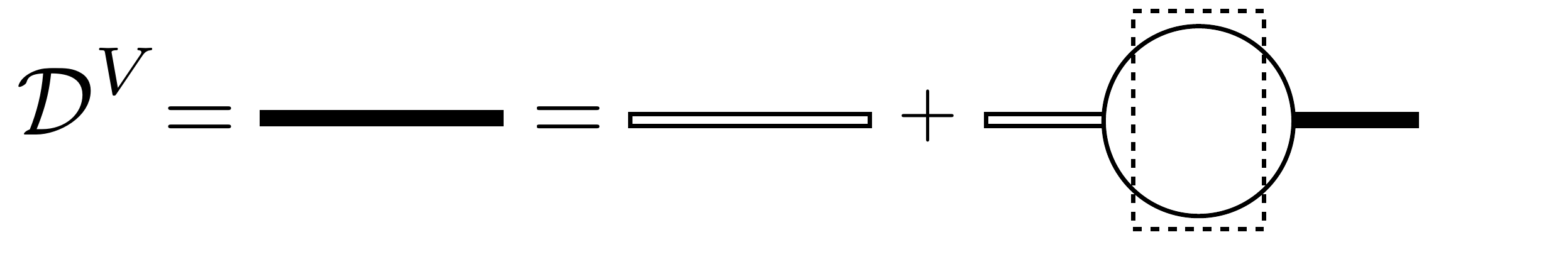}}
\subfigure[]{
\includegraphics[scale=0.235]{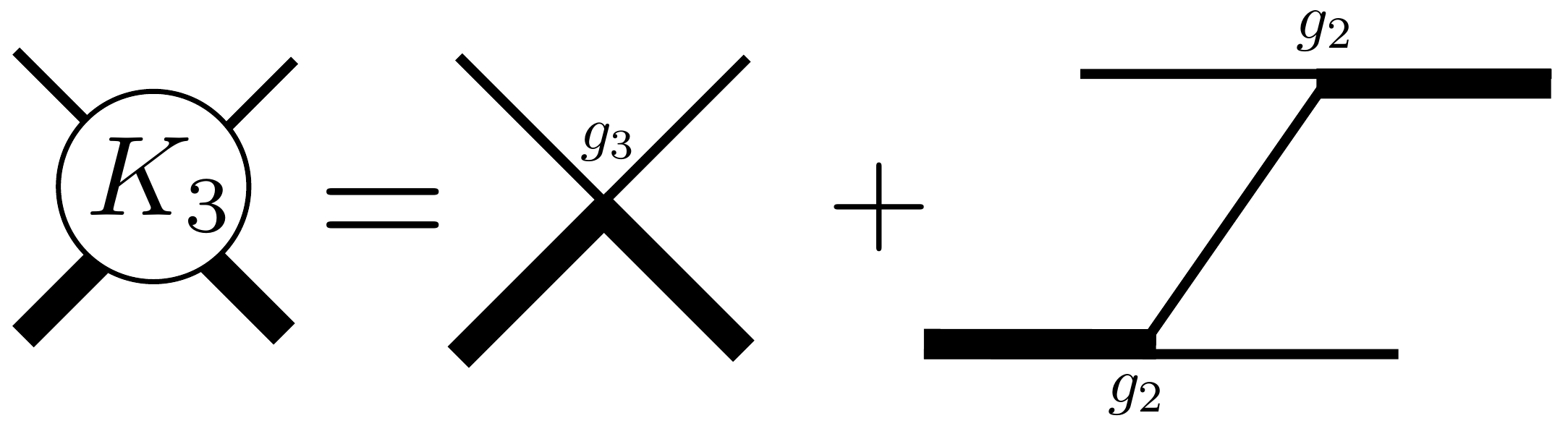}}
\caption{(a) Diagrammatic expansion of the three-body correlation function in a finite volume \cite{Briceno:2012rv}. $K_3$ denotes the three-body Bethe-Salpeter kernel. The leftmost (rightmost) circle corresponds to an interpolator that creates (annihilates) the three-particle initial (final) state with total energy $E$ and total momentum $\bm{P}$. Dashed rectangles indicate that all loop momenta on enclosed propagators are summed rather than integrated. (b) Diagrammatic equation satisfied by the full dimer propagator, $\mathcal{D}^{V}$, in a finite volume. The grey (black) band represents the full FV propagator while the double lines represent the bare propagator. c) The effective three-body Bethe-Salpeter kernel, $K_3$, is composed of a three-body contact interaction, characterized by the LEC $g_3$, as well as two-body contact interactions, described by the LEC $g_2$, via the exchange of a single boson.}
\label{fig:corrfunc}
\end{center}
\end{figure}

The kernel includes the three-body contact interaction as well as an exchange diagram with two insertions of the dimer-particle contact interactions, Fig. \ref{fig:corrfunc}. The poles of the FV correlation function correspond to the energy eigenvalues through an eigenvalue equation that has strong parallels with two-body coupled channel systems. The reason is that for each c.m. energy of the system, $E^*$, there are $N_{E^*}$ available boson-dimer eigenstates that can go on-shell inside the FV loops and contribute to power-law volume scaling of the QC. Unfortunately, the physical scattering amplitude does not directly show up in the QC, and can only be accessed by solving an integral equation, see Ref. \cite{Briceno:2012rv}.

An immediate implication of this formalism for bound state-particle systems below the bound-state breakup threshold can be obtained. The S-wave scattering phase shift of the bound state-particle can be extracted from the FV three-particle spectrum using the S-wave limit of the $2+1$ QC,
\begin{eqnarray}
\label{dbQC}
\qquad {q}\cot\delta_{Bd} &=&4\pi \ c^{\bm{d}}_{00}({q})+\eta\frac{e^{-\gamma_d L}}{L},
\end{eqnarray}
where ${q}=\sqrt{\frac{4}{3}\left(mE^*+\bar{q}^{2}\right)}$ is the relative momentum in the c.m. frame of the particle-bound state system and $\bar{q}$ is the relative momentum of the two particles comprising the bound states in the c.m. frame of the bound state. $m$ is the mass of the three identical particles, $E^*$ is the c.m. energy, $\gamma_d$ is the binding momentum of the particle in the infinite-volume limit,\footnote{To the order to which this calculation has been performed, one may choose to use the FV binding momentum.} $\delta_{Bd}$ is the scattering phase shift of the particle-bound state system, and the FV function $c^{\bm{d}}_{00}$ can be obtained from its general form in Eq. (\ref{Clm}) upon taking the appropriate limits. $\eta$ is an unknown coefficient that for sufficiently large volumes satisfies $\gamma_d L\gg1$ and is expected to be independent of the volume.  It therefore must be fitted when extrapolating results to the infinite volume. 
In other words, the extracted phase shifts from a two-body QC must be extrapolated to the infinite-volume limit using an exponential form -- a conjecture that had already been tested in LEFT calculations of the S-wave neutron-deuteron scattering by Bour, \emph{et al.} \cite{Bour:2012hn}. Systemic uncertainties of this formalism, in particular due to the use of an S-wave dimer in a FV cubic volume, are known and can be improved upon further investigations.\footnote{For a numerical  investigation of the $\pi-\rho$ scattering see Ref.~\cite{Roca:2012rx}, and for a benchmark LQCD calculation of the $a_1$ and $b_1$ resonances using this formalism consult Ref.~\cite{Lang:2014tia}. For an investigation of the implications of the three-particle spectrum using the so-called \emph{isobar approximation} see Ref.~\cite{Guo:2013qla}.}

Multi-particle scattering amplitudes from LQCD calculations are also the focus of theoretical investigations in particle physics. Despite significant progress in LQCD determinations of weak matrix elements in the two-body decay of the kaon, $K \rightarrow \pi\pi$ \cite{Blum:2011pu,Blum:2011ng,Blum:2012uk, Boyle:2012ys}, the three-body decay $K \rightarrow 3\pi$, crucial to studies of CP violation in the neutral kaon systems, has not been addressed via LQCD. For neutral D meson decays, the four-pion (six-pion, etc) channel couples to the two-pion decay mode of the $D$ meson along with several other channels. In order to understand the FV artifacts for studying these decays via LQCD, Hansen and Sharpe considered the scenario where the neutral D mesons primarily couples to two-particle final states, i.e., $D\rightarrow 2\pi,2K,2\eta$ in the $I=0$ channel in Ref.~\cite{Hansen:2012tf}. These authors have taken further steps in including multi-particle states in the FV analysis, focusing first on deriving a relativistic, model-independent $3 \rightarrow 3$ generalization of the L\"uscher formula for scalar particles.  Their ongoing work was presented and discussed in this workshop~\cite{Hansen:2013dla, Hansen:2014eka}.
\begin{figure}[h]
\begin{center}
\subfigure[]{
\includegraphics[scale=0.4]{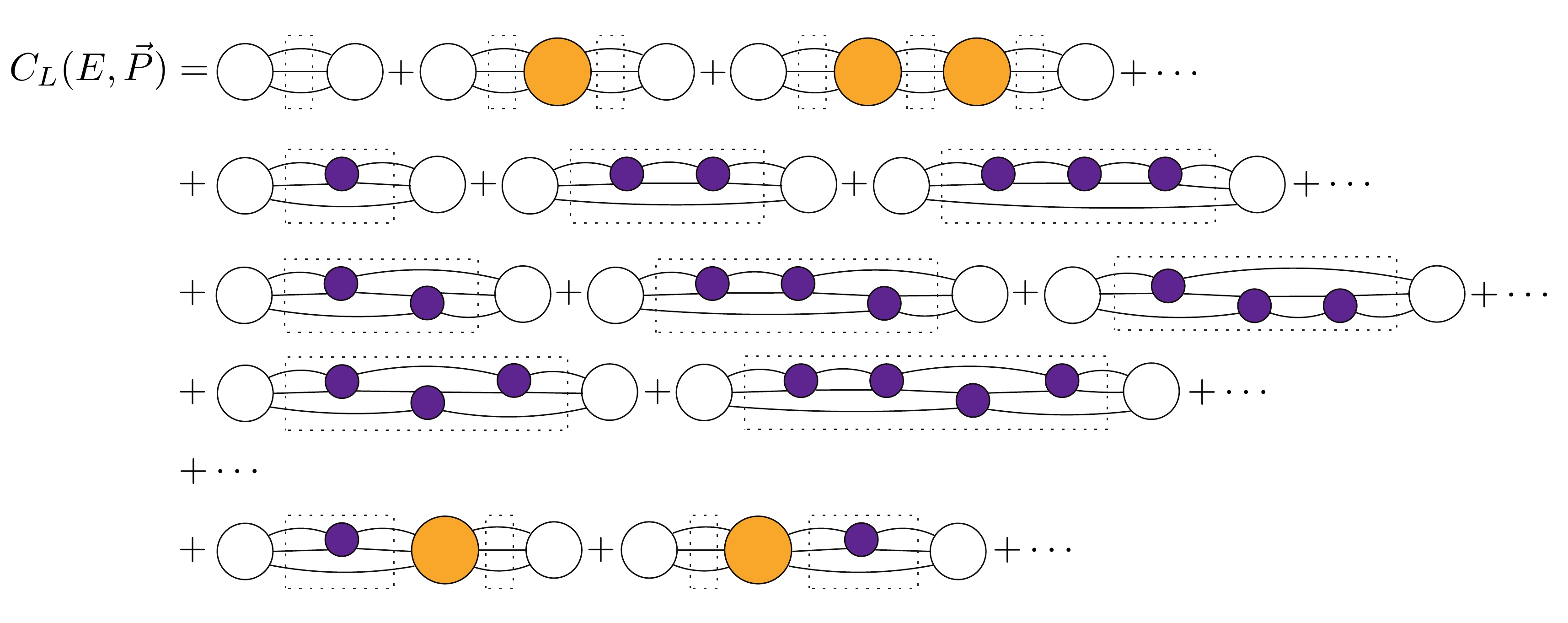}}
\subfigure[]{
\includegraphics[scale=0.4]{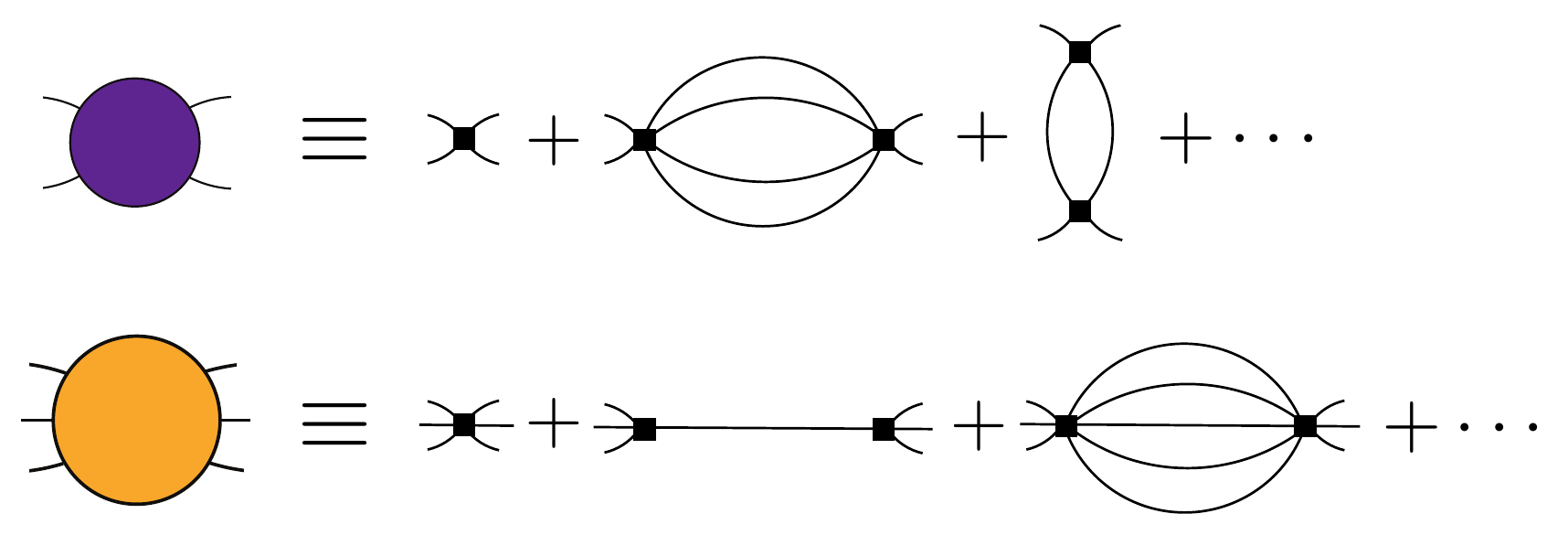}}
\caption{(a) Shown is the diagrammatic expansion of the FV three-particle correlation function without the dimer formalism \cite{Hansen:2013dla, Hansen:2014eka}. (b) Any insertion of circles between four (six) legs represents a two-to-two (three-to-three) Bethe-Salpeter kernel. All lines represent full propagators. The volume corrections due to dressing the propagators are exponential and can be neglected. The rest of the symbols are the same as in Fig. \ref{fig:corrfunc}. Figure is reproduced with the permission of Maxwell Hansen.}
\label{fig:corrfunc-HS}
\end{center}
\end{figure}

This formalism is pertinent for the mesonic sector with exact isospin symmetry where there are no $1 \rightarrow 2$ and  $2 \rightarrow 3$ transition amplitudes. Again, the connection between the three-particle spectrum and the infinite-volume scattering amplitudes can be found by looking at the pole structure of the three-particle correlation function, $C_{L}(E,\bm{P})$. Without a dimer-field approach, however, the diagrammatic expansion of the correlation function (shown in Fig. \ref{fig:corrfunc-HS}) is more involved. It is comprised of different classes of diagrams whose analytic structure of each must be examined carefully to identify the on-shell conditions, i.e. when the sums over loop momenta diverge. These are the contributions that scale with powers of $1/L$, and will form purely kinematic FV functions in the QC. The resulting QC 
is only related to the $3 \rightarrow 3$ scattering amplitude through the so-called \emph{divergence-free} scattering amplitude. This latter quantity does not suffer from divergences that arise due to $2 \rightarrow 2$ scatterings among pairs of three particles that can occur arbitrarily far from each other. Obtaining the physical amplitude from this quantity requires solving an integral equation. Similar to the result obtained using the dimer formalism, this QC is a matrix in the space of $N_{E^*}$ open  kinematic channels -- those that can go on-shell for any given c.m. energy. Additional subtleties associated with this QC are discussed in Ref. \cite{Hansen:2013dla, Hansen:2014eka}. The authors have performed a consistency check of this formalism in the weakly interacting limit of the three scalar particles and reproduced the ground-state energy up to $\mathcal{O}(1/L^6)$, previously calculated by Beane, Detmold and Savage \cite{Beane:2007qr} and by Tan \cite{Tan:2007bg}.

Another FV formula for three-particle systems that predates the above formalisms is 
due to Polejaeva and Rusetsky \cite{Polejaeva:2012ut}, which is derived using a non-relativistic potential model that describes scattering in coupled two- and three-particle channels. This formula confirms that the spectrum of three particles is completely determined by the S-matrix element for $2 \rightarrow 2$, $2 \rightarrow 3$ and $3 \rightarrow 3$ transitions. The QC presented by these authors show similar features to the ones discussed above. For example, in contrast to the algebraic two-particle QC, this three-particle QC is an integral equation that must be solved numerically. Earlier studies of the three-boson and three-nucleon systems in a finite volume using an EFT approach similar to the one presented here, have already investigated the volume dependence of three-particle bound-state energies such as the Triton's binding energy, see Refs. \cite{Kreuzer:2008bi,Kreuzer:2009jp, Kreuzer:2010ti, Kreuzer:2012sr}.\footnote{These works do not attempt to directly relate the spectrum to the three-body amplitude. Instead their QC is expressed in terms of the two- and three-body force parameters in the EFT that are matched to the infinite-volume physics.} 

The complexity of these QCs, in particular their non-algebraic form, suggests that a potential roadmap towards determining an infinite-volume scattering amplitude is to constrain the three-particle kernels. These kernels (LECs if a systematic EFT with a well-defined power counting can be used), can then be used to solve for scattering parameters as is customary now in applications of  chiral forces to nuclear many-body problems. One such study has been already carried out by Barnea, \emph{et al.} in Ref. \cite{Barnea:2013uqa} where the few-body observables are matched to the recent LQCD determination of the spectrum of the light nuclei at the SU(3) symmetric point by NPLQCD ~\cite{Beane:2012vq}, giving them a predictive power to eventually be able to construct the periodic table of the light nuclei. This possibility will be a new direction in studies of multi-nucleon systems and will most likely be pursued to constrain chiral multi-nucleon force parameters from LQCD calculations in the near future. In a closely related direction, by tuning the parameters of modern-day nuclear potentials, the LQCD energy eigenvalues can be matched to the energy eigenvalues evaluated using these potentials in a finite volume. The interplay between LQCD and such many-body nuclear calculations will be the subject of the next section of this review.

Another approach in extracting physical amplitudes from the FV spectrum, which intrinsically differs from the L\"uscher formalism, is the so-called \emph{potential} method, developed and extensively used by the HAL QCD collaboration \cite{Ishii:2006ec, Aoki:2009ji, Aoki:2009ji, Aoki:2012tk}. The working assumption is that the Nambu-Bethe-Salpeter (NBS) wavefunctions of two-nucleon systems can be extracted by forming the correlation functions of two-nucleon source and sink operators that are separated by a distance $\bm{r}$. As discussed in Sec.~\ref{subsec:FVMatrixelemts}, all correlation functions that overlap with two or more particle eigenstates depend on both on-shell and off-shell amplitude. Consequently, in general wavefunctions constructed from these will also depend on off-shell amplitudes. These wavefunctions should satisfy the Schr\"odinger equation,
\begin{eqnarray}
\left(\frac{\bm \nabla^2}{m_N} + \frac{k^2}{m_N}\right)\phi_k({\bm{r}})=\int d^3 r' V_k(\bm{r},\bm{r'}) \phi_k({\bm{r'}}),
\end{eqnarray}
where the potential $V_k(\bm{r},\bm{r'})$ is energy dependent and non local in general, and where $m_N$ and $k$ denote the nucleon mass and the magnitude of its three momentum in the c.m. frame of the two-nucleon system. For any nucleon separation within the range of the nuclear interactions, the extracted wavefunction depends on the source/sink interpolators used. However, for asymptotically large separations, there is a unique wavefunction that will satisfy the Helmholtz equation, from which the phase shifts of the infinite-volume theory can be extracted \cite{MJS,Beane:2010em}. Note that in this limit, the $\bm{r}$ dependence of the source/sink dependent factors in $ \phi_k({\bm{r}})$ will drop out and simultaneously $V_k(\bm{r},\bm{r'}) \rightarrow 0$. One can indeed extract the phase shifts from the asymptotic tail of the NBS wavefunction extracted from LQCD without constructing the potentials as is done in Ref. \cite{Aoki:2004wq} in the case of pion-pion scattering phase shifts. These extractions are equivalent to the L\"uscher method as described in L\"uscher's original papers \cite{Luscher:1986pf, Luscher:1990ux}. On the other hand, in the HAL QCD method, where the potentials are obtained as an  intermediate product,  certain assumptions are made. The main assumption is that the energy-independent non-local potentials at LO in a derivative expansion can be approximated by a local potential, $V(\bm{r})$. Then at the energy eigenvalues of the system $k_n$, $V(\bm{r})$ must, by construction, satisfy the Schr\"odinger equation with the source/sink dependent wavefunction $\phi_{k_n}({\bm{r}})$, and will lead to the infinite-volume phase shifts as true determinations based on QCD. However, away from the eigenenergies, the potential is not constrained to reproduce the phase shifts, and therefore the HAL QCD prediction for the phase shifts at other values of energies is no longer a model-independent prediction of QCD. These issues are the subject of an ongoing debate among experts on both sides and various publications are devoted to comparative studies of these two methods with no definitive consensus. One puzzling observation is the fact that although Refs.~\cite{Beane:2012vq, Beane:2013br, Yamazaki:2012hi} find a bound di-neutron for unphysical pion masses, the HAL QCD method does not lead to such a bound state. This could be due to insufficient statistics in the HAL QCD calculations which do not resolve  the long-range potential, among other issues as discussed above, see Refs. \cite{MJS, Beane:2010em, Walker-Loud:2014iea}.

An extension of the potential method to inelastic processes was presented in this workshop and some preliminary LQCD results were shown \cite{Aoki:2013zj, Aoki:2013cra, Aoki:2013kpa, Aoki:2012bb}. It is claimed that the asymptotic form of the n-body NBS wavefunctions will still depend on one phase shift, however the subtleties associated with the analytic structure of the multi-particle scattering amplitudes were not addressed. After proving the existence of the (coupled-channel) energy-independent potentials above the inelastic threshold, these potentials can be extracted from the multi-particle Schr\"odinger equation at the energy eigenvalues just as in the two-particle case. This multi-channel extension of the potential method has been used in studies of the H-dibaryon from LQCD, where by including the nearby channels $N \Xi$ and $\Sigma \Sigma$, it has been concluded that the H-dibaryon, although being a bound state at large pion masses, becomes a resonance at the physical point (with the physical quark masses) \cite{Inoue:2011ai} -- a prediction that needs to be confirmed by other collaborations with the use of other methods. HAL QCD has also published several results for two- and three-nucleon potentials at unphysical pion masses \cite{Doi:2012am, Aoki:2013zj}, that has enabled them to make predictions for the equation of state of nuclear matter (without including the three-nucleon forces) \cite{Inoue:2013nfe}. Given the discussion above regarding the tentatively large systematic errors of this methodology, it would be important to see if the observations made are backed by more formally sound approaches.

\section{LQCD and Nuclear Few- and Many-body Methods\label{few_many_body}} 
Despite a great deal of progress in studying many-body systems directly via LQCD, as partly mentioned in the previous sections, performing LQCD calculations of nuclei with $A>4$ with reliable level of precision will most likely be out of reach in the foreseeable future. Therefore, studies of nuclear reactions involving few and many nucleons will more likely continue via \emph{ab initio} methods. These refer to those calculations that take nucleons as the basic degrees of freedom -- where the high-energy degrees of freedom of QCD, quarks and gluons, have been integrated out. In the following, we divide these methods to two subsets.\footnote{This division does not reflect the extent and importance of individual methods and can be viewed as a division between continuum vs. lattice methods. The discussion of the chosen methods in Sec. \ref{Ab initio} follows the talks presented in the workshop.} The first one of which is what is customarily called \emph{ab initio} in literature, including NCSM (e.g., Refs.~\cite{Navratil:2000gs,Zhan:2004ct,Barrett:2012dr,Barrett:2013nh}), NCSMC \cite{Baroni:2013fe,Baroni:2012su}, Green's function Monte Carlo (GFMC) (e.g., Refs.~\cite{Pudliner:1997ck,Wiringa:2000gb,Pieper:2004qw}), CC methods (e.g., Refs.~\cite{Hagen:2007hi,Hagen:2008iw}), QMC calculations with the use of chiral forces \cite{Gezerlis:2013ipa, Gezerlis:2014zia, Lynn:2014zia, Roggero:2014lga}, as well as several other methods as named in the introduction of this review. The progress in applying some of these methods in studying nuclear few-(many-)body systems and their interplay with LQCD are briefly discussed in Sec. \ref{Ab initio}. The other popular methods are lattice EFTs (e.g., Refs.\cite{Lee:2008fa,Epelbaum:2010xt,Epelbaum:2011md,Lahde:2013kma}) where the evolution of multi-nucleon systems is calculated with an Euclidean time projection in a discretized FV spacetime, and where the EFT interactions among nucleons are included using an auxiliary field method. We briefly comment on the recent developments in this program and the role of LQCD in improving the calculated quantities in Sec. \ref{LEFT}. Finally, Sec. \ref{Matrix elements} has a concise discussion of the need for LQCD, along with the EFT and many-body techniques, to constrain the matrix elements of the current operators between single/multi-nucleonic states, in particular those that will refine the explorations of the beyond the Standard-Model scenarios.

\subsection{Ab initio methods for few- and many-body systems
\label{Ab initio}}
Nuclear few- and many-body calculations have progressed to a point where the uncertainties in calculations of light nuclei have less to do with algorithmic issues (e.g., numerical convergence, optimal oscillator parameter $b$, etc.), but more to do with the lack of understanding of particular aspects of the interactions between hadrons. Techniques such as the NCSMC, GFMC and CC readily calculate the low-lying spectrum of s- and p-shell nuclei \cite{Kamada:2001tv,Binder:2012mk,Hagen:2007hi}.   Coupled with the resonating group method, for example, NCSM calculations of certain nuclear reactions have been performed for both elastic and inelastic processes.  For example, elastic s-, p- and d-wave phase shifts have been performed for $d+{^3}{H}$ and $n+{^4}{He}$ scattering \cite{Quaglioni:2008sm,Navratil:2010jn}, and in the case of the inelastic radiative capture reaction $^7Be+p\rightarrow {^8B} + \gamma$, new resonances in the $0^+$, $1^+$, and $2^+$ channels have been predicted~\cite{Navratil:2011sa}.   Within our sun this latter reaction represents a vital step in a chain of reactions that ultimately produces neutrinos that are measured terrestrially. This reaction therefore has implications to solar neutrino measurements and our understanding of the Standard Solar Model.

Because of the complexity of many-body calculations for nuclear systems approaching the sd-shell, nuclear structure and reaction calculations utilize \emph{soft} two-body nuclear interactions, obtained through, for example, similarity renormalization group (SRG) transformations \cite{Bogner:2007rx,Tsukiyama:2010rj}.  Such soft interactions have momentum scales integrated out to scales as low as inverse 2.1 fm, thereby reducing the demand for large-model space calculations. However, for nuclei with A $>$ 8, momentum scales above this cutoff become dynamical and the applicability of SRG evolved potentials at the two-body level become questionable.\footnote{Recently SRG evolved potentials that include the induced 3-body interaction have been calculated \cite{Jurgenson:2010wy}.} Further, certain few-body (A$<$8) nuclear reactions and systems are very sensitive to three-body physics, such as $d+t$ fusion, $n+{^4}{He}$ reactions, and the maximal isospin few-body systems (e.g., three- and four-neutron systems).  The latter has direct implications to the composition and properties of neutron-rich matter.  For such systems, the use of chiral effective field theory-based interactions~\cite{Weinberg:1990rz, Kaplan:1998tg,Kaplan:1998we, Chen:1999tn, Kaplan:1996nv, Beane:2000fi, Kaplan:1998sz, vanKolck:1998bw,  Bedaque:1998kg, Bedaque:1998km, Bedaque:1999ve,Fleming:1999ee, Beane:2001bc, Nogga:2005hy, Birse:2005um, Epelbaum:2006pt, Birse:2007sx, Furnstahl:2008df} probably presents the most systematic and consistent approach to QCD, and therefore LQCD. This is because the interactions are based on the symmetries of QCD and exhibit a systematic hierarchy in interaction terms. Various successful implementations of chiral potentials in the QMC calculations of light and medium-mass nuclei as well as the neutron matter can be found in Refs. \cite{Gezerlis:2013ipa, Kruger:2013kua, Tews:2013wma, Gezerlis:2014zia, Lynn:2014zia, Roggero:2014lga}. A nice example of the power of such QMC techniques with systematic nuclear inclusion of interactions is shown in Fig. \ref{fig:NS-EOS}. By accounting for chiral NN and 3N forces up to next to NNLO (N$^3$LO) in a QMC calculation of the neutron matter \cite{Drischler:2013iza, Tews:2013wma}, and by the polytropic extensions to high densities \cite{Hebeler:2010jx, 2013ApJ...773...11H}, an uncertainty band corresponding to the mass vs. radius of the neutron stars can be predicted. This result is compared against the predictions of various phenomenological models.

In principle, the LECs that parametrize these interaction terms can be determined from experiment or from LQCD.  The three-body interaction terms up to N$^3$LO are shown in Fig. \ref{fig:3-body interaction}.  The coefficients $c_1$, $c_3$ and $c_4$ are extracted from $\pi N$ and $NN$ scattering~\cite{Epelbaum:2008ga,Machleidt:2011zz,Buettiker:1999ap,Fettes:1998ud,Entem:2001cg,Entem:2002sf,Entem:2003ft,Machleidt:2005uz,vanKolck:1994yi,Epelbaum:2002vt,Epelbaum:2006pt,Friar:2003yv,Friar:2004ca}, whereas the $c_D$ and $c_E$ coefficients are fit to the helion and helium systems~\cite{Gazit:2008ma, Maris:2012bt, Hammer:2012id}. In both cases the uncertainties are large and in the case of the latter the fits do not give unique solutions. These uncertainties negate the predictive power of key nuclear reaction calculations relevant to fusion and compact astrophysical objects. For example, the band presented in Fig. \ref{fig:NS-EOS} can be significantly reduced once more precise values of the relevant LECs are input into the many-body calculation \cite{Tews:2013wma}.

\begin{figure}
\centering
\includegraphics[width=.6\columnwidth,angle=0]{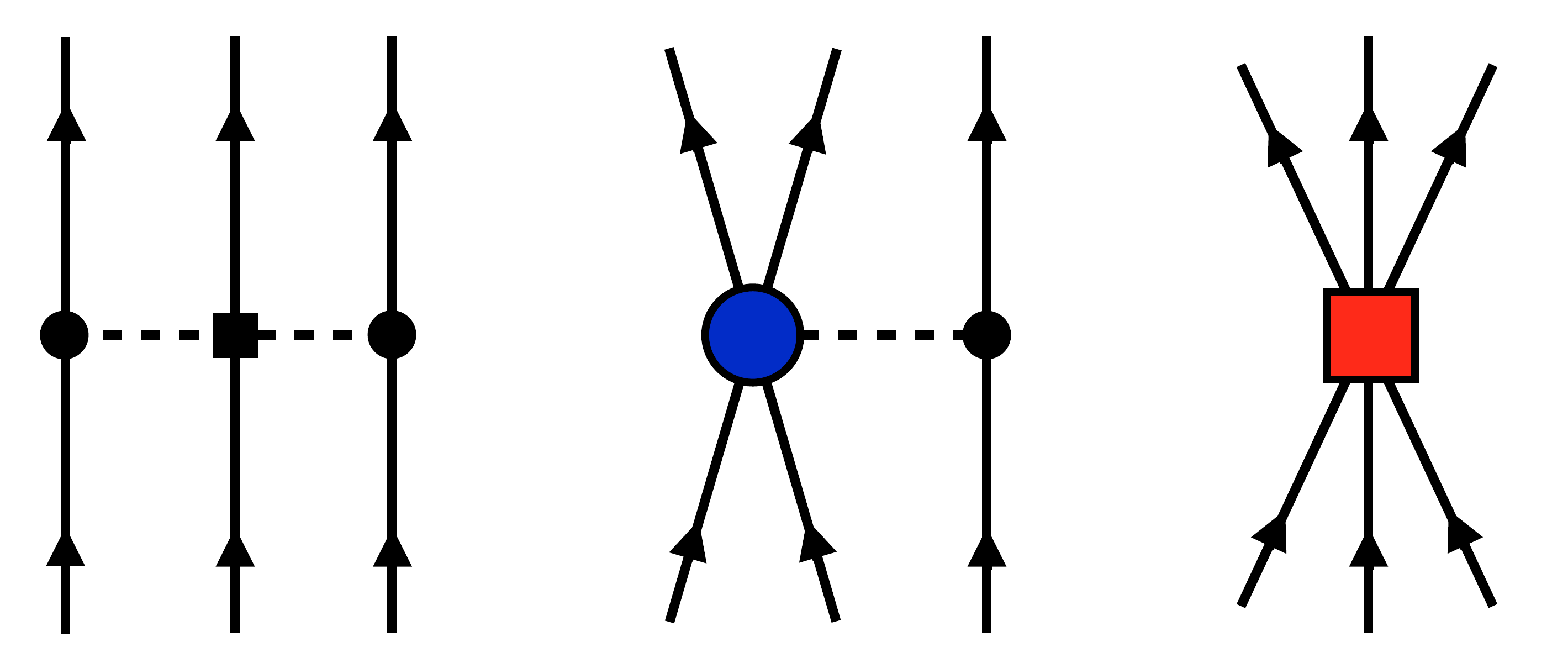}
\caption{The 3-nucleon interaction at N$^3$LO.  Solid lines represent the nucleons and dashed lines are the pions. The couplings for the left panel are $c_1$, $c_3$, and $c_4$.  The blue and red couplings are given by $c_D$ and $c_E$, respectively.
\label{fig:3-body interaction}}
\end{figure}
\begin{figure}
\centering
\includegraphics[width=.85\columnwidth,angle=0]{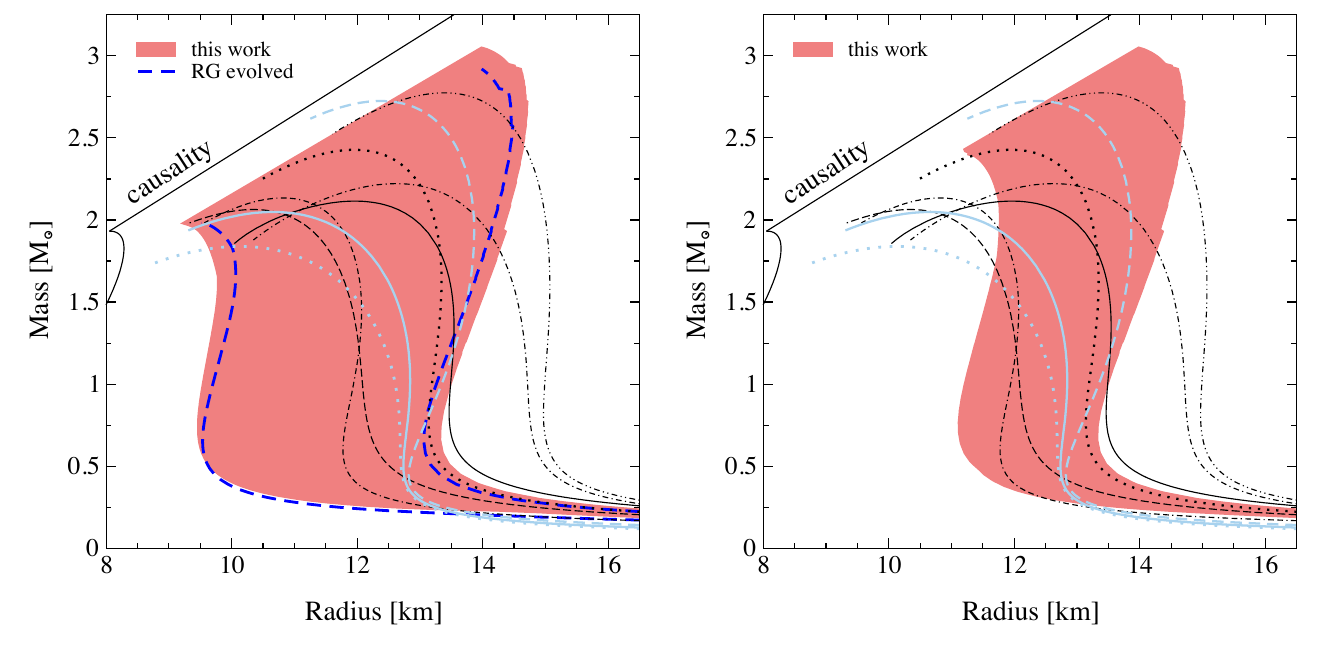}
\caption{Constraints on the mass-radius diagram of neutron stars obtained in Ref. \cite{Tews:2013wma} (red band) with the use of QMC with chiral EFT interactions at N$^3$LO, in comparison to the constraints from calculations based on RG-evolved NN interactions (thick dashed blue lines) \cite{2013ApJ...773...11H}. Shown in the plots are also the mass-radius relations obtained from the equations of state for core-collapse supernova simulations, See Ref. \cite{Tews:2013wma}. The left panel presents the band obtained with the constraint of a $1.97 M_{\odot}$ neutron star \cite{Kruger:2013kua}, while the right panel is based on constraint of a $2.4 M_{\odot}$ star. Figures are reproduced with the permission of Achim Schwenk.
\label{fig:NS-EOS}}
\end{figure}
With sufficient high-performance computing resources, LQCD can, in principle, calculate observables from which these LECs can also be extracted.  With appropriate boosting and usage of asymmetrical volumes, it is conceivable that certain LECs can be \emph{singled out}, or promoted to lower order so as to facilitate their extraction.  Such extractions will have an inherent uncertainty, since LQCD is a stochastic estimation at heart, but the uncertainties in this regard are controlled and can be reduced systematically.  In this manner, the connection between LQCD calculations and nuclear many-body calculations is made manifest through the nuclear forces.  

 At sufficiently high densities, such as those obtained within a proto-neutron star, the reactions of nucleons with hyperons (Y) can have profound consequences on the nuclear matter equation of state.  The recent observation of a 1.97 solar mass neutron star \cite{Demorest:2010bx} puts into question our understanding of hyper-nuclear forces, as most equation of states that incorporate NY interactions are too soft to allow such a large neutron star mass.  The inclusion of YYY and NYY interactions could possibly stiffen the equation of state, in an analogous manner demonstrated by 3N interactions.  The accounting of these 3-body hyperon interactions requires the study of reactions of hyper nuclei, a task that is very difficult experimentally but most likely best suited for LQCD studies.  As already mentioned above, for neutron-rich matter the isospin $I=\frac{3}{2}$ channel of the 3N force most likely plays a large role, yet it is poorly understood.  Again, LQCD provides the best opportunity to study these systems in a controlled and systematic fashion.

\subsection{Chiral effective field theory on the lattice
\label{LEFT}}
There has been effort to conjoin the methodology of lattice theories~\cite{Wilson:1974sk} and low-energy EFT for nuclear physics via the use of  {lattice} EFT. A \emph{lattice} theory can be any theory whose spacetime is discretized and placed on a finite grid. This allows one to non-perturbatively evaluate physical observables of the theory. Lattice methods have been extremely successful in the study of strongly interacting systems (see Refs.~\cite{Lee:2008fa, Drut:2012md} for reviews on the topic). Most recently, this technique has been applied to nuclear EFTs with zero range forces~\cite{Bulgac:2005pj, Borasoy:2005yc, Magierski:2008wa, Endres:2012cw, Endres:2011er, Bulgac:2008zz, Bour:2011ef,Bour:2012hn, Rokash:2013xda, Pine:2013zja, Rupak:2013aue} and more realistic chiral interactions \cite{Lee:2004si, Borasoy:2005yc, Borasoy:2006qn}. In going from the very first quantum lattice study of nuclear matter~\cite{Brockmann:1992in} to the first many-body lattice calculation using chiral effective field theory~\cite{Lee:2004si}, there were several benchmark calculations that eventually allowed for a well-defined chiral perturbation theory with lattice regularization (e.g., see Refs.~\cite{Muller:1999cp, PhysRevB.66.140504, Chandrasekharan:2003ub, Shushpanov:1998ms, Lewis:2000cc, Borasoy:2003pg}). 

Section~\ref{Sec:LQCD} discussed, in great detail, challenges associated with the determination of scattering phase shifts of two-particle systems in a finite cubic volume. By taking advantage of the fact that the degrees of freedom of LEFT calculations are nucleons, two-nucleon scattering phase shifts can be more efficiently determined by imposing a hard spherical wall at some fixed large radius \cite{Borasoy:2007vy}. This approach allows for the determination of the channel phase shifts with no partial-wave mixing from the energies of the nearly-spherical standing waves. For channels with partial-wave mixing this method leads to a set of coupled equations that can be solved to extract scattering phase shift and mixing angles. The advantage of this method is that it reduces the partial-wave mixing that is inherent in calculations in a cubic finite volume (see Table~\ref{table:irreps}). This method was implemented to extract scattering parameters for systems with $J\leq4$ including isospin-breaking and electromagnetic effects \cite{Borasoy:2007vy, Epelbaum:2010xt}. Figure~\ref{fig:3S1_eps_LEFT} shows the calculation of $^3S_1$ phase shift and the $J=1$ mixing angle, $\epsilon_1$, determined in Ref.~\cite{Epelbaum:2010xt} and compared to their experimental values~\cite{NIJMEGEN, Stoks:1993tb}. 

%%%%%%%%%%%%%%%%%%%%%%%%%%%
\begin{figure}[h!] 
\begin{center} 
{
\includegraphics[scale=0.7]{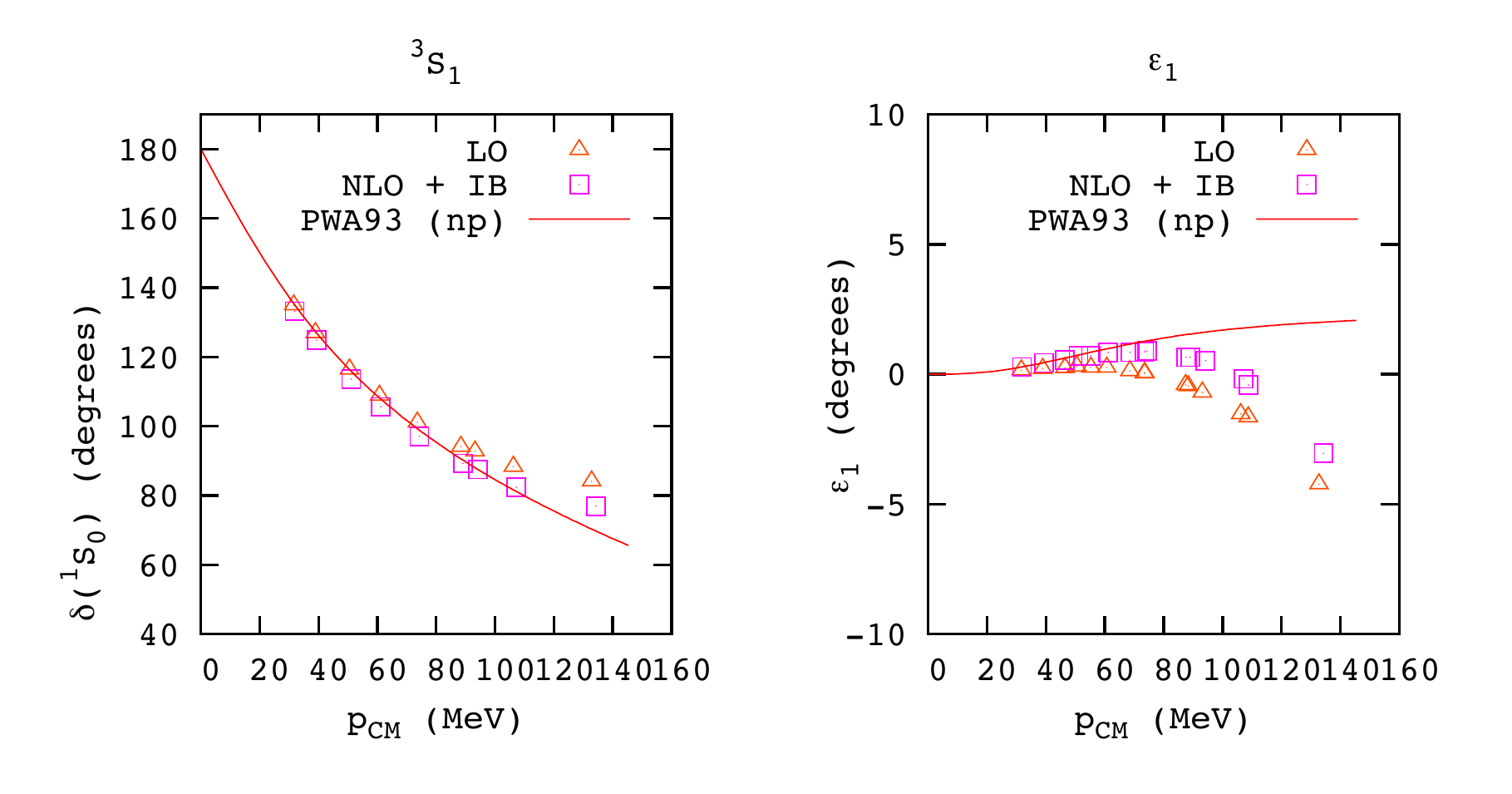}
\label{fig:3S1_eps_LEFT} }
\caption{ 
The neutron-proton $^1S_0$ phase shift and $J=1$ mixing angle as a function of  c.m. momentum determined in Ref.~\cite{Epelbaum:2010xt} and compared to their experimental values~\cite{NIJMEGEN, Stoks:1993tb}.
The definition of LO, NLO and isospin-breaking (IB) contributions are given in Ref.~\cite{Epelbaum:2010xt}. Figures are reproduced with the permission of Dean Lee.
} \end{center} \end{figure}
%%%%%%%%%%%%%%%%%%%%%%%%%%%

Just like LQCD calculations, LEFT calculations are performed in a finite Euclidean spacetime volume. Therefore the formal challenges associated with determination of bound states, scattering cross sections and matrix elements of electroweak operators discussed in Sec.~\ref{Sec:LQCD} are also present in LEFT calculations. In this regard, LEFT has additional advantages to circumvent these challenges, as for instance, typical LEFT calculations are performed at physical volumes in the order of 10~fm or larger. This feature controls the FV effects and has made the studies of bound states rather successful~\cite{Epelbaum:2009pd}, including, most recently, the determination of the ground-state energies of alpha nuclei from $^4$He up to $^{28}$Si with moderate accuracy~\cite{Lahde:2013uqa}.

Determining the $m_\pi$-dependence of nuclear observables is not only essential to have an accurate description of the nuclear force, but also gives further insight into questions of \emph{fine tunning} in the Universe. This is a topic that has generated a great deal of interest in recent years \cite{Beane:2002xf, Beane:2002vs, Epelbaum:2002gb, Braaten:2003eu, Bedaque:2010hr, Cheoun:2011yn, Ali:2012dj, Coc:2012xk,  Epelbaum:2012iu, Epelbaum:2012qn, Berengut:2013nh, Beane:2013br}. For instance, as mentioned already in Sec. \ref{Sec:Introduction}, a LQCD calculation by the NPLQCD Collaboration~\cite{Beane:2013br} suggests that the ratio of scattering length to effective range, $a/r$, for the deuteron remains unnaturally large ($>$1) for unphysical quark masses (see Fig. \ref{a-to-r}). A recent LEFT calculation by  Epelbaum \emph{et al.}~\cite{Epelbaum:2012iu} explored the stability of the Hoyle state as a function of the light quark mass. The Hoyle state is the name given to the spinless even-parity $^{12}$C resonance that resides near the $^8$Be-alpha thresholds and plays an essential role in the triple-alpha process that is responsible for the abundance of carbon in nature, an amount that is sufficient to support life ~\cite{Hoyle:1954zz}. Epelbaum \emph{et al.} found that the stability of this process is most sensitive to the $m_\pi$-dependence of spin singlet $(a_s)$ and spin-triplet $(a_t)$ scattering lengths, and in particular the derivatives with respect to $m_\pi$ of their respective inverses,
\begin{eqnarray}
\bar{A}_s\equiv\left.\frac{\partial a_s^{-1}}{\partial m_\pi}\right|_{m_\pi^{phys}},\hspace{1cm}
\bar{A}_t\equiv\left.\frac{\partial a_t^{-1}}{\partial m_\pi}\right|_{m_\pi^{phys}}.
\end{eqnarray}
Their findings are best illustrated by Òsurvivability bandsÓ shown in Fig.~\ref{fig:hoyle}. Shown are the estimate of the values of $\bar{A}_s$ and $\bar{A}_t$ for which the production of carbon would be sufficiently high to support life given a variation on the quark masses of $1\%$ and $5\%$, which is compared with the NNLO values of these quantities. As future LQCD calculations are performed in increasingly smaller values of the light quark masses, these will allow for the overall reduction of the uncertainties of the $m_\pi$-dependence of physical observables. 

\begin{figure}[h] 
\begin{center} 
\includegraphics[scale=0.42]{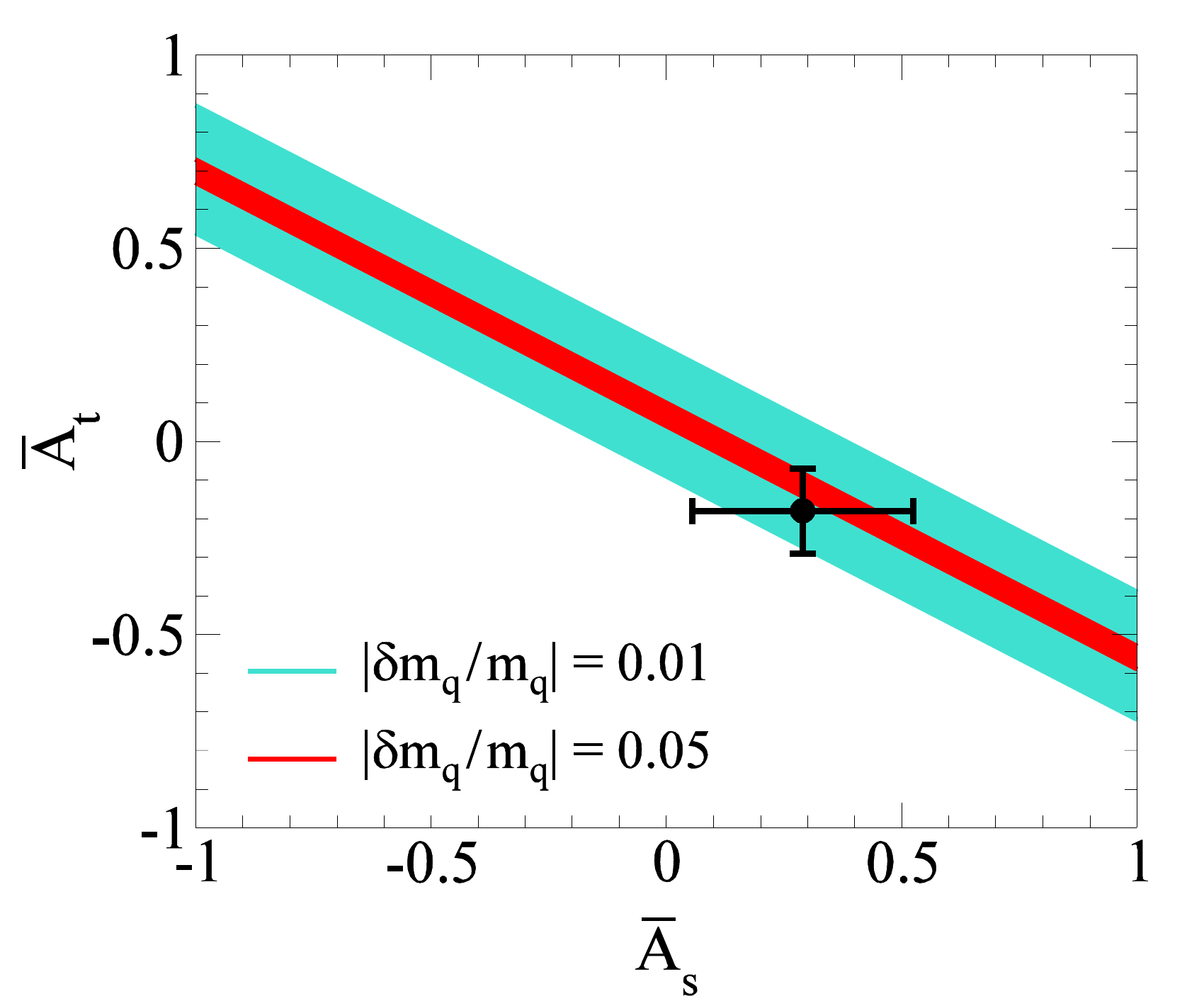}

\caption{Shown are the estimate of the values of $\bar{A}_s$ and $\bar{A}_t$ for which the production of carbon would be sufficiently high to support life given a variation on the quark masses of $1\%$ and $5\%$, which is compared with the  NNLO values of these quantities~\cite{Epelbaum:2012iu,Epelbaum:2013wla}.  
The figure are reproduced with the permission of Dean Lee.
\label{fig:hoyle}}
\end{center}
\end{figure}
\begin{figure}[h] 
\begin{center}  
{\includegraphics[scale=0.23]{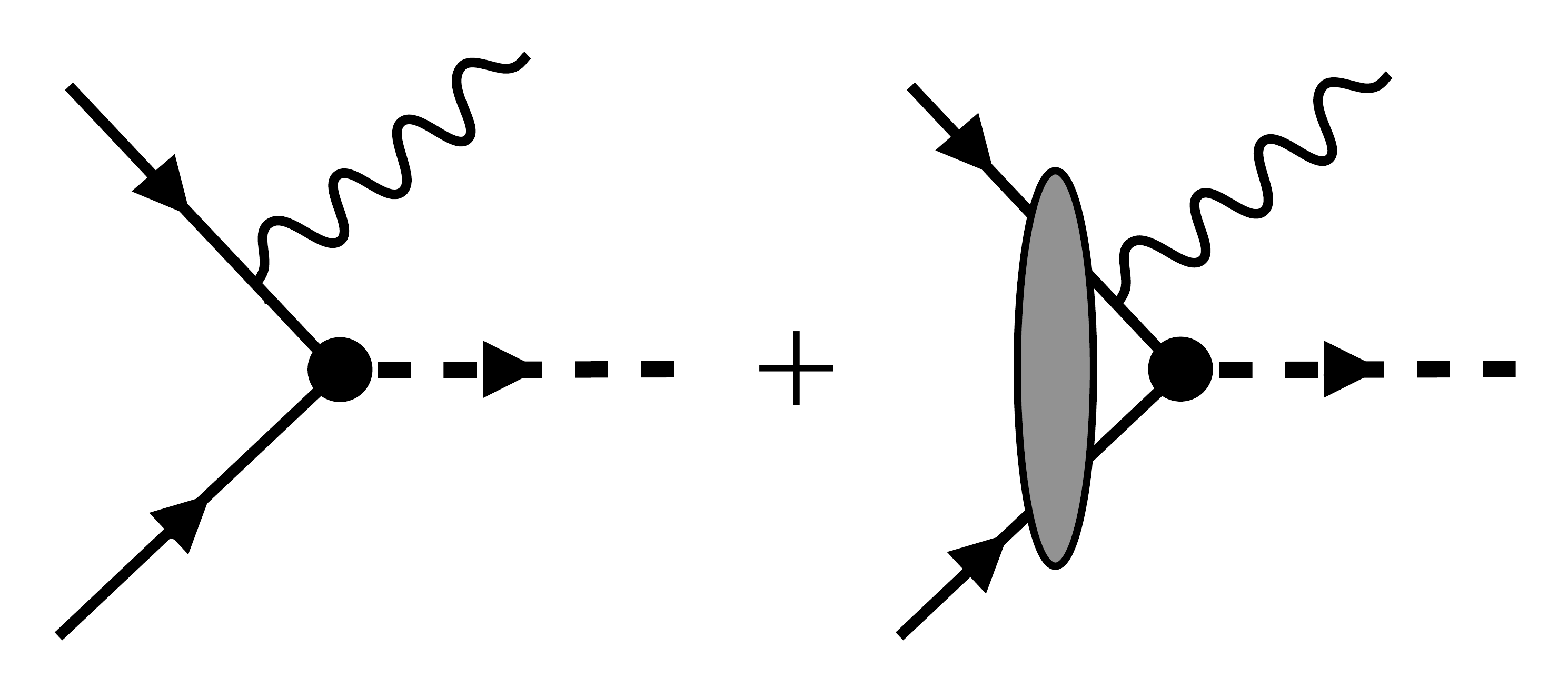} 
\label{fig:E1CaptureC} } 
\caption{ Leading order contributions to radiative capture process $p(n,\gamma)d$. Wavy line represents the photon, dashed line the deuteron, and blob the set of possible initial state interactions.
} \end{center}
\end{figure}
Recently, Rupak and Lee~\cite{Rupak:2013aue} presented a method for studying the radiative capture process $p(n,\gamma)d$, depicted in Fig.~\ref{fig:E1CaptureC}, via LEFT. This is an important process in nuclear physics as it provides stringent bounds on the primordial deuterium abundance. The non-perturbative contribution to the reduced matrix elements of Fig.~\ref{fig:E1CaptureC} can be evaluated in terms of the position space two-particle Green's function
\begin{eqnarray}
\qquad 
G(E;\mathbf{x},\mathbf{y})=\langle  \mathbf{y}|\frac{1}{E-\hat{H}_{s}+i\epsilon} | \mathbf{x}\rangle,
\end{eqnarray}
 where $H_s$ is the strong interaction Hamiltonian of the incoming spin-singlet channel and $\epsilon$ serves as an infrared regulator. Rupak and Lee evaluated this object numerically using the pionless EFT approximation at LO and recovered the well-known continuum result~\cite{Chen:1999bg, Rupak:1999rk} in the limit that the infrared regulator is taken to zero. This is a very promising technique that may prove to be useful in future studies of more challenging matrix elements in the presence of a more realistic nuclear Hamiltonian.

%%%%%%%%%%%%%%%%%%%%%%%%%%%
\subsection{Nuclear matrix elements
\label{Matrix elements}}
Beyond the Standard Model (BSM) of particle physics are reactions that potentially occur between ordinary matter (as given by the Standard Model) and exotic matter.  For example, a dark matter Weakly Interacting Massive Particle (WIMP), because of its large mass, could produce a Higgs boson that subsequently couples to the quarks within a nucleon, as depicted in the left panel of Fig. \ref{fig:wimp}.  At zero momentum transfer, the cross section for the spin independent elastic WIMP-nucleon ($\chi N$) is
\begin{equation}
\qquad \qquad 
\sigma_{\chi N}\sim | \sum_f G_f(m^2_{\xi}) f\,\,|^2\ ,
\end{equation}
where
\begin{equation}
\qquad \qquad
f=\frac{m_f}{m_N}\langle N|\bar{q}_fq_f|N\rangle\ ,
\end{equation}
and the function $G_f$ depends on several parameters related to the BSM theory and the sum is over flavors $f$ of the theory.
Such a reaction is difficult to detect because of its small size and is dominated by the standard strong QCD hadronic matrix element, $\langle N|\bar{q}_fq_f|N\rangle$ \cite{Gasser:1980sb,Borasoy:1996bx}, which must be understood well for extraction of BSM parameters. 
\begin{figure}[t]
\centering
\includegraphics[width=.45\columnwidth,angle=0]{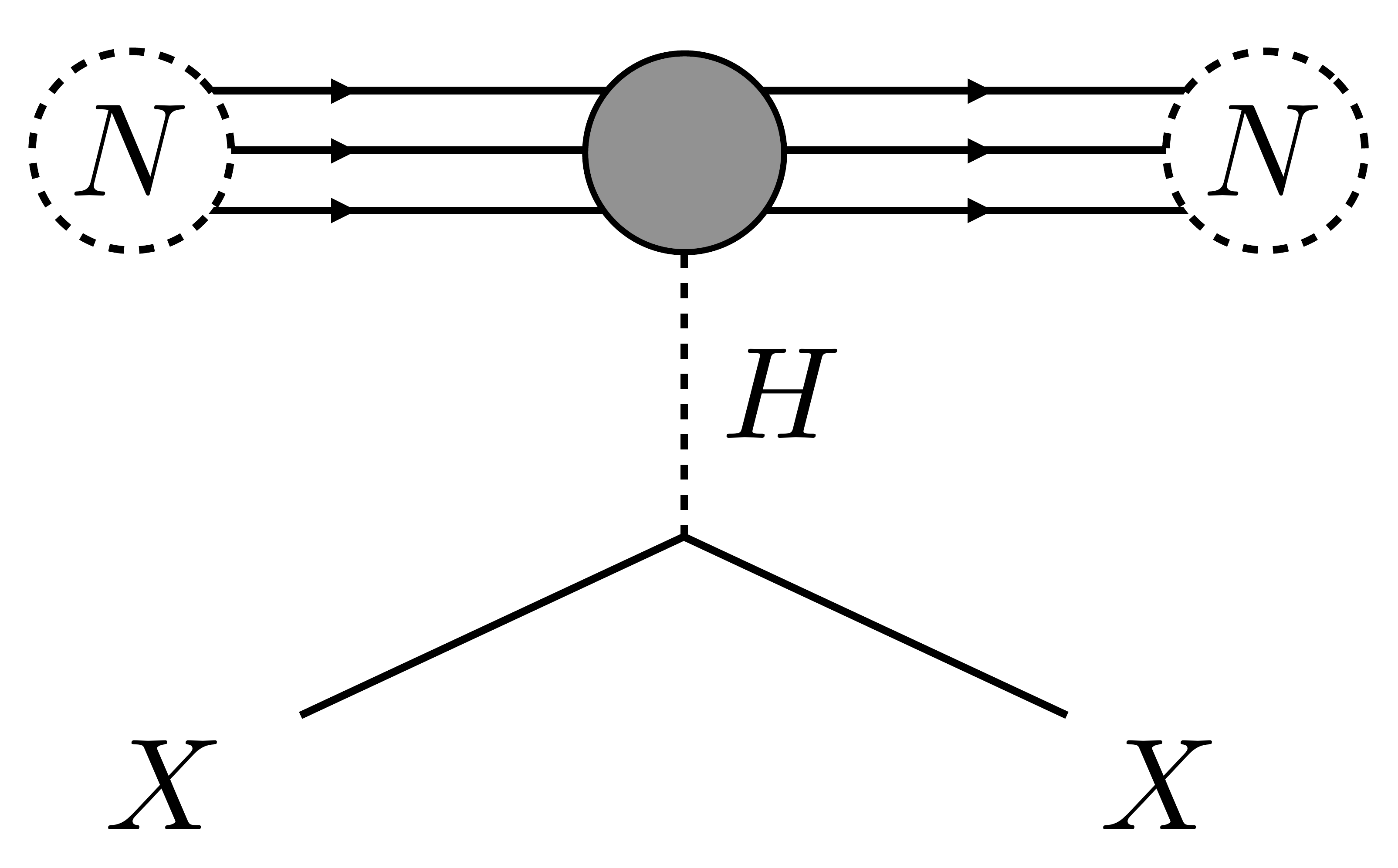}
\caption{The coupling of a WIMP particle, X, with a nucleon $N$ via Higgs (H) exchange with the internal quarks of the nucleon. 
\label{fig:wimp}}
\end{figure}
The scalar strange-quark matrix element of the nucleon has received a great deal of attention from the LQCD community in recent years~\cite{Toussaint:2009pz,Young:2009zb,Takeda:2010cw,Bali:2011ks,Dinter:2012tt,Gong:2012nw,Jung:2012rz,Oksuzian:2012rzb,Engelhardt:2012gd,Freeman:2012ry,Oksuzian:2012rzb,Durr:2011mp,Horsley:2011wr,Semke:2012gs,Shanahan:2012wh,Ren:2012aj, Junnarkar:2013ac}, and a \emph{lattice average} has been calculated to be $\frac{m_s}{m_N}\langle N|\bar{s}s|N\rangle_{latt.}=0.043(11)$~\cite{Junnarkar:2013ac}.  Any future precision terrestrial measurements of similar BSM observables will also require an equally precise understanding of its strong interaction QCD component.  In Ref. \cite{Menendez:2012tm} it was shown that the spin-dependent WIMP-nucleon reaction is very sensitive to nuclear structure (see also Ref. \cite{Fitzpatrick:2012ix} for a detailed EFT study of nuclear responses relevant to the dark-matter detections).  The coherent scattering of this process in many-nucleon systems is one of the main experimental methods (e.g., XENON100 experiment) for detecting dark matter candidates.  Reducing the uncertainties of this process for single-nucleon systems is therefore paramount, and can be done via LQCD calculations, which would involve understanding the sea quark content of nucleons.  This calculation is difficult because it involves disconnected diagrams, in addition to the usual suspects, but should be computationally manageable with next generation high-performance computing resources. A preliminary study by Beane, \emph{et al.} \cite{Beane:2013kca} uses the few available LQCD data for the binding energy of the light-nuclei at unphysical quark masses, and in combination with the physical values, gives an estimate of the significance of two-body currents in the calculations of the leading scalar-isoscalar WIMP-nucleus interactions beyond the impulse (single-nucleon) approximation, conjecturing them to be small. This result can be readily refined once further calculations at more unphysical pion masses are performed, and already demonstrates the role of LQCD in constraining the the important nuclear matrix elements relevant to this program.

An explanation of the current matter/anti-matter asymmetry also requires physics that is BSM.  In addition to lepton-number violating interactions, it is conceivable for baryon-number violating interactions to also exist.  For example, Grand Unified Theories (GUTs) commonly have a $\Delta B=1$ violating interaction that enables, for example, proton decay.  Also easily included in GUTs are $\Delta B=2$ violating interactions.  Such interactions would be responsible for $N$-$\bar{N}$ oscillations.  From an EFT point of view, there are 24 general $\Delta B=2$ violating operators that can be written down~\cite{schrock1982}, however, this number can be further reduced to 6 linearly independent operators by application of symmetries~\cite{Buchoff:2012bm}.  These operators can be incorporated in a LQCD study to determine the hadronic component of these $\Delta B=2$ matrix elements and this is the subject on an ongoing program~\cite{Buchoff:2012bm}.

%%%%%%%%%%%%%%%%%%%%%%%%%%%%%%%%%%%%%%%%%%%%%%%%%%%%%%%%%%

%%%%%%%%%%%%%%%%%%%%%%%%%%%%%%%%%%%%%%%%%%%%%%%%%%%%%%%%%%
%%%%%%%%%%%%%%%%%%%%%%%%%%%%%%%%%%%%%%%%%%%%%%%%%%%%%%%%%%

%%%%%%%%%%%%%%%%%%%%%%%%%%%%%%%%%%%%%%%%%%%%%%%%%%%%%%%%%%
%\newpage

%%%%%%%%%%%%%%%%%%%%%%%%%%%%%%%%%

\section{Future Outlook \label{Sec:ConclusionFutureOutlook}}

The workshop in ``nuclear reactions from lattice QCD'' presented a great opportunity for an effective communication among participants with expertise in quantum many-body calculations, EFT and LEFT, and LQCD for nuclear and particle physics. As a result of talks and discussions, the goals were set for the future of the field with well-defined plans. It is clear that although there has been a great deal of progress in the development of formalism pertinent for LQCD, there is still a need for much more progress in this area. In order to reliably perform LQCD studies of nuclear few-body systems, there are \emph{arguably} five major formal topics of research that must be addressed.
\\
\emph{\indent 1. The signal/noise problem}:  The poor signal/noise problem that is inherent in performing LQCD calculations with finite baryon density and/or chemical potential continues to pose a challenge, despite many efforts to elucidate and alleviate this problem~\cite{Lepage89, MJSsign, Lee:2011sm, Endres:2011jm, Endres:2011mm, Grabowska:2012ik, Detmold:2014hla}. In particular, this has been the main reason why the binding energy of light nuclei have only been calculated via LQCD for pion masses as heavy as $\sim 500~{\rm MeV}$~\cite{Beane:2012vq, Yamazaki:2009ua}. As an example, in the most recent work by Yamazaki \emph{et al.}~\cite{Yamazaki:2013rna}, the authors reported that with approximately 200,000 measurements they have not been able to reliably determine the light-nuclei spectra and more statistics were needed. Furthermore, excited states essentially have poorer signal, but determining these with high precision is crucial to be able to determine scattering parameters directly from LQCD. 
\\
\indent \emph{2. Multi-particle operators}: As discussed in Sec.~\ref{Sec:Introduction}, there has been much progress in the development of lattice operators for multi-particle systems~\cite{Beane:2005rj, Beane:2006mx, Beane:2006gf, Beane:2007es, Detmold:2008fn, Beane:2009py, Thomas:2011rh, Beane:2011sc, Basak:2005ir, Peardon:2009gh, Dudek:2010wm, Edwards:2011jj, Dudek:2012ag, Dudek:2012gj, Shi:2011mr, Detmold:2012wc, Detmold:2010au, Doi:2012xd, Detmold:2012eu, Detmold:2013gua, Detmold:2012eu}. There has been also some progress in  construction of multi-particle operators for systems involving half-integer spin particles that do not reside in a trivial irrep of the corresponding little group \cite{Morningstar:2013bda, Foley:2012wb}. This is essential in order to properly interpret the FV spectrum, since different irreps encode different information regarding the infinite-volume scattering parameters, as illustrated by Fig.~\ref{deut_tet}. Furthermore, as one approaches the physical quark masses, larger volumes will be needed in order for $m_\pi L\gg1$. This will lead to energy levels that are increasingly denser and therefore are much more challenging to numerically resolve. In order to disentangle the spectra for nuclear systems in such conditions, one will have to generate a wide basis of operators from which one can properly obtain the optimal basis as was done for the I=0 $\pi\pi$ system by Dudek \emph{et al.}~\cite{Dudek:2012xn}.
\\
\indent \emph{ 3. Finite-volume spectrum for multi-particle systems}: As discussed in Sec.~\ref{Sec:LQCD}, it is fair to claim that the implication of the one- and two-body spectrum determined via LQCD is formally understood, as manifested by the compact quantization condition of the two-body spectrum, Eq.~(\ref{eq:QC}). There are several sectors of the formalism pertinent for few-body physics that have not reached this level of maturity. The first necessary step in being able to properly understand few-body systems requires constructing a ~\emph{non-perturbative, model-independent, relativistic} framework for the three-body sector, e.g., Ref. \cite{Hansen:2014eka}, and moving towards its implementation. Although there is already an existing formalism for studying perturbatively weak N particles in a finite volume~\cite{Smigielski:2008pa}, having a non-perturbative result is essential, because, as illustrated by Fig.~\ref{deut_tet}, FV artifacts may upgrade small infinite-volume effects to be of significant size. This will be of great relevance when generalizing the formalism discussed in Sec.~\ref{eq:QC} for three-nucleon systems, since the nuclear sector exhibits both non-perturbative and seemingly finely tuned interactions. Furthermore, it will be necessary for the FV formalism to accommodate the fact that the number of particles is in general not conserved in inelastic processes. 
\\
\indent \emph{ 4. Matrix elements involving few-body systems}: The determination of electroweak matrix elements involving two or more particles in either the initial and/or final state is both technically and formally challenging.  
Despite progress in studies of $1\rightarrow 2$ processes via a FV formalism \cite{Lellouch:2000pv, Agadjanov:2014kha, Briceno:2014uqa}, to this day, for processes of the form $2\rightarrow 2$ where the external current has both a one-body and two-body contribution, no universal expression has been reached~\cite{Detmold:2004qn, Bernard:2012bi, Briceno:2012yi}. For these processes, what has been obtained are expressions that in general depend on the form of the one and two-body currents written in the low-energy Lagrangian that appropriately describe the low-energy degrees of freedom of the system of interest. To extend the range of validity of such formalisms to beyond EFTs, obtaining a universal relation between the transition amplitudes and the FV matrix elements is essential.
\\
\color{black}
\indent \emph{ 5. Other sources of systematics}: 
As emphasized in the introduction, to quote the final result for a LQCD calculation that is directly comparable with experiment one must take into account all sources of systematics, namely unphysical quark masses, finite volume, discretization effects and electromagnetic effects.

\noindent
\emph{a) Unphysical quark masses:} As was already discussed previously in the context of two-nucleon systems~\cite{Beane:2013br}~(see Fig.~\ref{a-to-r}) and the Hoyle state~\cite{Epelbaum:2012iu}~(see Fig.~\ref{fig:hoyle}), the $m_\pi$-dependence of physical observables is interesting in its own right (also see Refs.~\cite{Beane:2002xf, Beane:2002vs, Epelbaum:2002gb, Braaten:2003eu, Bedaque:2010hr, Cheoun:2011yn, Ali:2012dj, Coc:2012xk, Epelbaum:2012qn, Berengut:2013nh}). Nevertheless, designing the optimal extrapolation to the physical pion mass will in general depend on the observable of interest. For low-energy two-body observables one can envision utilizing $\chi PT$ or some $\chi PT$-based model, which has been extensively implemented.  For states lying above inelastic thresholds the pathway is not as obvious and further work is needed.

\noindent
\emph{b) Finite-volume effects associated with the finite range of hadronic forces:} We have emphasized the need for the determination of power-law FV effects of physical observables as they are closely related to physical amplitudes. It is expected that these can be written in a universal form, such as Eq.~\ref{eq:QC}, as they originate from the kinematic on-shell conditions. These universal expressions are correct up to corrections that are suppressed by $\mathcal{O}(e^{-m_\pi L})$. It is safe to say that due to moderately large volumes and (unphysically) heavy quark masses of the present-day two-body calculations, these corrections have not exceeded the achieved level of precision of these calculations. Furthermore, as the signal/noise ratio of calculations typically improves with the increased volume, the goal is to implement larger volumes where these systematics effects are further suppressed. Nevertheless, in order to have a rigorous estimate of systematic errors it will be necessary to determine the functional form of these corrections, which are not universal. As the calculations are moving towards physical values for the light quark masses, these corrections will become significant. Such corrections have already been studied for $\pi\pi$~\cite{Bedaque:2006yi} and $NN$ systems~\cite{Sato:2007ms} in an S-wave, as well as the $\pi\pi$ system in a P-wave in Refs.~\cite{Chen:2012rp, Albaladejo:2013bra} with the use of applicable EFTs. 
\\
\noindent
\emph{c) Discretization effects:} To this date the majority of exploratory few-body calculations are performed at a single lattice spacing. However it is expected that discretization effects are well below the level of precision reached in these calculations. Nevertheless, the PACS-CS collaboration reported that the $m_\pi$-dependence of their light meson-meson scattering lengths was describable by Wilson chiral perturbation theory (W$\chi$PT) but not by $\chi$PT~\cite{Sasaki:2013vxa}. W$\chi$PT is a low-energy EFT that includes discretization effects from Wilson fermions~\cite{Sharpe:1998xm}. Again, discretization effects are not universal and depend on the action used in the calculation. There has been a great deal of work in constructing EFTs that encapsulate the finite lattice spacing effects, see e.g., Refs.~\cite{Sharpe:1998xm, Rupak:2002sm,Aoki:2003yv, Bar:2003mh, Sharpe:2004ny}, and these have been used to analytically determine discretization corrections for two-body systems~\cite{Chen:2005ab, Chen:2006wf,  Aoki:2008gy, Hansen:2011mc, Sasaki:2013vxa}. Thus far these have only considered corrections to near threshold states. As one enters an era where systems under study can have arbitrary momenta in the vicinity of any number of inelasticities, these works will need to be extended. 
\\
\noindent
\emph{d) Electromagnetic effects:}
Some LQCD calculations have recently incorporated QED interactions in an attempt to determine the electromagnetic mass splittings among several neutral and charged hadrons ~\cite{Blum:2007cy, Blum:2010ym, Aoki:2012st, deDivitiis:2013xla, Borsanyi:2014jba}, and to determine the hadronic contributions to the muon anomalous magnetic moment \cite{Hayakawa:2005eq,Blum:2014oka,Benayoun:2014tra}. The large (power-law) FV effects due to the infinite-range of QED interactions have been controlled in several cases \cite{Duncan:1996xy, Hayakawa:2008an, Davoudi:2014qua, Borsanyi:2014jba}, while further investigation of the volume effects must be undertaken for a wide range of single-particle quantities. Given the progress in generating QCD+QED gauge configurations, it is conceivable that few-body lattice calculations including QED will become a reality in the near future. For energies where two particles or more can become kinematically allowed, one has to deal with the challenging task of connecting observables in presence of both QED and QCD interactions to the FV spectra. A recent progress in this problem has been made by Beane and Savage in Ref. \cite{Beane:2014qha} where it is shown that  due to an infrared cutoff in a system of two charged particles in a finite volume, the FV QED corrections to the scattering amplitudes can be obtained in a perturbative approach; thus avoiding the challenging task of  re-summation of the QED ladder diagrams at low energies in a finite volume. To be able to properly map out the Coulomb-modified scattering phase shifts in these systems, further efforts must be devoted to generalization and implementation of this formalism.

\color{black}

In summary, it is reasonable to expect that in the next two years, LQCD calculations of multi-baryons systems with $A=2-4$  be performed at lighter pion masses, although such calculations at the physical pion mass might only become viable in five years or more. To include electromagnetism as well as isospin-breaking effects in these calculations is also within reach in the next 10 years. The study of multi-neutron systems was also planned for the next few years. Three-body and four-body forces are estimated to be extracted in the next 5 years, and as a result the systematic matching between LQCD and quantum many-body calculations will  become a possibility in foreseeable future. This program is rapidly moving towards the ultimate goal of founding the field of nuclear physics upon the standard model of particles and interactions which directly impacts the precision calculations in nuclear reaction and fusion research, nuclear astrophysics and physics beyond the Standard Model.
%%%%%%%%%%%%%%%%%%%%%%%%%%%%%%%%%%%%%%%%%%%%%%%%%%%%%%%%%% 
 
%%%%%%%%%%%%%%%%%%%%%%%%%%%%%%%%%%%%%%%%%%%%%%%%%%%%%%%%%%%%
\ack
We gratefully acknowledge discussions with, and supports from, Martin Savage and David Kaplan. We would like to thank all the participants of the workshop for their contributions, and for promoting inspiring discussions, and the help from the extremely kind staff of the Institute for Nuclear Theory. We appreciate useful discussions with Jozef Dudek, Peng Guo, Maxwell Hansen, Dean Lee and Achim Schwenk and their permission to use the corresponding figures. We also would like to thank Achim Schwenk and Michael Wagman for comments on the initial draft of this review. RB acknowledges support from the U.S. Department of Energy contract DE-AC05-06OR23177, under which Jefferson Science Associates, LLC, manages and operates the Jefferson Laboratory. ZD was supported in part by DOE grant No. DE-FG02-97ER41014 and by DOE grant No. DE-FG02-00ER41132. The work of TL was supported by the DFG and NSFC through CRC110.

%%%%%%%%%%%%%%%%%%%%%%%%%%%%%%%%%%%%%%%%%%%%%%%%%%%%%%%%%%%%

\section*{References}

\bibliographystyle{unsrt.bst}
\bibliography{bibi}

\providecommand{\newblock}{}
\begin{thebibliography}{100}
\expandafter\ifx\csname url\endcsname\relax
  \def\url#1{{\tt #1}}\fi
\expandafter\ifx\csname urlprefix\endcsname\relax\def\urlprefix{URL }\fi
\providecommand{\eprint}[2][]{\url{#2}}

\bibitem{Duncan:1996xy}
Duncan A, Eichten E and Thacker H 1996 {\em Phys.Rev.Lett.\/} {\bf 76}
  3894--3897 (\textit{Preprint} \eprint{hep-lat/9602005})

\bibitem{Hayakawa:2005eq}
Hayakawa M, Blum T, Izubuchi T and Yamada N 2006 {\em PoS\/} {\bf LAT2005} 353
  (\textit{Preprint} \eprint{hep-lat/0509016})

\bibitem{Blum:2007cy}
Blum T, Doi T, Hayakawa M, Izubuchi T and Yamada N 2007 {\em Phys.Rev.\/} {\bf
  D76} 114508 (\textit{Preprint} \eprint{0708.0484})

\bibitem{Basak:2008na}
Basak S {\em et~al.\/} (MILC Collaboration) 2008 {\em PoS\/} {\bf LATTICE2008}
  127 (\textit{Preprint} \eprint{0812.4486})

\bibitem{Blum:2010ym}
Blum T, Zhou R, Doi T, Hayakawa M, Izubuchi T {\em et~al.\/} 2010 {\em
  Phys.Rev.\/} {\bf D82} 094508 (\textit{Preprint} \eprint{1006.1311})

\bibitem{Portelli:2010yn}
Portelli A {\em et~al.\/} (Budapest-Marseille-Wuppertal Collaboration) 2010
  {\em PoS\/} {\bf LATTICE2010} 121 (\textit{Preprint} \eprint{1011.4189})

\bibitem{Portelli:2012pn}
Portelli A, Duerr S, Fodor Z, Frison J, Hoelbling C {\em et~al.\/} 2011 {\em
  PoS\/} {\bf LATTICE2011} 136 (\textit{Preprint} \eprint{1201.2787})

\bibitem{Aoki:2012st}
Aoki S, Ishikawa K, Ishizuka N, Kanaya K, Kuramashi Y {\em et~al.\/} 2012 {\em
  Phys.Rev.\/} {\bf D86} 034507 (\textit{Preprint} \eprint{1205.2961})

\bibitem{deDivitiis:2013xla}
de~Divitiis G, Frezzotti R, Lubicz V, Martinelli G, Petronzio R {\em et~al.\/}
  2013 {\em Phys.Rev.\/} {\bf D87} 114505 (\textit{Preprint}
  \eprint{1303.4896})

\bibitem{Borsanyi:2013lga}
Borsanyi S, Duerr S, Fodor Z, Frison J, Hoelbling C {\em et~al.\/} 2013 {\em
  Phys.Rev.Lett.\/} {\bf 111} 252001 (\textit{Preprint} \eprint{1306.2287})

\bibitem{Drury:2013sfa}
Drury S, Blum T, Hayakawa M, Izubuchi T, Sachrajda C {\em et~al.\/} 2013
  (\textit{Preprint} \eprint{1312.0477})

\bibitem{Borsanyi:2014jba}
Borsanyi S, Durr S, Fodor Z, Hoelbling C, Katz S {\em et~al.\/} 2014
  (\textit{Preprint} \eprint{1406.4088})

\bibitem{Wilson:1974sk}
Wilson K~G 1974 {\em Phys.Rev.\/} {\bf D10} 2445--2459

\bibitem{Beane:2012vq}
Beane S~R, Chang E, Cohen S~D, Detmold W, Lin H {\em et~al.\/} 2013 {\em
  Phys.Rev.\/} {\bf D87} 034506 (\textit{Preprint} \eprint{1206.5219})

\bibitem{Beane:2005rj}
Beane S~R, Bedaque P~F, Orginos K and Savage M~J (NPLQCD Collaboration) 2006
  {\em Phys.Rev.\/} {\bf D73} 054503 (\textit{Preprint}
  \eprint{hep-lat/0506013})

\bibitem{Beane:2006mx}
Beane S, Bedaque P, Orginos K and Savage M 2006 {\em Phys.Rev.Lett.\/} {\bf 97}
  012001 (\textit{Preprint} \eprint{hep-lat/0602010})

\bibitem{Beane:2006gf}
Beane S~R {\em et~al.\/} (NPLQCD Collaboration) 2007 {\em Nucl.Phys.\/} {\bf
  A794} 62--72 (\textit{Preprint} \eprint{hep-lat/0612026})

\bibitem{Beane:2007es}
Beane S~R, Detmold W, Luu T~C, Orginos K, Savage M~J {\em et~al.\/} 2008 {\em
  Phys.Rev.Lett.\/} {\bf 100} 082004 (\textit{Preprint} \eprint{0710.1827})

\bibitem{Detmold:2008fn}
Detmold W, Savage M~J, Torok A, Beane S~R, Luu T~C {\em et~al.\/} 2008 {\em
  Phys.Rev.\/} {\bf D78} 014507 (\textit{Preprint} \eprint{0803.2728})

\bibitem{Beane:2009py}
Beane S~R {\em et~al.\/} (NPLQCD Collaboration) 2010 {\em Phys.Rev.\/} {\bf
  D81} 054505 (\textit{Preprint} \eprint{0912.4243})

\bibitem{Thomas:2011rh}
Thomas C~E, Edwards R~G and Dudek J~J 2012 {\em Phys.Rev.\/} {\bf D85} 014507
  (\textit{Preprint} \eprint{1107.1930})

\bibitem{Beane:2011sc}
Beane S {\em et~al.\/} (NPLQCD Collaboration) 2012 {\em Phys.Rev.\/} {\bf D85}
  034505 (\textit{Preprint} \eprint{1107.5023})

\bibitem{Basak:2005ir}
Basak S {\em et~al.\/} (Lattice Hadron Physics Collaboration (LHPC)) 2005 {\em
  Phys.Rev.\/} {\bf D72} 074501 (\textit{Preprint} \eprint{hep-lat/0508018})

\bibitem{Peardon:2009gh}
Peardon M {\em et~al.\/} (Hadron Spectrum Collaboration) 2009 {\em Phys.Rev.\/}
  {\bf D80} 054506 (\textit{Preprint} \eprint{0905.2160})

\bibitem{Dudek:2010wm}
Dudek J~J, Edwards R~G, Peardon M~J, Richards D~G and Thomas C~E 2010 {\em
  Phys.Rev.\/} {\bf D82} 034508 (\textit{Preprint} \eprint{1004.4930})

\bibitem{Edwards:2011jj}
Edwards R~G, Dudek J~J, Richards D~G and Wallace S~J 2011 {\em Phys.Rev.\/}
  {\bf D84} 074508 (\textit{Preprint} \eprint{1104.5152})

\bibitem{Dudek:2012ag}
Dudek J~J and Edwards R~G 2012 {\em Phys.Rev.\/} {\bf D85} 054016
  (\textit{Preprint} \eprint{1201.2349})

\bibitem{Dudek:2012gj}
Dudek J~J, Edwards R~G and Thomas C~E 2012 {\em Phys.Rev.\/} {\bf D86} 034031
  (\textit{Preprint} \eprint{1203.6041})

\bibitem{Yamazaki:2009ua}
Yamazaki T, Kuramashi Y and Ukawa A (PACS-CS Collaboration) 2010 {\em
  Phys.Rev.\/} {\bf D81} 111504 (\textit{Preprint} \eprint{0912.1383})

\bibitem{Yamazaki:2012hi}
Yamazaki T, Ishikawa K~i, Kuramashi Y and Ukawa A 2012 {\em Phys.Rev.\/} {\bf
  D86} 074514 (\textit{Preprint} \eprint{1207.4277})

\bibitem{Shi:2011mr}
Shi Z and Detmold W 2011 {\em PoS\/} {\bf LATTICE2011} 328 (\textit{Preprint}
  \eprint{1111.1656})

\bibitem{Detmold:2012wc}
Detmold W, Orginos K and Shi Z 2012 {\em Phys.Rev.\/} {\bf D86} 054507
  (\textit{Preprint} \eprint{1205.4224})

\bibitem{Detmold:2010au}
Detmold W and Savage M~J 2010 {\em Phys.Rev.\/} {\bf D82} 014511
  (\textit{Preprint} \eprint{1001.2768})

\bibitem{Doi:2012xd}
Doi T and Endres M~G 2013 {\em Comput. Phys. Commun.\/} {\bf 184} 117
  (\textit{Preprint} \eprint{1205.0585})

\bibitem{Detmold:2012eu}
Detmold W and Orginos K 2012  (\textit{Preprint} \eprint{1207.1452})

\bibitem{Detmold:2013gua}
Detmold W and Nicholson A~N 2013 {\em Phys.Rev.\/} {\bf D88} 074501
  (\textit{Preprint} \eprint{1308.5186})

\bibitem{Aoki:2013ldr}
Aoki S, Aoki Y, Bernard C, Blum T, Colangelo G {\em et~al.\/} 2013
  (\textit{Preprint} \eprint{1310.8555})

\bibitem{Beane:2007xs}
Beane S~R, Luu T~C, Orginos K, Parreno A, Savage M~J {\em et~al.\/} 2008 {\em
  Phys.Rev.\/} {\bf D77} 014505 (\textit{Preprint} \eprint{0706.3026})

\bibitem{Feng:2009ij}
Feng X, Jansen K and Renner D~B 2010 {\em Phys.Lett.\/} {\bf B684} 268--274
  (\textit{Preprint} \eprint{0909.3255})

\bibitem{Yamazaki:2004qb}
Yamazaki T {\em et~al.\/} (CP-PACS Collaboration) 2004 {\em Phys.Rev.\/} {\bf
  D70} 074513 (\textit{Preprint} \eprint{hep-lat/0402025})

\bibitem{Dudek:2010ew}
Dudek J~J, Edwards R~G, Peardon M~J, Richards D~G and Thomas C~E 2011 {\em
  Phys.Rev.\/} {\bf D83} 071504 (\textit{Preprint} \eprint{1011.6352})

\bibitem{Dudek:2012xn}
Dudek J~J, Edwards R~G and Thomas C~E 2013 {\em Phys.Rev.\/} {\bf D87} 034505
  (\textit{Preprint} \eprint{1212.0830})

\bibitem{Luscher:1986pf}
Luscher M 1986 {\em Commun.Math.Phys.\/} {\bf 105} 153--188

\bibitem{Luscher:1990ux}
Luscher M 1991 {\em Nucl.Phys.\/} {\bf B354} 531--578

\bibitem{Rummukainen:1995vs}
Rummukainen K and Gottlieb S~A 1995 {\em Nucl. Phys.\/} {\bf B450} 397--436
  (\textit{Preprint} \eprint{hep-lat/9503028})

\bibitem{Beane:2003yx}
Beane S, Bedaque P, Parreno A and Savage M 2005 {\em Nucl.Phys.\/} {\bf A747}
  55--74 (\textit{Preprint} \eprint{nucl-th/0311027})

\bibitem{Beane:2003da}
Beane S~R, Bedaque P, Parreno A and Savage M 2004 {\em Phys.Lett.\/} {\bf B585}
  106--114 (\textit{Preprint} \eprint{hep-lat/0312004})

\bibitem{Li:2003jn}
Li X and Liu C 2004 {\em Phys.Lett.\/} {\bf B587} 100--104 (\textit{Preprint}
  \eprint{hep-lat/0311035})

\bibitem{Detmold:2004qn}
Detmold W and Savage M~J 2004 {\em Nucl.Phys.\/} {\bf A743} 170--193
  (\textit{Preprint} \eprint{hep-lat/0403005})

\bibitem{Feng:2004ua}
Feng X, Li X and Liu C 2004 {\em Phys.Rev.\/} {\bf D70} 014505
  (\textit{Preprint} \eprint{hep-lat/0404001})

\bibitem{Liu:2005kr}
Liu C, Feng X and He S 2006 {\em Int.J.Mod.Phys.\/} {\bf A21} 847--850
  (\textit{Preprint} \eprint{hep-lat/0508022})

\bibitem{Bedaque:2004kc}
Bedaque P~F 2004 {\em Phys.Lett.\/} {\bf B593} 82--88 (\textit{Preprint}
  \eprint{nucl-th/0402051})

\bibitem{Christ:2005gi}
Christ N~H, Kim C and Yamazaki T 2005 {\em Phys.Rev.\/} {\bf D72} 114506
  (\textit{Preprint} \eprint{hep-lat/0507009})

\bibitem{He:2005ey}
He S, Feng X and Liu C 2005 {\em JHEP\/} {\bf 0507} 011 (\textit{Preprint}
  \eprint{hep-lat/0504019})

\bibitem{Kim:2005gf}
Kim C, Sachrajda C and Sharpe S~R 2005 {\em Nucl.Phys.\/} {\bf B727} 218--243
  (\textit{Preprint} \eprint{hep-lat/0507006})

\bibitem{Bernard:2008ax}
Bernard V, Lage M, Meissner U~G and Rusetsky A 2008 {\em JHEP\/} {\bf 0808} 024
  (\textit{Preprint} \eprint{0806.4495})

\bibitem{Lage:2009zv}
Lage M, Meissner U~G and Rusetsky A 2009 {\em Phys.Lett.\/} {\bf B681} 439--443
  (\textit{Preprint} \eprint{0905.0069})

\bibitem{Bour:2011ef}
Bour S, Koenig S, Lee D, Hammer H~W and Meissner U~G 2011 {\em Phys.Rev.\/}
  {\bf D84} 091503 (\textit{Preprint} \eprint{1107.1272})

\bibitem{Davoudi:2011md}
Davoudi Z and Savage M~J 2011 {\em Phys.Rev.\/} {\bf D84} 114502
  (\textit{Preprint} \eprint{1108.5371})

\bibitem{Fu:2011xz}
Fu Z 2012 {\em Phys.Rev.\/} {\bf D85} 014506 (\textit{Preprint}
  \eprint{1110.0319})

\bibitem{Leskovec:2012gb}
Leskovec L and Prelovsek S 2012 {\em Phys.Rev.\/} {\bf D85} 114507
  (\textit{Preprint} \eprint{1202.2145})

\bibitem{Gockeler:2012yj}
Gockeler M, Horsley R, Lage M, Meissner U~G, Rakow P {\em et~al.\/} 2012 {\em
  Phys.Rev.\/} {\bf D86} 094513 (\textit{Preprint} \eprint{1206.4141})

\bibitem{Hansen:2012tf}
Hansen M~T and Sharpe S~R 2012 {\em Phys.Rev.\/} {\bf D86} 016007
  (\textit{Preprint} \eprint{1204.0826})

\bibitem{Briceno:2012yi}
Briceno R~A and Davoudi Z 2013 {\em Phys. Rev. D. 88,\/} {\bf 094507}
  (\textit{Preprint} \eprint{1204.1110})

\bibitem{Li:2012bi}
Li N and Liu C 2013 {\em Phys.Rev.\/} {\bf D87} 014502 (\textit{Preprint}
  \eprint{1209.2201})

\bibitem{Guo:2012hv}
Guo P, Dudek J, Edwards R and Szczepaniak A~P 2013 {\em Phys.Rev.\/} {\bf D88}
  014501 (\textit{Preprint} \eprint{1211.0929})

\bibitem{Ishizuka:2009bx}
Ishizuka N 2009 {\em PoS\/} {\bf LAT2009} 119 (\textit{Preprint}
  \eprint{0910.2772})

\bibitem{Bernard:2010fp}
Bernard V, Lage M, Meissner U~G and Rusetsky A 2011 {\em JHEP\/} {\bf 1101} 019
  (\textit{Preprint} \eprint{1010.6018})

\bibitem{Briceno:2013rwa}
Briceno R~A 2013  (\textit{Preprint} \eprint{1311.6032})

\bibitem{Briceno:2013lba}
Briceno R~A, Davoudi Z and Luu T~C 2013 {\em Phys.Rev.\/} {\bf D88} 034502
  (\textit{Preprint} \eprint{1305.4903})

\bibitem{Briceno:2013bda}
Briceno R~A, Davoudi Z, Luu T and Savage M~J 2013  (\textit{Preprint}
  \eprint{1309.3556})

\bibitem{Briceno:2013hya}
Briceno R~A, Davoudi Z, Luu T~C and Savage M~J 2013 {\em Phys.Rev.\/} {\bf D89}
  074509 (\textit{Preprint} \eprint{1311.7686})

\bibitem{Li:2014wga}
Li N, Li S~Y and Liu C 2014  (\textit{Preprint} \eprint{1401.5569})

\bibitem{Beane:2013br}
Beane S, Chang E, Cohen S, Detmold W, Junnarkar P {\em et~al.\/} 2013 {\em
  Phys.Rev.\/} {\bf C88} 024003 (\textit{Preprint} \eprint{1301.5790})

\bibitem{Dudek:2009qf}
Dudek J~J, Edwards R~G, Peardon M~J, Richards D~G and Thomas C~E 2009 {\em
  Phys.Rev.Lett.\/} {\bf 103} 262001 (\textit{Preprint} \eprint{0909.0200})

\bibitem{Dudek:2011tt}
Dudek J~J, Edwards R~G, Joo B, Peardon M~J, Richards D~G {\em et~al.\/} 2011
  {\em Phys.Rev.\/} {\bf D83} 111502 (\textit{Preprint} \eprint{1102.4299})

\bibitem{Dudek:2011bn}
Dudek J~J 2011 {\em Phys.Rev.\/} {\bf D84} 074023 (\textit{Preprint}
  \eprint{1106.5515})

\bibitem{Dudek:2013yja}
Dudek J~J, Edwards R~G, Guo P and Thomas C~E 2013 {\em Phys.Rev.\/} {\bf D88}
  094505 (\textit{Preprint} \eprint{1309.2608})

\bibitem{Li:2007ey}
Li X {\em et~al.\/} (CLQCD Collaboration) 2007 {\em JHEP\/} {\bf 0706} 053
  (\textit{Preprint} \eprint{hep-lat/0703015})

\bibitem{Aoki:2007rd}
Aoki S {\em et~al.\/} (CP-PACS Collaboration) 2007 {\em Phys.Rev.\/} {\bf D76}
  094506 (\textit{Preprint} \eprint{0708.3705})

\bibitem{Beane:2010hg}
Beane S~R {\em et~al.\/} (NPLQCD Collaboration) 2011 {\em Phys.Rev.Lett.\/}
  {\bf 106} 162001 (\textit{Preprint} \eprint{1012.3812})

\bibitem{Beane:2011xf}
Beane S~R, Chang E, Detmold W, Joo B, Lin H {\em et~al.\/} 2011 {\em
  Mod.Phys.Lett.\/} {\bf A26} 2587--2595 (\textit{Preprint} \eprint{1103.2821})

\bibitem{Beane:2011iw}
Beane S~R {\em et~al.\/} (NPLQCD Collaboration) 2012 {\em Phys.Rev.\/} {\bf
  D85} 054511 (\textit{Preprint} \eprint{1109.2889})

\bibitem{Beane:2012ey}
Beane S, Chang E, Cohen S, Detmold W, Lin H~W {\em et~al.\/} 2012 {\em
  Phys.Rev.Lett.\/} {\bf 109} 172001 (\textit{Preprint} \eprint{1204.3606})

\bibitem{Pelissier:2011ib}
Pelissier C~S, Alexandru A and Lee F~X 2011 {\em PoS\/} {\bf LATTICE2011} 134
  (\textit{Preprint} \eprint{1111.2314})

\bibitem{Lang:2011mn}
Lang C, Mohler D, Prelovsek S and Vidmar M 2011 {\em Phys.Rev.\/} {\bf D84}
  054503 (\textit{Preprint} \eprint{1105.5636})

\bibitem{Pelissier:2012pi}
Pelissier C and Alexandru A 2013 {\em Phys.Rev.\/} {\bf D87} 014503
  (\textit{Preprint} \eprint{1211.0092})

\bibitem{Ozaki:2012ce}
Ozaki S and Sasaki S 2013 {\em Phys.Rev.\/} {\bf D87} 014506 (\textit{Preprint}
  \eprint{1211.5512})

\bibitem{Buchoff:2012ja}
Buchoff M~I, Luu T~C and Wasem J 2012 {\em Phys.Rev.\/} {\bf D85} 094511
  (\textit{Preprint} \eprint{1201.3596})

\bibitem{Mohler:2013rwa}
Mohler D, Lang C, Leskovec L, Prelovsek S and Woloshyn R 2013 {\em
  Phys.Rev.Lett.\/} {\bf 111} 222001 (\textit{Preprint} \eprint{1308.3175})

\bibitem{Lang:2014tia}
Lang C, Leskovec L, Mohler D and Prelovsek S 2014  (\textit{Preprint}
  \eprint{1401.2088})

\bibitem{Briceno:2013pda}
Briceno R~A 2013  (\textit{Preprint} \eprint{1309.7923})

\bibitem{Roca:2012rx}
Roca L and Oset E 2012 {\em Phys.Rev.\/} {\bf D85} 054507 (\textit{Preprint}
  \eprint{1201.0438})

\bibitem{Polejaeva:2012ut}
Polejaeva K and Rusetsky A 2012 {\em Eur.Phys.J.\/} {\bf A48} 67
  (\textit{Preprint} \eprint{1203.1241})

\bibitem{Briceno:2012rv}
Briceno R~A and Davoudi Z 2012 {\em Phys.Rev.\/} {\bf D87} 094507
  (\textit{Preprint} \eprint{1212.3398})

\bibitem{Hansen:2013dla}
Hansen M~T and Sharpe S~R 2013  (\textit{Preprint} \eprint{1311.4848})

\bibitem{Hansen:2014eka}
Hansen M~T and Sharpe S~R 2014  (\textit{Preprint} \eprint{1408.5933})

\bibitem{Aoki:2009ji}
Aoki S, Hatsuda T and Ishii N 2010 {\em Prog.Theor.Phys.\/} {\bf 123} 89--128
  (\textit{Preprint} \eprint{0909.5585})

\bibitem{HALQCD:2012aa}
Ishii N {\em et~al.\/} (HAL QCD Collaboration) 2012 {\em Phys.Lett.\/} {\bf
  B712} 437--441 (\textit{Preprint} \eprint{1203.3642})

\bibitem{Aoki:2012tk}
Aoki S {\em et~al.\/} (HAL QCD Collaboration) 2012 {\em PTEP\/} {\bf 2012}
  01A105 (\textit{Preprint} \eprint{1206.5088})

\bibitem{Murano:2013xxa}
Murano K, Ishii N, Aoki S, Doi T, Hatsuda T {\em et~al.\/} 2013
  (\textit{Preprint} \eprint{1305.2293})

\bibitem{Aoki:2013cra}
Aoki S, Ishii N, Doi T, Ikeda Y and Inoue T 2013 {\em Phys.Rev.\/} {\bf D88}
  014036 (\textit{Preprint} \eprint{1303.2210})

\bibitem{Zheng:1993qx}
Zheng D~C, Barrett B~R, Jaqua L, Vary J~P and Mccarthy R~J 1993 {\em
  Phys.Rev.\/} {\bf C48} 1083--1091 (\textit{Preprint}
  \eprint{nucl-th/9304025})

\bibitem{Jaqua:1993zz}
Jaqua L, Zheng D~C, Barrett B~R and Vary J~P 1993 {\em Phys.Rev.\/} {\bf C48}
  1765--1769

\bibitem{Zheng:1994zza}
Zheng D~C, Barrett B~R, Vary J~P and Mccarthy R~J 1994 {\em Phys.Rev.\/} {\bf
  C49} 1999--2004

\bibitem{Navratil:1996vm}
Navratil P and Barrett B~R 1996 {\em Phys.Rev.\/} {\bf C54} 2986--2995
  (\textit{Preprint} \eprint{nucl-th/9609046})

\bibitem{Navratil:1998uf}
Navratil P and Barrett B~R 1998 {\em Phys.Rev.\/} {\bf C57} 3119--3128
  (\textit{Preprint} \eprint{nucl-th/9804014})

\bibitem{Navratil:2000gs}
Navratil P, Vary J~P and Barrett B~R 2000 {\em Phys.Rev.\/} {\bf C62} 054311

\bibitem{Zhan:2004ct}
Zhan H, Nogga A, Barrett B, Vary J and Navratil P 2004 {\em Phys.Rev.\/} {\bf
  C69} 034302 (\textit{Preprint} \eprint{nucl-th/0401047})

\bibitem{Barrett:2012dr}
Barrett B~R 2012 {\em J.Phys.Conf.Ser.\/} {\bf 403} 012013

\bibitem{Barrett:2013nh}
Barrett B~R, Navratil P and Vary J~P 2013 {\em Prog.Part.Nucl.Phys.\/} {\bf 69}
  131--181

\bibitem{Baroni:2013fe}
Baroni S, Navratil P and Quaglioni S 2013 {\em Phys.Rev.\/} {\bf C87} 034326
  (\textit{Preprint} \eprint{1301.3450})

\bibitem{Baroni:2012su}
Baroni S, Navratil P and Quaglioni S 2013 {\em Phys.Rev.Lett.\/} {\bf 110}
  022505 (\textit{Preprint} \eprint{1210.1897})

\bibitem{Hagen:2007hi}
Hagen G, Dean D, Hjorth-Jensen M, Papenbrock T and Schwenk A 2007 {\em
  Phys.Rev.\/} {\bf C76} 044305 (\textit{Preprint} \eprint{0707.1516})

\bibitem{Hagen:2008iw}
Hagen G, Papenbrock T, Dean D and Hjorth-Jensen M 2008 {\em Phys.Rev.Lett.\/}
  {\bf 101} 092502 (\textit{Preprint} \eprint{0806.3478})

\bibitem{Gezerlis:2013ipa}
Gezerlis A, Tews I, Epelbaum E, Gandolfi S, Hebeler K {\em et~al.\/} 2013 {\em
  Phys.Rev.Lett.\/} {\bf 111} 032501 (\textit{Preprint} \eprint{1303.6243})

\bibitem{Gezerlis:2014zia}
Gezerlis A, Tews I, Epelbaum E, Freunek M, Gandolfi S {\em et~al.\/} 2014
  (\textit{Preprint} \eprint{1406.0454})

\bibitem{Lynn:2014zia}
Lynn J, Carlson J, Epelbaum E, Gandolfi S, Gezerlis A {\em et~al.\/} 2014
  (\textit{Preprint} \eprint{1406.2787})

\bibitem{Roggero:2014lga}
Roggero A, Mukherjee A and Pederiva F 2014 {\em Phys.Rev.Lett.\/} {\bf 112}
  221103 (\textit{Preprint} \eprint{1402.1576})

\bibitem{Otsuka:2009cs}
Otsuka T, Suzuki T, Holt J~D, Schwenk A and Akaishi Y 2010 {\em
  Phys.Rev.Lett.\/} {\bf 105} 032501 (\textit{Preprint} \eprint{0908.2607})

\bibitem{Holt:2010yb}
Holt J~D, Otsuka T, Schwenk A and Suzuki T 2012 {\em J.Phys.\/} {\bf G39}
  085111 (\textit{Preprint} \eprint{1009.5984})

\bibitem{Gallant:2012as}
Gallant A, Bale J, Brunner T, Chowdhury U, Ettenauer S {\em et~al.\/} 2012 {\em
  Phys.Rev.Lett.\/} {\bf 109} 032506 (\textit{Preprint} \eprint{1204.1987})

\bibitem{Wienholtz:2013nya}
Wienholtz F, Beck D, Blaum K, Borgmann C, Breitenfeldt M {\em et~al.\/} 2013
  {\em Nature\/} {\bf 498} 346--349

\bibitem{Holt:2014aya}
Holt J, Menendez J, Simonis J and Schwenk A 2014 {\em Phys.Rev.\/} {\bf C90}
  024312 (\textit{Preprint} \eprint{1405.7602})

\bibitem{Holt:2012fr}
Holt J, Menendez J and Schwenk A 2013 {\em Phys.Rev.Lett.\/} {\bf 110} 022502
  (\textit{Preprint} \eprint{1207.1509})

\bibitem{Pudliner:1997ck}
Pudliner B, Pandharipande V, Carlson J, Pieper S~C and Wiringa R~B 1997 {\em
  Phys.Rev.\/} {\bf C56} 1720--1750 (\textit{Preprint}
  \eprint{nucl-th/9705009})

\bibitem{Wiringa:2000gb}
Wiringa R~B, Pieper S~C, Carlson J and Pandharipande V 2000 {\em Phys.Rev.\/}
  {\bf C62} 014001 (\textit{Preprint} \eprint{nucl-th/0002022})

\bibitem{Pieper:2004qw}
Pieper S~C, Wiringa R~B and Carlson J 2004 {\em Phys.Rev.\/} {\bf C70} 054325
  (\textit{Preprint} \eprint{nucl-th/0409012})

\bibitem{Cipollone:2013zma}
Cipollone A, Barbieri C and Navrátil P 2013 {\em Phys.Rev.Lett.\/} {\bf 111}
  062501 (\textit{Preprint} \eprint{1303.4900})

\bibitem{Soma:2013xha}
Somà V, Cipollone A, Barbieri C, Navrátil P and Duguet T 2014 {\em Phys.Rev.\/}
  {\bf C89} 061301 (\textit{Preprint} \eprint{1312.2068})

\bibitem{Tsukiyama:2010rj}
Tsukiyama K, Bogner S and Schwenk A 2011 {\em Phys.Rev.Lett.\/} {\bf 106}
  222502 (\textit{Preprint} \eprint{1006.3639})

\bibitem{Tsukiyama:2012sm}
Tsukiyama K, Bogner S and Schwenk A 2012 {\em Phys.Rev.\/} {\bf C85} 061304
  (\textit{Preprint} \eprint{1203.2515})

\bibitem{Hergert:2012nb}
Hergert H, Bogner S, Binder S, Calci A, Langhammer J {\em et~al.\/} 2013 {\em
  Phys.Rev.\/} {\bf C87} 034307 (\textit{Preprint} \eprint{1212.1190})

\bibitem{Hergert:2013uja}
Hergert H, Binder S, Calci A, Langhammer J and Roth R 2013 {\em
  Phys.Rev.Lett.\/} {\bf 110} 242501 (\textit{Preprint} \eprint{1302.7294})

\bibitem{Bogner:2014baa}
Bogner S, Hergert H, Holt J, Schwenk A, Binder S {\em et~al.\/} 2014
  (\textit{Preprint} \eprint{1402.1407})

\bibitem{Lee:2004si}
Lee D, Borasoy B and Schaefer T 2004 {\em Phys.Rev.\/} {\bf C70} 014007
  (\textit{Preprint} \eprint{nucl-th/0402072})

\bibitem{Borasoy:2005yc}
Borasoy B, Krebs H, Lee D and Meissner U~G 2006 {\em Nucl.Phys.\/} {\bf A768}
  179--193 (\textit{Preprint} \eprint{nucl-th/0510047})

\bibitem{Borasoy:2006qn}
Borasoy B, Epelbaum E, Krebs H, Lee D and Meissner U~G 2007 {\em Eur.Phys.J.\/}
  {\bf A31} 105--123 (\textit{Preprint} \eprint{nucl-th/0611087})

\bibitem{Tiburzi:2005hg}
Tiburzi B~C 2005 {\em Phys.Lett.\/} {\bf B617} 40--48 (\textit{Preprint}
  \eprint{hep-lat/0504002})

\bibitem{Jiang:2006gna}
Jiang F~J and Tiburzi B 2007 {\em Phys.Lett.\/} {\bf B645} 314--321
  (\textit{Preprint} \eprint{hep-lat/0610103})

\bibitem{Boyle:2007wg}
Boyle P, Flynn J, Juttner A, Sachrajda C and Zanotti J 2007 {\em JHEP\/} {\bf
  0705} 016 (\textit{Preprint} \eprint{hep-lat/0703005})

\bibitem{Simula:2007fa}
Simula S (ETMC Collaboration) 2007 {\em PoS\/} {\bf LAT2007} 371
  (\textit{Preprint} \eprint{0710.0097})

\bibitem{Boyle:2008yd}
Boyle P, Flynn J, Juttner A, Kelly C, de~Lima H~P {\em et~al.\/} 2008 {\em
  JHEP\/} {\bf 0807} 112 (\textit{Preprint} \eprint{0804.3971})

\bibitem{Aoki:2008gv}
Aoki S, Horsley R, Izubuchi T, Nakamura Y, Pleiter D {\em et~al.\/} 2008
  (\textit{Preprint} \eprint{0808.1428})

\bibitem{Boyle:2012nb}
Boyle P~A, Flynn J~M, Juttner A, Sachrajda C, Sivalingam K {\em et~al.\/} 2012
  {\em PoS\/} {\bf LATTICE2012} 112 (\textit{Preprint} \eprint{1212.3188})

\bibitem{Brandt:2013mb}
Brandt B~B, Juttner A and Wittig H 2012 {\em PoS\/} {\bf ConfinementX} 112
  (\textit{Preprint} \eprint{1301.3513})

\bibitem{Osterwalder:1973dx}
Osterwalder K and Schrader R 1973 {\em Commun.Math.Phys.\/} {\bf 31} 83--112

\bibitem{Maiani:1990ca}
Maiani L and Testa M 1990 {\em Phys.Lett.\/} {\bf B245} 585--590

\bibitem{Huang:1957im}
Huang K and Yang C 1957 {\em Phys.Rev.\/} {\bf 105} 767--775

\bibitem{Luscher:1985dn}
Luscher M 1986 {\em Commun.Math.Phys.\/} {\bf 104} 177

\bibitem{Bedaque:2006yi}
Bedaque P~F, Sato I and Walker-Loud A 2006 {\em Phys.Rev.\/} {\bf D73} 074501
  (\textit{Preprint} \eprint{hep-lat/0601033})

\bibitem{Sato:2007ms}
Sato I and Bedaque P~F 2007 {\em Phys.Rev.\/} {\bf D76} 034502
  (\textit{Preprint} \eprint{hep-lat/0702021})

\bibitem{PhysRevLett.7.46}
Byers N and Yang C~N 1961 {\em Phys. Rev. Lett.\/} {\bf 7}(2) 46--49

\bibitem{deDivitiis:2004rf}
de~Divitiis G~M and Tantalo N 2004  (\textit{Preprint}
  \eprint{hep-lat/0409154})

\bibitem{Sachrajda:2004mi}
Sachrajda C and Villadoro G 2005 {\em Phys.Lett.\/} {\bf B609} 73--85
  (\textit{Preprint} \eprint{hep-lat/0411033})

\bibitem{Jiang:2008ja}
Jiang F~J and Tiburzi B 2008 {\em Phys.Rev.\/} {\bf D78} 114505
  (\textit{Preprint} \eprint{0810.1495})

\bibitem{Bedaque:2004ax}
Bedaque P~F and Chen J~W 2005 {\em Phys.Lett.\/} {\bf B616} 208--214
  (\textit{Preprint} \eprint{hep-lat/0412023})

\bibitem{Agadjanov:2013wqa}
Agadjanov D, Meissner U~G and Rusetsky A 2013  (\textit{Preprint}
  \eprint{1310.7875})

\bibitem{Briceno:2014oea}
Briceno R~A 2014  (\textit{Preprint} \eprint{1401.3312})

\bibitem{Shepherd:2009zz}
Shepherd M (GLUEX Collaboration) 2009 {\em AIP Conf.Proc.\/} {\bf 1182}
  816--819

\bibitem{Zihlmann:2010zz}
Zihlmann B (GlueX Collaboration) 2010 {\em AIP Conf.Proc.\/} {\bf 1257}
  116--120

\bibitem{Somov:2011zz}
Somov A (GlueX Collaboration) 2011 {\em AIP Conf.Proc.\/} {\bf 1374} 282--285

\bibitem{Smith:2012ch}
Smith E~S (GlueX Collaboration) 2012 {\em EPJ Web Conf.\/} {\bf 37} 01026

\bibitem{Hansen:2012bj}
Hansen M~T and Sharpe S~R 2012 {\em PoS\/} {\bf LATTICE2012} 127
  (\textit{Preprint} \eprint{1211.0511})

\bibitem{Dresselhaus}
Dresselhaus M~S, Dresselhaus G and Jorio A 2008 {\em Applications of Group
  Theory to the Physics of Solids\/} 1st ed (Springer)

\bibitem{Moore:2005dw}
Moore D~C and Fleming G~T 2006 {\em Phys.Rev.\/} {\bf D73} 014504
  (\textit{Preprint} \eprint{hep-lat/0507018})

\bibitem{Moore:2006ng}
Moore D~C and Fleming G~T 2006 {\em Phys.Rev.\/} {\bf D74} 054504
  (\textit{Preprint} \eprint{hep-lat/0607004})

\bibitem{Luu:2011ep}
Luu T and Savage M~J 2011 {\em Phys.Rev.\/} {\bf D83} 114508 (\textit{Preprint}
  \eprint{1101.3347})

\bibitem{Mandula:1983wb}
Mandula J~E and Shpiz E 1984 {\em Nucl.Phys.\/} {\bf B232} 180

\bibitem{Johnson:1982yq}
Johnson R 1982 {\em Phys.Lett.\/} {\bf B114} 147

\bibitem{Basak:2005aq}
Basak S, Edwards R, Fleming G, Heller U, Morningstar C {\em et~al.\/} 2005 {\em
  Phys.Rev.\/} {\bf D72} 094506 (\textit{Preprint} \eprint{hep-lat/0506029})

\bibitem{Berkowitz:2012xq}
Berkowitz E, Cohen T~D and Jefferson P 2012  (\textit{Preprint}
  \eprint{1211.2261})

\bibitem{Guo:2013vsa}
Guo P 2013 {\em Phys.Rev.\/} {\bf D88} 014507 (\textit{Preprint}
  \eprint{1304.7812})

\bibitem{Doring:2011vk}
Doring M, Meissner U~G, Oset E and Rusetsky A 2011 {\em Eur.Phys.J.\/} {\bf
  A47} 139 (\textit{Preprint} \eprint{1107.3988})

\bibitem{Doring:2011ip}
Doring M, Haidenbauer J, Meissner U~G and Rusetsky A 2011 {\em Eur.Phys.J.\/}
  {\bf A47} 163 (\textit{Preprint} \eprint{1108.0676})

\bibitem{Doring:2012eu}
Doring M, Meissner U, Oset E and Rusetsky A 2012 {\em Eur.Phys.J.\/} {\bf A48}
  114 (\textit{Preprint} \eprint{1205.4838})

\bibitem{Wu:2014vma}
Wu J~J, Lee T~S~H, Thomas A and Young R 2014  (\textit{Preprint}
  \eprint{1402.4868})

\bibitem{Dudek:2014qha}
Dudek J~J, Edwards R~G, Thomas C~E and Wilson D~J 2014  (\textit{Preprint}
  \eprint{1406.4158})

\bibitem{Hayakawa:2008an}
Hayakawa M and Uno S 2008 {\em Prog.Theor.Phys.\/} {\bf 120} 413--441
  (\textit{Preprint} \eprint{0804.2044})

\bibitem{Davoudi:2014qua}
Davoudi Z and Savage M~J 2014 {\em Phys.Rev.\/} {\bf D90} 054503
  (\textit{Preprint} \eprint{1402.6741})

\bibitem{Lellouch:2000pv}
Lellouch L and Luscher M 2001 {\em Commun.Math.Phys.\/} {\bf 219} 31--44
  (\textit{Preprint} \eprint{hep-lat/0003023})

\bibitem{Blum:2011pu}
Blum T, Boyle P, Christ N, Garron N, Goode E {\em et~al.\/} 2011 {\em
  Phys.Rev.\/} {\bf D84} 114503 (\textit{Preprint} \eprint{1106.2714})

\bibitem{Blum:2011ng}
Blum T, Boyle P, Christ N, Garron N, Goode E {\em et~al.\/} 2012 {\em
  Phys.Rev.Lett.\/} {\bf 108} 141601 (\textit{Preprint} \eprint{1111.1699})

\bibitem{Blum:2012uk}
Blum T, Boyle P, Christ N, Garron N, Goode E {\em et~al.\/} 2012 {\em
  Phys.Rev.\/} {\bf D86} 074513 (\textit{Preprint} \eprint{1206.5142})

\bibitem{Boyle:2012ys}
Boyle P {\em et~al.\/} (RBC, UKQCD) 2013 {\em Phys.Rev.Lett.\/} {\bf 110}
  152001 (\textit{Preprint} \eprint{1212.1474})

\bibitem{Meyer:2013dxa}
Meyer H~B 2013  (\textit{Preprint} \eprint{1303.0138})

\bibitem{Meyer:2012wk}
Meyer H~B 2012  (\textit{Preprint} \eprint{1202.6675})

\bibitem{Bernard:2012bi}
Bernard V, Hoja D, Meissner U~G and Rusetsky A 2012 {\em JHEP\/} {\bf 1209} 023
  (\textit{Preprint} \eprint{1205.4642})

\bibitem{Descotes-Genon:2013wba}
Descotes-Genon S, Matias J and Virto J 2013 {\em Phys.Rev.\/} {\bf D88} 074002
  (\textit{Preprint} \eprint{1307.5683})

\bibitem{Altmannshofer:2013foa}
Altmannshofer W and Straub D~M 2013 {\em Eur.Phys.J.\/} {\bf C73} 2646
  (\textit{Preprint} \eprint{1308.1501})

\bibitem{Bobeth:2012vn}
Bobeth C, Hiller G and van Dyk D 2013 {\em Phys.Rev.\/} {\bf D87} 034016
  (\textit{Preprint} \eprint{1212.2321})

\bibitem{vanDyk:2013uaa}
van Dyk D 2013  (\textit{Preprint} \eprint{1307.6699})

\bibitem{Hambrock:2013zya}
Hambrock C, Hiller G, Schacht S and Zwicky R 2013  (\textit{Preprint}
  \eprint{1308.4379})

\bibitem{Beaujean:2013soa}
Beaujean F, Bobeth C and van Dyk D 2013  (\textit{Preprint} \eprint{1310.2478})

\bibitem{Lyon:2014hpa}
Lyon J and Zwicky R 2014  (\textit{Preprint} \eprint{1406.0566})

\bibitem{Lyon:2013gba}
Lyon J and Zwicky R 2013 {\em Phys.Rev.\/} {\bf D88} 094004 (\textit{Preprint}
  \eprint{1305.4797})

\bibitem{Bowler:1993rz}
Bowler K {\em et~al.\/} (UKQCD Collaboration) 1994 {\em Phys.Rev.Lett.\/} {\bf
  72} 1398--1401 (\textit{Preprint} \eprint{hep-lat/9311004})

\bibitem{Bernard:1993yt}
Bernard C~W, Hsieh P and Soni A 1994 {\em Phys.Rev.Lett.\/} {\bf 72} 1402--1405
  (\textit{Preprint} \eprint{hep-lat/9311010})

\bibitem{Burford:1995fc}
Burford D {\em et~al.\/} (UKQCD Collaboration) 1995 {\em Nucl.Phys.\/} {\bf
  B447} 425--440 (\textit{Preprint} \eprint{hep-lat/9503002})

\bibitem{Abada:2002ie}
Abada A {\em et~al.\/} (SPQcdR collaboration) 2003 {\em
  Nucl.Phys.Proc.Suppl.\/} {\bf 119} 625--628 (\textit{Preprint}
  \eprint{hep-lat/0209116})

\bibitem{Bowler:2004zb}
Bowler K, Gill J, Maynard C and Flynn J (UKQCD Collaboration) 2004 {\em JHEP\/}
  {\bf 0405} 035 (\textit{Preprint} \eprint{hep-lat/0402023})

\bibitem{Becirevic:2006nm}
Becirevic D, Lubicz V and Mescia F 2007 {\em Nucl.Phys.\/} {\bf B769} 31--43
  (\textit{Preprint} \eprint{hep-ph/0611295})

\bibitem{Abada:1995fa}
Abada A {\em et~al.\/} (APE Collaboration) 1996 {\em Phys.Lett.\/} {\bf B365}
  275--284 (\textit{Preprint} \eprint{hep-lat/9503020})

\bibitem{Horgan:2013hoa}
Horgan R~R, Liu Z, Meinel S and Wingate M 2013  (\textit{Preprint}
  \eprint{1310.3722})

\bibitem{Horgan:2013pva}
Horgan R~R, Liu Z, Meinel S and Wingate M 2013  (\textit{Preprint}
  \eprint{1310.3887})

\bibitem{Christ:2012se}
Christ N, Izubuchi T, Sachrajda C, Soni A and Yu J (RBC and UKQCD
  Collaborations) 2013 {\em Phys.Rev.\/} {\bf D88} 014508 (\textit{Preprint}
  \eprint{1212.5931})

\bibitem{Bai:2014cva}
Bai Z, Christ N, Izubuchi T, Sachrajda C, Soni A {\em et~al.\/} 2014
  (\textit{Preprint} \eprint{1406.0916})

\bibitem{Agadjanov:2014kha}
Agadjanov A, Bernard V, Mei§ner U~G and Rusetsky A 2014  (\textit{Preprint}
  \eprint{1405.3476})

\bibitem{Wasem:2011zz}
Wasem J 2012 {\em Phys.Rev.\/} {\bf C85} 022501 (\textit{Preprint}
  \eprint{1108.1151})

\bibitem{Briceno:2014uqa}
Brice\~no R~A, Hansen M~T and Walker-Loud A 2014  (\textit{Preprint}
  \eprint{1406.5965})

\bibitem{Kaplan:1996nv}
Kaplan D~B 1997 {\em Nucl. Phys.\/} {\bf B494} 471--484 (\textit{Preprint}
  \eprint{nucl-th/9610052})

\bibitem{Beane:2000fi}
Beane S~R and Savage M~J 2001 {\em Nucl. Phys.\/} {\bf A694} 511--524
  (\textit{Preprint} \eprint{nucl-th/0011067})

\bibitem{Bedaque:1997qi}
Bedaque P~F and Van~Kolck U 1998 {\em Phys.Lett.\/} {\bf B428} 221--226
  (\textit{Preprint} \eprint{nucl-th/9710073})

\bibitem{Bedaque:1998mb}
Bedaque P~F, Hammer H~W and Van~Kolck U 1998 {\em Phys.Rev.\/} {\bf C58}
  641--644 (\textit{Preprint} \eprint{nucl-th/9802057})

\bibitem{Bedaque:1998kg}
Bedaque P~F, Hammer H~W and Van~Kolck U 1999 {\em Phys.Rev.Lett.\/} {\bf 82}
  463--467 (\textit{Preprint} \eprint{nucl-th/9809025})

\bibitem{Bedaque:1998km}
Bedaque P~F, Hammer H~W and Van~Kolck U 1999 {\em Nucl.Phys.\/} {\bf A646}
  444--466 (\textit{Preprint} \eprint{nucl-th/9811046})

\bibitem{Gabbiani:1999yv}
Gabbiani F, Bedaque P~F and Griesshammer H~W 2000 {\em Nucl.Phys.\/} {\bf A675}
  601--620 (\textit{Preprint} \eprint{nucl-th/9911034})

\bibitem{Bedaque:1999vb}
Bedaque P~F and Griesshammer H~W 2000 {\em Nucl.Phys.\/} {\bf A671} 357--379
  (\textit{Preprint} \eprint{nucl-th/9907077})

\bibitem{Bedaque:1999ve}
Bedaque P~F, Hammer H~W and Van~Kolck U 2000 {\em Nucl.Phys.\/} {\bf A676}
  357--370 (\textit{Preprint} \eprint{nucl-th/9906032})

\bibitem{Bedaque:2000ft}
Bedaque P~F, Braaten E and Hammer H~W 2000 {\em Phys.Rev.Lett.\/} {\bf 85}
  908--911 (\textit{Preprint} \eprint{cond-mat/0002365})

\bibitem{Bour:2012hn}
Bour S, Hammer H~W, Lee D and Meissner U~G 2012 {\em Phys.Rev.\/} {\bf C86}
  034003 (\textit{Preprint} \eprint{1206.1765})

\bibitem{Guo:2013qla}
Guo P 2013  (\textit{Preprint} \eprint{1303.3349})

\bibitem{Beane:2007qr}
Beane S~R, Detmold W and Savage M~J 2007 {\em Phys.Rev.\/} {\bf D76} 074507
  (\textit{Preprint} \eprint{0707.1670})

\bibitem{Tan:2007bg}
Tan S 2008 {\em Phys.Rev.\/} {\bf A78} 013636 (\textit{Preprint}
  \eprint{0709.2530})

\bibitem{Kreuzer:2008bi}
Kreuzer S and Hammer H~W 2009 {\em Phys.Lett.\/} {\bf B673} 260--263
  (\textit{Preprint} \eprint{0811.0159})

\bibitem{Kreuzer:2009jp}
Kreuzer S and Hammer H~W 2010 {\em Eur.Phys.J.\/} {\bf A43} 229--240
  (\textit{Preprint} \eprint{0910.2191})

\bibitem{Kreuzer:2010ti}
Kreuzer S and Hammer H~W 2011 {\em Phys.Lett.\/} {\bf B694} 424--429
  (\textit{Preprint} \eprint{1008.4499})

\bibitem{Kreuzer:2012sr}
Kreuzer S and Griesshammer H~W 2012 {\em Eur.Phys.J.\/} {\bf A48} 93
  (\textit{Preprint} \eprint{1205.0277})

\bibitem{Barnea:2013uqa}
Barnea N, Contessi L, Gazit D, Pederiva F and van Kolck U 2013
  (\textit{Preprint} \eprint{1311.4966})

\bibitem{Ishii:2006ec}
Ishii N, Aoki S and Hatsuda T 2007 {\em Phys.Rev.Lett.\/} {\bf 99} 022001
  (\textit{Preprint} \eprint{nucl-th/0611096})

\bibitem{MJS}
Savage M~J 2007  {Unpublished notes.}

\bibitem{Beane:2010em}
Beane S~R, Detmold W, Orginos K and Savage M 2011 {\em Prog.Part.Nucl.Phys.\/}
  {\bf 66} 1--40 (\textit{Preprint} \eprint{1004.2935})

\bibitem{Aoki:2004wq}
Aoki S {\em et~al.\/} (CP-PACS Collaboration) 2005 {\em
  Nucl.Phys.Proc.Suppl.\/} {\bf 140} 305--307 (\textit{Preprint}
  \eprint{hep-lat/0409063})

\bibitem{Walker-Loud:2014iea}
Walker-Loud A 2014  (\textit{Preprint} \eprint{1401.8259})

\bibitem{Aoki:2013zj}
Aoki S, Balog J, Doi T, Inoue T and Weisz P 2013 {\em Int.J.Mod.Phys.\/} {\bf
  E22} 1330012 (\textit{Preprint} \eprint{1302.0185})

\bibitem{Aoki:2013kpa}
Aoki S (HAL QCD Collaboration) 2013 {\em PoS\/} {\bf LATTICE2013} 222
  (\textit{Preprint} \eprint{1312.7194})

\bibitem{Aoki:2012bb}
Aoki S, Charron B, Doi T, Hatsuda T, Inoue T {\em et~al.\/} 2013 {\em
  Phys.Rev.\/} {\bf D87} 034512 (\textit{Preprint} \eprint{1212.4896})

\bibitem{Inoue:2011ai}
Inoue T {\em et~al.\/} (HAL QCD Collaboration) 2012 {\em Nucl.Phys.\/} {\bf
  A881} 28--43 (\textit{Preprint} \eprint{1112.5926})

\bibitem{Doi:2012am}
Doi T (HAL QCD Collaboration) 2012  (\textit{Preprint} \eprint{1212.1606})

\bibitem{Inoue:2013nfe}
Inoue T {\em et~al.\/} (HAL QCD collaboration) 2013 {\em Phys.Rev.Lett.\/} {\bf
  111} 112503 (\textit{Preprint} \eprint{1307.0299})

\bibitem{Lee:2008fa}
Lee D 2009 {\em Prog.Part.Nucl.Phys.\/} {\bf 63} 117--154 (\textit{Preprint}
  \eprint{0804.3501})

\bibitem{Epelbaum:2010xt}
Epelbaum E, Krebs H, Lee D and Meissner U~G 2010 {\em Eur.Phys.J.\/} {\bf A45}
  335--352 (\textit{Preprint} \eprint{1003.5697})

\bibitem{Epelbaum:2011md}
Epelbaum E, Krebs H, Lee D and Meissner U~G 2011 {\em Phys.Rev.Lett.\/} {\bf
  106} 192501 (\textit{Preprint} \eprint{1101.2547})

\bibitem{Lahde:2013kma}
Laehde T~A, Epelbaum E, Krebs H, Lee D, Meissner U~G {\em et~al.\/} 2013
  (\textit{Preprint} \eprint{1311.1968})

\bibitem{Kamada:2001tv}
Kamada H, Nogga A, Gloeckle W, Hiyama E, Kamimura M {\em et~al.\/} 2001 {\em
  Phys.Rev.\/} {\bf C64} 044001 (\textit{Preprint} \eprint{nucl-th/0104057})

\bibitem{Binder:2012mk}
Binder S, Langhammer J, Calci A, Navratil P and Roth R 2013 {\em Phys.Rev.\/}
  {\bf C87} 021303 (\textit{Preprint} \eprint{1211.4748})

\bibitem{Quaglioni:2008sm}
Quaglioni S and Navratil P 2008 {\em Phys.Rev.Lett.\/} {\bf 101} 092501
  (\textit{Preprint} \eprint{0804.1560})

\bibitem{Navratil:2010jn}
Navratil P, Roth R and Quaglioni S 2010 {\em Phys.Rev.\/} {\bf C82} 034609
  (\textit{Preprint} \eprint{1007.0525})

\bibitem{Navratil:2011sa}
Navratil P, Roth R and Quaglioni S 2011 {\em Phys.Lett.\/} {\bf B704} 379--383
  (\textit{Preprint} \eprint{1105.5977})

\bibitem{Bogner:2007rx}
Bogner S, Furnstahl R, Maris P, Perry R, Schwenk A {\em et~al.\/} 2008 {\em
  Nucl.Phys.\/} {\bf A801} 21--42 (\textit{Preprint} \eprint{0708.3754})

\bibitem{Jurgenson:2010wy}
Jurgenson E, Navratil P and Furnstahl R 2011 {\em Phys.Rev.\/} {\bf C83} 034301
  (\textit{Preprint} \eprint{1011.4085})

\bibitem{Weinberg:1990rz}
Weinberg S 1990 {\em Phys.Lett.\/} {\bf B251} 288--292

\bibitem{Kaplan:1998tg}
Kaplan D~B, Savage M~J and Wise M~B 1998 {\em Phys. Lett.\/} {\bf B424}
  390--396 (\textit{Preprint} \eprint{nucl-th/9801034})

\bibitem{Kaplan:1998we}
Kaplan D~B, Savage M~J and Wise M~B 1998 {\em Nucl. Phys.\/} {\bf B534}
  329--355 (\textit{Preprint} \eprint{nucl-th/9802075})

\bibitem{Chen:1999tn}
Chen J~W, Rupak G and Savage M~J 1999 {\em Nucl.Phys.\/} {\bf A653} 386--412
  (\textit{Preprint} \eprint{nucl-th/9902056})

\bibitem{Kaplan:1998sz}
Kaplan D~B, Savage M~J and Wise M~B 1999 {\em Phys. Rev.\/} {\bf C59} 617--629
  (\textit{Preprint} \eprint{nucl-th/9804032})

\bibitem{vanKolck:1998bw}
van Kolck U 1999 {\em Nucl.Phys.\/} {\bf A645} 273--302 (\textit{Preprint}
  \eprint{nucl-th/9808007})

\bibitem{Fleming:1999ee}
Fleming S, Mehen T and Stewart I~W 2000 {\em Nucl.Phys.\/} {\bf A677} 313--366
  (\textit{Preprint} \eprint{nucl-th/9911001})

\bibitem{Beane:2001bc}
Beane S, Bedaque P~F, Savage M and van Kolck U 2002 {\em Nucl.Phys.\/} {\bf
  A700} 377--402 (\textit{Preprint} \eprint{nucl-th/0104030})

\bibitem{Nogga:2005hy}
Nogga A, Timmermans R and van Kolck U 2005 {\em Phys.Rev.\/} {\bf C72} 054006
  (\textit{Preprint} \eprint{nucl-th/0506005})

\bibitem{Birse:2005um}
Birse M~C 2006 {\em Phys.Rev.\/} {\bf C74} 014003 (\textit{Preprint}
  \eprint{nucl-th/0507077})

\bibitem{Epelbaum:2006pt}
Epelbaum E and Meissner U~G 2013 {\em Few Body Syst.\/} {\bf 54} 2175--2190
  (\textit{Preprint} \eprint{nucl-th/0609037})

\bibitem{Birse:2007sx}
Birse M~C 2007 {\em Phys.Rev.\/} {\bf C76} 034002 (\textit{Preprint}
  \eprint{0706.0984})

\bibitem{Furnstahl:2008df}
Furnstahl R~J, Rupak G and Schaefer T 2008 {\em Ann.Rev.Nucl.Part.Sci.\/} {\bf
  58} 1--25 (\textit{Preprint} \eprint{0801.0729})

\bibitem{Kruger:2013kua}
KrŸger T, Tews I, Hebeler K and Schwenk A 2013 {\em Phys.Rev.\/} {\bf C88}
  025802 (\textit{Preprint} \eprint{1304.2212})

\bibitem{Tews:2013wma}
Tews I, Krüger T, Gezerlis A, Hebeler K and Schwenk A 2013  (\textit{Preprint}
  \eprint{1310.3643})

\bibitem{Drischler:2013iza}
Drischler C, Soma V and Schwenk A 2014 {\em Phys.Rev.\/} {\bf C89} 025806
  (\textit{Preprint} \eprint{1310.5627})

\bibitem{Hebeler:2010jx}
Hebeler K, Lattimer J, Pethick C and Schwenk A 2010 {\em Phys.Rev.Lett.\/} {\bf
  105} 161102 (\textit{Preprint} \eprint{1007.1746})

\bibitem{2013ApJ...773...11H}
{Hebeler} K, {Lattimer} J~M, {Pethick} C~J and {Schwenk} A 2013 {\em Astrophys.
  J.\/} {\bf 773} 11 (\textit{Preprint} \eprint{1303.4662})

\bibitem{Epelbaum:2008ga}
Epelbaum E, Hammer H~W and Meissner U~G 2009 {\em Rev.Mod.Phys.\/} {\bf 81}
  1773--1825 (\textit{Preprint} \eprint{0811.1338})

\bibitem{Machleidt:2011zz}
Machleidt R and Entem D 2011 {\em Phys.Rept.\/} {\bf 503} 1--75
  (\textit{Preprint} \eprint{1105.2919})

\bibitem{Buettiker:1999ap}
Buettiker P and Meissner U~G 2000 {\em Nucl.Phys.\/} {\bf A668} 97--112
  (\textit{Preprint} \eprint{hep-ph/9908247})

\bibitem{Fettes:1998ud}
Fettes N, Meissner U~G and Steininger S 1998 {\em Nucl.Phys.\/} {\bf A640}
  199--234 (\textit{Preprint} \eprint{hep-ph/9803266})

\bibitem{Entem:2001cg}
Entem D and Machleidt R 2002 {\em Phys.Lett.\/} {\bf B524} 93--98
  (\textit{Preprint} \eprint{nucl-th/0108057})

\bibitem{Entem:2002sf}
Entem D and Machleidt R 2002 {\em Phys.Rev.\/} {\bf C66} 014002
  (\textit{Preprint} \eprint{nucl-th/0202039})

\bibitem{Entem:2003ft}
Entem D and Machleidt R 2003 {\em Phys.Rev.\/} {\bf C68} 041001
  (\textit{Preprint} \eprint{nucl-th/0304018})

\bibitem{Machleidt:2005uz}
Machleidt R and Entem D 2005 {\em J.Phys.\/} {\bf G31} S1235--S1244
  (\textit{Preprint} \eprint{nucl-th/0503025})

\bibitem{vanKolck:1994yi}
van Kolck U 1994 {\em Phys.Rev.\/} {\bf C49} 2932--2941

\bibitem{Epelbaum:2002vt}
Epelbaum E, Nogga A, Gloeckle W, Kamada H, Meissner U~G {\em et~al.\/} 2002
  {\em Phys.Rev.\/} {\bf C66} 064001 (\textit{Preprint}
  \eprint{nucl-th/0208023})

\bibitem{Friar:2003yv}
Friar J~L, van Kolck U, Payne G and Coon S 2003 {\em Phys.Rev.\/} {\bf C68}
  024003 (\textit{Preprint} \eprint{nucl-th/0303058})

\bibitem{Friar:2004ca}
Friar J~L, van Kolck U, Rentmeester M and Timmermans R 2004 {\em Phys.Rev.\/}
  {\bf C70} 044001 (\textit{Preprint} \eprint{nucl-th/0406026})

\bibitem{Gazit:2008ma}
Gazit D, Quaglioni S and Navratil P 2009 {\em Phys.Rev.Lett.\/} {\bf 103}
  102502 (\textit{Preprint} \eprint{0812.4444})

\bibitem{Maris:2012bt}
Maris P, Vary J~P and Navratil P 2013 {\em Phys.Rev.\/} {\bf C87} 014327
  (\textit{Preprint} \eprint{1205.5686})

\bibitem{Hammer:2012id}
Hammer H~W, Nogga A and Schwenk A 2013 {\em Rev.Mod.Phys.\/} {\bf 85} 197
  (\textit{Preprint} \eprint{1210.4273})

\bibitem{Demorest:2010bx}
Demorest P, Pennucci T, Ransom S, Roberts M and Hessels J 2010 {\em Nature\/}
  {\bf 467} 1081--1083 (\textit{Preprint} \eprint{1010.5788})

\bibitem{Drut:2012md}
Drut J~E and Nicholson A~N 2013 {\em J.Phys.\/} {\bf G40} 043101
  (\textit{Preprint} \eprint{1208.6556})

\bibitem{Bulgac:2005pj}
Bulgac A, Drut J~E and Magierski P 2006 {\em Phys.Rev.Lett.\/} {\bf 96} 090404
  (\textit{Preprint} \eprint{cond-mat/0505374})

\bibitem{Magierski:2008wa}
Magierski P, Wlazlowski G, Bulgac A and Drut J~E 2008  (\textit{Preprint}
  \eprint{0801.1504})

\bibitem{Endres:2012cw}
Endres M~G, Kaplan D~B, Lee J~W and Nicholson A~N 2013 {\em Phys.Rev.\/} {\bf
  A87} 023615 (\textit{Preprint} \eprint{1203.3169})

\bibitem{Endres:2011er}
Endres M~G, Kaplan D~B, Lee J~W and Nicholson A~N 2011 {\em Phys.Rev.\/} {\bf
  A84} 043644 (\textit{Preprint} \eprint{1106.5725})

\bibitem{Bulgac:2008zz}
Bulgac A, Drut J~E and Magierski P 2008 {\em Phys.Rev.\/} {\bf A78} 023625
  (\textit{Preprint} \eprint{0803.3238})

\bibitem{Rokash:2013xda}
Rokash A, Epelbaum E, Krebs H, Lee D and Meissner U~G 2013 {\em J.Phys.\/} {\bf
  G41} 015105 (\textit{Preprint} \eprint{1308.3386})

\bibitem{Pine:2013zja}
Pine M, Lee D and Rupak G 2013 {\em Eur.Phys.J.\/} {\bf A49} 151
  (\textit{Preprint} \eprint{1309.2616})

\bibitem{Rupak:2013aue}
Rupak G and Lee D 2013 {\em Phys.Rev.Lett.\/} {\bf 111} 032502
  (\textit{Preprint} \eprint{1302.4158})

\bibitem{Brockmann:1992in}
Brockmann R and Frank J 1992 {\em Phys.Rev.Lett.\/} {\bf 68} 1830--1833

\bibitem{Muller:1999cp}
Muller H, Koonin S, Seki R and van Kolck U 2000 {\em Phys.Rev.\/} {\bf C61}
  044320 (\textit{Preprint} \eprint{nucl-th/9910038})

\bibitem{PhysRevB.66.140504}
Sewer A, Zotos X and Beck H 2002 {\em Phys. Rev. B\/} {\bf 66}(14) 140504
  (\textit{Preprint} \eprint{cond-mat/0204053})
  \urlprefix\url{http://link.aps.org/doi/10.1103/PhysRevB.66.140504}

\bibitem{Chandrasekharan:2003ub}
Chandrasekharan S, Pepe M, Steffen F and Wiese U 2004 {\em
  Nucl.Phys.Proc.Suppl.\/} {\bf 129} 507--509 (\textit{Preprint}
  \eprint{hep-lat/0309093})

\bibitem{Shushpanov:1998ms}
Shushpanov I and Smilga A~V 1999 {\em Phys.Rev.\/} {\bf D59} 054013
  (\textit{Preprint} \eprint{hep-ph/9807237})

\bibitem{Lewis:2000cc}
Lewis R and Ouimet P~P~A 2001 {\em Phys.Rev.\/} {\bf D64} 034005
  (\textit{Preprint} \eprint{hep-ph/0010043})

\bibitem{Borasoy:2003pg}
Borasoy B, Lewis R and Ouimet P~P~A 2004 {\em Nucl.Phys.Proc.Suppl.\/} {\bf
  128} 141--147 (\textit{Preprint} \eprint{hep-lat/0310054})

\bibitem{Borasoy:2007vy}
Borasoy B, Epelbaum E, Krebs H, Lee D and Meissner U~G 2007 {\em Eur.Phys.J.\/}
  {\bf A34} 185--196 (\textit{Preprint} \eprint{0708.1780})

\bibitem{NIJMEGEN}
  Nijmegen Phase Shift Analysis, http://nn-online.org/

\bibitem{Stoks:1993tb}
Stoks V, Kompl R, Rentmeester M and de~Swart J 1993 {\em Phys.Rev.\/} {\bf C48}
  792--815

\bibitem{Epelbaum:2009pd}
Epelbaum E, Krebs H, Lee D and Meissner U~G 2010 {\em Phys.Rev.Lett.\/} {\bf
  104} 142501 (\textit{Preprint} \eprint{0912.4195})

\bibitem{Lahde:2013uqa}
Laehde T~A, Epelbaum E, Krebs H, Lee D, Meissner U~G {\em et~al.\/} 2014 {\em
  Phys.Lett.\/} {\bf B732} 110 (\textit{Preprint} \eprint{1311.0477})

\bibitem{Beane:2002xf}
Beane S~R and Savage M~J 2003 {\em Nucl.Phys.\/} {\bf A717} 91--103
  (\textit{Preprint} \eprint{nucl-th/0208021})

\bibitem{Beane:2002vs}
Beane S~R and Savage M~J 2003 {\em Nucl.Phys.\/} {\bf A713} 148--164
  (\textit{Preprint} \eprint{hep-ph/0206113})

\bibitem{Epelbaum:2002gb}
Epelbaum E, Meissner U~G and Gloeckle W 2003 {\em Nucl.Phys.\/} {\bf A714}
  535--574 (\textit{Preprint} \eprint{nucl-th/0207089})

\bibitem{Braaten:2003eu}
Braaten E and Hammer H 2003 {\em Phys.Rev.Lett.\/} {\bf 91} 102002
  (\textit{Preprint} \eprint{nucl-th/0303038})

\bibitem{Bedaque:2010hr}
Bedaque P~F, Luu T and Platter L 2011 {\em Phys.Rev.\/} {\bf C83} 045803
  (\textit{Preprint} \eprint{1012.3840})

\bibitem{Cheoun:2011yn}
Cheoun M~K, Kajino T, Kusakabe M and Mathews G~J 2011 {\em Phys.Rev.\/} {\bf
  D84} 043001 (\textit{Preprint} \eprint{1104.5547})

\bibitem{Ali:2012dj}
Hossain~Ali M, Jakir~Hossain M and Tariq A~S~B 2013 {\em Phys.Rev.\/} {\bf D88}
  034001 (\textit{Preprint} \eprint{1212.2753})

\bibitem{Coc:2012xk}
Coc A, Descouvemont P, Olive K~A, Uzan J~P and Vangioni E 2012 {\em
  Phys.Rev.\/} {\bf D86} 043529 (\textit{Preprint} \eprint{1206.1139})

\bibitem{Epelbaum:2012iu}
Epelbaum E, Krebs H, Laehde T~A, Lee D and Meissner U~G 2013 {\em Phys. Rev.
  Lett. 110,\/} {\bf 112502} (\textit{Preprint} \eprint{1212.4181})

\bibitem{Epelbaum:2012qn}
Epelbaum E, Krebs H, Laehde T~A, Lee D and Meissner U~G 2012 {\em
  Phys.Rev.Lett.\/} {\bf 109} 252501 (\textit{Preprint} \eprint{1208.1328})

\bibitem{Berengut:2013nh}
Berengut J, Epelbaum E, Flambaum V, Hanhart C, Meissner U~G {\em et~al.\/} 2013
  {\em Phys.Rev.\/} {\bf D87} 085018 (\textit{Preprint} \eprint{1301.1738})

\bibitem{Hoyle:1954zz}
Hoyle F 1954 {\em Astrophys.J.Suppl.\/} {\bf 1} 121--146

\bibitem{Epelbaum:2013wla}
Epelbaum E, Krebs H, Laehde T~A, Lee D and Meissner U~G 2013 {\em
  Eur.Phys.J.\/} {\bf A49} 82 (\textit{Preprint} \eprint{1303.4856})

\bibitem{Chen:1999bg}
Chen J~W and Savage M~J 1999 {\em Phys.Rev.\/} {\bf C60} 065205
  (\textit{Preprint} \eprint{nucl-th/9907042})

\bibitem{Rupak:1999rk}
Rupak G 2000 {\em Nucl.Phys.\/} {\bf A678} 405--423 (\textit{Preprint}
  \eprint{nucl-th/9911018})

\bibitem{Gasser:1980sb}
Gasser J 1981 {\em Annals Phys.\/} {\bf 136} 62

\bibitem{Borasoy:1996bx}
Borasoy B and Meissner U~G 1997 {\em Annals Phys.\/} {\bf 254} 192--232
  (\textit{Preprint} \eprint{hep-ph/9607432})

\bibitem{Toussaint:2009pz}
Toussaint D and Freeman W (MILC Collaboration) 2009 {\em Phys.Rev.Lett.\/} {\bf
  103} 122002 (\textit{Preprint} \eprint{0905.2432})

\bibitem{Young:2009zb}
Young R and Thomas A 2010 {\em Phys.Rev.\/} {\bf D81} 014503 (\textit{Preprint}
  \eprint{0901.3310})

\bibitem{Takeda:2010cw}
Takeda K {\em et~al.\/} (JLQCD collaboration) 2011 {\em Phys.Rev.\/} {\bf D83}
  114506 (\textit{Preprint} \eprint{1011.1964})

\bibitem{Bali:2011ks}
Bali G~S {\em et~al.\/} (QCDSF Collaboration) 2012 {\em Phys.Rev.\/} {\bf D85}
  054502 (\textit{Preprint} \eprint{1111.1600})

\bibitem{Dinter:2012tt}
Dinter S {\em et~al.\/} (ETM Collaboration) 2012 {\em JHEP\/} {\bf 1208} 037
  (\textit{Preprint} \eprint{1202.1480})

\bibitem{Gong:2012nw}
Gong M, Li A, Alexandru A, Draper T and Liu K (xQCD Collaboration) 2011 {\em
  PoS\/} {\bf LATTICE2011} 156 (\textit{Preprint} \eprint{1204.0685})

\bibitem{Jung:2012rz}
Jung C (RBC Collaboration, UKQCD Collaboration) 2012 {\em PoS\/} {\bf
  LATTICE2012} 164 (\textit{Preprint} \eprint{1301.5397})

\bibitem{Oksuzian:2012rzb}
Ohki H {\em et~al.\/} (JLQCD Collaboration) 2013 {\em Phys.Rev.\/} {\bf D87}
  034509 (\textit{Preprint} \eprint{1208.4185})

\bibitem{Engelhardt:2012gd}
Engelhardt M 2012 {\em Phys.Rev.\/} {\bf D86} 114510 (\textit{Preprint}
  \eprint{1210.0025})

\bibitem{Freeman:2012ry}
Freeman W and Toussaint D (MILC Collaboration) 2013 {\em Phys.Rev.\/} {\bf D88}
  054503 (\textit{Preprint} \eprint{1204.3866})

\bibitem{Durr:2011mp}
Duerr S, Fodor Z, Hemmert T, Hoelbling C, Frison J {\em et~al.\/} 2012 {\em
  Phys.Rev.\/} {\bf D85} 014509 (\textit{Preprint} \eprint{1109.4265})

\bibitem{Horsley:2011wr}
Horsley R {\em et~al.\/} (QCDSF-UKQCD Collaborations) 2012 {\em Phys.Rev.\/}
  {\bf D85} 034506 (\textit{Preprint} \eprint{1110.4971})

\bibitem{Semke:2012gs}
Semke A and Lutz M 2012 {\em Phys.Lett.\/} {\bf B717} 242--247
  (\textit{Preprint} \eprint{1202.3556})

\bibitem{Shanahan:2012wh}
Shanahan P, Thomas A and Young R 2013 {\em Phys.Rev.\/} {\bf D87} 074503
  (\textit{Preprint} \eprint{1205.5365})

\bibitem{Ren:2012aj}
Ren X~L, Geng L, Martin~Camalich J, Meng J and Toki H 2012 {\em JHEP\/} {\bf
  1212} 073 (\textit{Preprint} \eprint{1209.3641})

\bibitem{Junnarkar:2013ac}
Junnarkar P and Walker-Loud A 2013 {\em Phys.Rev.\/} {\bf D87} 114510
  (\textit{Preprint} \eprint{1301.1114})

\bibitem{Menendez:2012tm}
Menendez J, Gazit D and Schwenk A 2012 {\em Phys.Rev.\/} {\bf D86} 103511
  (\textit{Preprint} \eprint{1208.1094})

\bibitem{Fitzpatrick:2012ix}
Fitzpatrick A~L, Haxton W, Katz E, Lubbers N and Xu Y 2013 {\em JCAP\/} {\bf
  1302} 004 (\textit{Preprint} \eprint{1203.3542})

\bibitem{Beane:2013kca}
Beane S, Cohen S, Detmold W, Lin H~W and Savage M 2013  (\textit{Preprint}
  \eprint{1306.6939})

\bibitem{schrock1982}
Rao S and Shrock R 1982 {\em Phys. Lett. B\/} {\bf 116} 238

\bibitem{Buchoff:2012bm}
Buchoff M~I, Schroeder C and Wasem J 2012 {\em PoS\/} {\bf LATTICE2012} 128
  (\textit{Preprint} \eprint{1207.3832})

\bibitem{Lepage89}
Lepage G~P 1989  Invited lectures given at TASI'89 Summer School, Boulder, CO,
  Jun 4-30, 1989

\bibitem{MJSsign}
Savage M~J 2010  {private communication.}

\bibitem{Lee:2011sm}
Lee J~W, Endres M~G, Kaplan D~B and Nicholson A~N 2011 {\em PoS\/} {\bf
  LATTICE2011} 203 (\textit{Preprint} \eprint{1111.3793})

\bibitem{Endres:2011jm}
Endres M~G, Kaplan D~B, Lee J~W and Nicholson A~N 2011 {\em Phys.Rev.Lett.\/}
  {\bf 107} 201601 (\textit{Preprint} \eprint{1106.0073})

\bibitem{Endres:2011mm}
Endres M~G, Kaplan D~B, Lee J~W and Nicholson A~N 2011 {\em PoS\/} {\bf
  LATTICE2011} 017 (\textit{Preprint} \eprint{1112.4023})

\bibitem{Grabowska:2012ik}
Grabowska D, Kaplan D~B and Nicholson A~N 2013 {\em Phys.Rev.\/} {\bf D87}
  014504 (\textit{Preprint} \eprint{1208.5760})

\bibitem{Detmold:2014hla}
Detmold W and Endres M~G 2014  (\textit{Preprint} \eprint{1404.6816})

\bibitem{Yamazaki:2013rna}
Yamazaki T, Ishikawa K~I, Kuramashi Y and Ukawa A 2013  (\textit{Preprint}
  \eprint{1310.5797})

\bibitem{Morningstar:2013bda}
Morningstar C, Bulava J, Fahy B, Foley J, Jhang Y {\em et~al.\/} 2013 {\em
  Phys.Rev.\/} {\bf D88} 014511 (\textit{Preprint} \eprint{1303.6816})

\bibitem{Foley:2012wb}
Foley J, Bulava J, Jhang Y~C, Juge K~J, Lenkner D {\em et~al.\/} 2011 {\em
  PoS\/} {\bf LATTICE2011} 120 (\textit{Preprint} \eprint{1205.4223})

\bibitem{Smigielski:2008pa}
Smigielski B and Wasem J 2009 {\em Phys.Rev.\/} {\bf D79} 054506
  (\textit{Preprint} \eprint{0811.4392})

\bibitem{Chen:2012rp}
Chen H~X and Oset E 2013 {\em Phys.Rev.\/} {\bf D87} 016014 (\textit{Preprint}
  \eprint{1202.2787})

\bibitem{Albaladejo:2013bra}
Albaladejo M, Rios G, Oller J and Roca L 2013  (\textit{Preprint}
  \eprint{1307.5169})

\bibitem{Sasaki:2013vxa}
Sasaki K, Ishizuka N, Oka M and Yamazaki T (PACS-CS Collaboration) 2014 {\em
  Phys.Rev.\/} {\bf D89} 054502 (\textit{Preprint} \eprint{1311.7226})

\bibitem{Sharpe:1998xm}
Sharpe S~R and Singleton Robert~L J 1998 {\em Phys.Rev.\/} {\bf D58} 074501
  (\textit{Preprint} \eprint{hep-lat/9804028})

\bibitem{Rupak:2002sm}
Rupak G and Shoresh N 2002 {\em Phys.Rev.\/} {\bf D66} 054503
  (\textit{Preprint} \eprint{hep-lat/0201019})

\bibitem{Aoki:2003yv}
Aoki S 2003 {\em Phys.Rev.\/} {\bf D68} 054508 (\textit{Preprint}
  \eprint{hep-lat/0306027})

\bibitem{Bar:2003mh}
Bar O, Rupak G and Shoresh N 2004 {\em Phys.Rev.\/} {\bf D70} 034508
  (\textit{Preprint} \eprint{hep-lat/0306021})

\bibitem{Sharpe:2004ny}
Sharpe S~R and Wu J~M 2005 {\em Phys.Rev.\/} {\bf D71} 074501
  (\textit{Preprint} \eprint{hep-lat/0411021})

\bibitem{Chen:2005ab}
Chen J~W, O'Connell D, Van~de Water R~S and Walker-Loud A 2006 {\em
  Phys.Rev.\/} {\bf D73} 074510 (\textit{Preprint} \eprint{hep-lat/0510024})

\bibitem{Chen:2006wf}
Chen J~W, O'Connell D and Walker-Loud A 2007 {\em Phys.Rev.\/} {\bf D75} 054501
  (\textit{Preprint} \eprint{hep-lat/0611003})

\bibitem{Aoki:2008gy}
Aoki S, Bar O and Biedermann B 2008 {\em Phys.Rev.\/} {\bf D78} 114501
  (\textit{Preprint} \eprint{0806.4863})

\bibitem{Hansen:2011mc}
Hansen M~T and Sharpe S~R 2012 {\em Phys.Rev.\/} {\bf D85} 054504
  (\textit{Preprint} \eprint{1112.3998})

\bibitem{Blum:2014oka}
Blum T, Chowdhury S, Hayakawa M and Izubuchi T 2014  (\textit{Preprint}
  \eprint{1407.2923})

\bibitem{Benayoun:2014tra}
Benayoun M, Bijnens J, Blum T, Caprini I, Colangelo G {\em et~al.\/} 2014
  (\textit{Preprint} \eprint{1407.4021})

\bibitem{Beane:2014qha}
Beane S~R and Savage M~J 2014  (\textit{Preprint} \eprint{1407.4846})

\end{thebibliography}

%%%%%%%%%%%%%%%%%%%%%%%%%%%%%%%%%%%%%%%%%%%%%%%%%%%%%%%%%%%%

\end{document}